\RequirePackage{fix-cm}
\documentclass[smallextended]{svjour3}    % onecolumn (ditto)
\smartqed  % flush right qed marks, e.g. at end of proof
\usepackage{graphicx}
\usepackage{url}
\usepackage{latexsym}
\usepackage{color}
\usepackage[round]{natbib}
\usepackage{txfonts}
\usepackage{amssymb}
\newcommand{\fpath}{./}
\definecolor{aurometalsaurus}{rgb}{0.43, 0.5, 0.5}
\definecolor{ao(english)}{rgb}{0.0, 0.5, 0.0}
\definecolor{seagreen}{rgb}{0.180392,0.545098,0.341176}
\definecolor{forestgreen}{rgb}{0.133333,0.545098,0.133333}

\newcommand{\white}{\color{white}}

\newcommand{\mod}{}
\newcommand{\ie}{i.\,e.}
\newcommand{\eg}{e.\,g.}
\newcommand{\etal}{et al.}

\DeclareMathAlphabet{\mathitbf}{OML}{cmm}{b}{it}

\newcommand{\dotscp}{DOT}

\newcommand{\goes}{{\it GOES}}
\newcommand{\hinode}{{\it Hinode}}

\newcommand{\nso}{NSO}
\newcommand{\kpno}{KPNO}
\newcommand{\skylab}{{\it Skylab}}
\newcommand{\soho}{{\it SOHO}}
\newcommand{\sdo}{{\it SDO}}
\newcommand{\solis}{{\it SOLIS}}
\newcommand{\so}{{\it Solar Orbiter}}
\newcommand{\sola}{{\it Solar-A}}
\newcommand{\solb}{{\it Solar-B}}
\newcommand{\solc}{{\it Solar-C}}
\newcommand{\sunrise}{{\it Sunrise}}

\newcommand{\stereo}{{\it STEREO}}
\newcommand{\trace}{{\it TRACE}}
\newcommand{\wind}{{\it WIND}}

\newcommand{\yohkoh}{{\it Yohkoh}}
\newcommand{\farcs}{\mbox{\ensuremath{.\!\!^{\prime\prime}}}}%  % fractional arcsecond symbol: 0.''0
\newcommand{\arcs}{\mbox{\ensuremath{^{\prime\prime}}}}
\newcommand{\asr}{Adv. Space Res.}

\newcommand{\araa}{Ann. Rev. Astron. Astrophys.}

\newcommand{\aap}{Astron. Astrophys.}
\newcommand{\aapr}{Astron. Astrophys. Rev.}
\newcommand{\aaps}{Astron. Astrophys. Suppl. Ser.}

\newcommand{\arep}{Astron. Reports}
\newcommand{\ajp}{Australian J. Phys.}
\newcommand{\apj}{Astrophys.~J.}
\newcommand{\apjl}{Astrophys.~J. Lett.}

\newcommand{\apss}{Astrophys. Space Sci.}
\newcommand{\aspcs}{Astron. Soc. Pac. Conf. Ser.}

\newcommand{\esasp}{ESA Special Publication}

\newcommand{\gafd}{Geophys. Astrophys. Fluid Dyn.}

\newcommand{\grl}{Geophys. Res. Lett.}

\newcommand{\jfm}{J. Fluid Mech.}
\newcommand{\jgr}{J. Geophys. Res.}
\newcommand{\jgrsp}{J. Geophys. Res. (Space Phys.)}

\newcommand{\mnras}{Mon. Not. Roy. Astr. Soc.}

\newcommand{\nat}{Nature}

\newcommand{\lrsp}{Living Rev. Sol. Phys.}
\newcommand{\pasj}{Pub. Astron. Soc.~Jap.}
\newcommand{\pasp}{Pub. Astron. Soc. Pac.}

\newcommand{\pht}{Phil. Trans. Roy. Soc. Lond. A}
\newcommand{\pnas}{Proc. Nat. Acad. Sci.}

\newcommand{\prl}{Phys. Rev. Lett.}

\newcommand{\rmp}{Rev. Mod. Phys.}
\newcommand{\rpp}{Rep. Prog. Phys.}
\newcommand{\raa}{Res. Astron. Astrophys.}
\newcommand{\solphys}{Sol. Phys.}
\newcommand{\sci}{Science}
\newcommand{\ssr}{Space Sci. Rev.}

\journalname{Astron. Astrophys. Rev.}
\begin{document}

\title{The Magnetic Field in the Solar Atmosphere}
% \subtitle{Do you have a subtitle?\\ If so, write it here}
\author{Thomas Wiegelmann  \and Julia K.~Thalmann \and Sami K.~Solanki}
% \authorrunning{Wiegelmann \& Thalmann et al.} % if too long for running head
\institute{Thomas Wiegelmann \at
Max-Planck-Institut f\"ur Sonnensystemforschung, Justus-von-Liebig-Weg 3, 37077 G\"ottingen, Germany \\
 Tel.: +49-551-384-979-155\\
 Fax: +49-551-384-979-240\\
\email{wiegelmann@mps.mpg.de} \and Julia K. Thalmann \at
Institute for Physics/IGAM, University of Graz, Universit\"atsplatz 5/II, 8010, Graz, Austria \\
Tel.: +43-316-3808599\\
Fax: +49-316-3809825\\
\email{julia.thalmann@uni-graz.at} \and Sami K. Solanki \at
Max-Planck-Institut f\"ur Sonnensystemforschung, Justus-von-Liebig-Weg 3, 37077 G\"ottingen, Germany \\
Tel.: +49-551-384-979-552\\
Fax: +49-551-384-979-190\\
\email{solanki@mps.mpg.de} \\
and \at
School of Space Research, Kyung Hee University,
Yongin, Gyeonggi 446-701, Republic of Korea
 }

\date{Draft version: \today}

\maketitle

\begin{abstract}
This publication provides an overview of magnetic fields in
the solar atmosphere with the focus lying on the corona. The solar magnetic field
couples the solar interior with the visible surface of the Sun and with its
atmosphere. It is also responsible for all solar activity in its numerous
manifestations. Thus, dynamic phenomena such as coronal mass ejections and
flares are magnetically driven. In addition, the field also plays a crucial
role in heating the solar chromosphere and corona as well as in accelerating
the solar wind. Our main emphasis is the magnetic field in the upper solar
atmosphere so that photospheric and chromospheric magnetic  structures
are mainly discussed where relevant for higher solar layers. Also, the
discussion of the solar atmosphere and activity is limited to those topics of direct
relevance to the magnetic field. After giving a brief overview about the
solar magnetic field in general and its global structure, we discuss in more
detail the magnetic field in active regions, the quiet Sun and
coronal holes.
\end{abstract}

\keywords{Sun \and Photosphere \and Chromosphere \and Corona  \and Magnetic
Field \and Active Region \and Quiet Sun \and Coronal Holes}

\tableofcontents

\section{Introduction}
\label{s:intro}

In order to understand the physical processes in the solar interior,
its atmosphere as well as the interplanetary environment (including
``space weather'' close to Earth), a detailed knowledge of the temporal and
spatial properties of the magnetic field is essential.
This is because the magnetic field is the
link between everything, from the Sun's interior to the outer edges of
our solar system. The magnetic field is created in the solar interior, can be
measured with highest accuracy on the Sun's visible surface (the photosphere)
and controls most physical processes in the solar atmosphere.
Within this review, we aim to give an overview of the magnetic
coupling from the solar surface to the Sun's upper atmosphere, with
special emphasis on the structure and evolution of the coronal magnetic
field. Magnetic features in the photosphere
are discussed if they cause a coronal response.

The techniques and challenges of
measuring the magnetic field throughout
the atmosphere are not discussed here, but are covered by earlier
reviews \citep[][]{rao_05,whi_05,vdg_cul_09,car_09,2013A&ARv..21...66S}.
Outside the scope of this paper is
the generation of the solar magnetic field by dynamo
processes \citep[for comprehensive reviews see][]{oss_03b,lrsp-2010-3}.
For an in-depth discussion of the observational patterns
resulting at photospheric levels from the dynamics in the Sun's
convection zone we refer to
\cite{1987ARA&A..25...83Z} and \cite{1994smf..conf..301S}.
We also do not discuss the role of the magnetic field and related
physical processes far away from the Sun
(beyond the solar corona) and its transport to those places.
Here, we refer the interested reader to specialized reviews on the solar wind
\citep[][]{mar_06,ofm_10,bru_car_13}, space weather \citep[][]{schw_06},
and the heliospheric magnetic field \citep[][]{owe_for_13}.

We start our review by giving an introduction to
 the most important magnetic aspects of the lower solar atmosphere,
including the photosphere (section~\ref{ss:photo}), chromosphere
(section~\ref{ss:chromo}) and the corona (section~\ref{ss:corona}).
The magnetic coupling between these layers is discussed in
section~\ref{ss:coupling}.
An overview on the currently most widely used local and global model
approaches to assess the coronal magnetic field is given in
section~\ref{s:modeling}. In the remaining
sections, we provide more detailed descriptions of what we know to date about
the coronal magnetic field's structure
in different parts of the Sun's atmosphere, starting with the magnetic field
on global scales (section~\ref{s:global}),
in active and quiet-Sun regions (sections~\ref{s:active} and \ref{s:quiet},
respectively). Finally, we review the magnetic aspects of
coronal holes in section~\ref{s:chs} and provide a summary and outlook in
section~{\ref{s:summary}}.

In most cases, we restrict ourselves to mentioning whether the discussed results
were obtained from the analysis of directly
measured magnetic fields or inferred from modelling.
For further reading, we want to draw the reader's attention to classical overviews
of the theoretical aspects of solar magnetism by \cite{1979cmft.book.....P}
and \cite{pri_82,2014masu.book.....P},
as well as previous descriptions dedicated to aspects of the
magnetic properties of the Sun's magnetic field by \cite{sol_inh_06}.
We also refer to \cite{2000ssma.book.....S} for a comparative work on the
magnetic activity of the Sun and other stars.

Abbreviations used throughout this manuscript are defined in
Appendix~\ref{app:abbreviations}.

\subsection{Photosphere}
\label{ss:photo}

The photosphere contains the visible solar surface and vertically spans
about 500~km of the solar atmosphere, where the temperature
decreases from about 6\,000~K at the bottom of the photosphere to
about 4\,000~K \citep[temperature minimum;][]{fou_04}.
In these layers, due to the momentum gained on its journey towards
the surface, the convective material of the Sun's interior overshoots
into the solar atmosphere, which is stable against thermal convection.
% The momentum gained by the upwelling subsurface material on its
% journey towards the surface makes such material overshoot
% into the atmosphere, which is stable against thermal convection.
Only after passing a distance comparable to the density scale height does it
eventually turn over to form lanes of down flowing material
(see reviews by \citet{nor_ste_09} and \citet{2012LRSP....9....4S}).
As a consequence, the photosphere reveals a granular
pattern comprised of ascending warmer gas in the centers of the granules
and descending cooler gas in the intergranular lanes separating them.
In contrast to the layers below the solar surface, in the atmosphere the energy is
dominantly transported by radiation rather than convection.

\subsubsection{Magnetic flux emergence}
\label{sss:emergence}

A significant part of the properties of the photospheric magnetic features
is determined by the amount of magnetic flux carried by the
${\rm\Omega}$-loops that rise through the convection zone towards the
solar surface.
The largest of these loops may form large bipolar ARs that
harbour sunspots or sunspot groups \citep[][]{dur_88}. Large sunspots
and sunspot groups have magnetic fluxes of $\approx$\,$10^{21}$~Mx and
$10^{22}$~Mx, respectively
\citep[][]{pri_82,2014masu.book.....P},
and are responsible for a great part of the Sun's activity
(see section \ref{s:active} for details).
Much of the flux in ARs that is not in the form of sunspots is
organized in magnetic concentrations (much) smaller than spots. Either
in the form of pores or, most commonly, magnetic elements.
Magnetic pores, sunspot-like features that are
characterized by the absence of a penumbra, carry fluxes of some
$10^{20}$~Mx to $10^{21}$~Mx \citep[][]{tho_wei_04,sob_dmo_12}.
Magnetic elements within ARs carry fluxes of $10^{18}$~Mx to $10^{20}$~Mx
\citep[][]{abr_lon_05}. Note that it is unclear, however, whether the larger flux
features observed by \cite{abr_lon_05} are indeed bright magnetic elements, or possibly
darker features such as protopores.

Smaller rising ${\rm\Omega}$-loops result in the formation of smaller ARs until a lower limit of
roughly $10^{20}$~Mx. Below that we generally speak of ``ephemeral regions''
($10^{18}$~Mx to $10^{20}$~Mx). Even smaller are the smallest
so far resolved bipolar features, the internetwork magnetic loops
\citep[][]{mar_col_07,cen_soc_07,mar_bel_09,dan_bee_10} which
emerge throughout the QS (although preferring a meso-scale pattern).
They have fluxes of roughly $10^{16}$~Mx to $10^{17}$~Mx
\citep[][]{lin_rim_99} and display in general weak
equipartition (that is, the magnetic energy density is similar to
the kinetic energy density of the convective flows)
intrinsic fields.
% of approximately 300~G to 500~G.
Occasionally, these weak fields may be intensified
% to form kG features (1kG=1\,000~G)
due to a convective collapse \citep{par_78b,spr_79}.
The latter amplifies the magnetic field in intergranular
downflow regions due to the combined effect of enhanced cooling
of the intergranular plasma (due to the transport of flux by the horizontal
granular flows into this region) and the super-adiabatic stratification of
the ambient plasma. In small flux concentrations, however, radiative
energy exchange may be able to considerably slow down the cooling of
the downflow material so that the collapse is prohibited and the gross part
of this field remains relatively weak
\citep[see][and section~\ref{s:quiet} for further details]
{ven_86,sol_zuf_96,gro_sch_98}.

It is interesting to note that although each AR typically carries 100 times as
much flux as an ephemeral region, the number of ephemeral regions
appearing on the solar surface over a solar cycle outnumbers that of
ARs by a factor of $10^4$, so that the ephemeral regions bring
roughly 100 times more magnetic flux to the solar surface than ARs.
Similarly, ephemeral regions carry roughly 100 times as much flux as a
typical internetwork feature but all internetwork features appearing over
a solar cycle together provide roughly 100 times more magnetic flux
\citep[][and note that this is partly offset by the much lower lifetime of
the smaller magnetic bipolar features]{zir_87}.
Altogether, the number of magnetic features with a certain amount of flux
follows a power law distribution with an exponent of $-1.85$ \citep{par_def_09},
which is close to $-2.0$ found by \cite{har_zwa_93}.
The latter means that, at any given time, small and
large magnetic regions contribute a similar magnetic flux.

\subsubsection{Spatial properties of magnetic features}
\label{sss:prop_photo}

The different types of bipolar features have rather different properties.
The ARs are largely restricted to the activity belts
\citep[\ie, within approximately $\pm30^\circ$ around the solar equator;
see][]{hal_nic_25}.
Their constituent sunspots are more or
less E-W aligned with a certain tilt, with respect to the exact E-W direction
\citep[corresponding to Joy's law; see][]{hal_ell_19}. This tilt increases with
increasing latitude \citep[][]{1995ApJ...441..886C,2012ApJ...758..115L}
and seems to be inversely correlated to the strength
of the upcoming solar cycle \citep[][]{2010A&A...518A...7D}.
Variation of the number of sunspots with time is often used as a measure
of the solar cycle.
Lifetimes of sunspots vary over a range of periods, with the larger
ones living for months \citep[][]{pet_vdg_97}.
ARs have been reported to have a tendency to emerge near existing ARs
forming so-called active longitudes
\citep[][]{iva_07}, although there has been
controversy regarding their reality
(see section~\ref{s:global} for a more detailed discussion).

Despite being preferentially concentrated around the activity belts
\citep[][]{har_mar_73,mar_88}, ephemeral
regions appear over a much larger fraction
of the solar surface \citep[][]{yan_zha_14},
indicating that they may be generated by a local rather than global dynamo
process. Without observations of the poles, however, this claim is not tenable
% (see section \ref{pchs} for polar observations with \hinode).
(see section \ref{pchs} for further details). They live
for hours to days and display a much tendency to align with the E-W orientation
than ARs.
They may even not have such a trend at all \citep[][]{hag_schr_03,yan_zha_14}.
Their number varies much less over a solar cycle than that of ARs and there
are inconsistent results regarding whether their number varies in phase or in anti-phase
with the solar cycle \citep[][]{mar_har_79,mar_88,hag_schr_03}.

Whereas the location of ARs and ephemeral regions are determined
mainly by the latitudes and longitudes of emergence, the spatial distribution of
other magnetic features, such as the magnetic network, is also influenced by the
transport of magnetic flux at the solar surface by a variety of flows.
The properties of the magnetic network changes in the course of the solar cycle:
around solar minimum it is weak and consists mainly of mixed polarities,
except near the poles which are essentially unipolar regions (and with each
pole having a different polarity).
Around solar maximum the mixed polarity regions are
augmented by large unipolar regions up to solar latitudes of about 60$^\circ$
which are the decay products of old ARs.
Finally, the internetwork fields appear all over the Sun, including also the
interior of ARs. Individual internetwork elements live only for minutes to hours
and they show no preference for any particular orientation
\citep[][and references therein]{dew_ste_09}. They display no
dependence on the solar cycle to the extent that can be tested so far
\citep[][]{2013A&A...555A..33B}.

\subsubsection{Origin of internetwork fields}
\label{qs:dynamo}

There has been considerable debate concerning the origin of internetwork
fields. One proposal regarding their origin is that they are either
the consequence of a recycling of magnetic flux from
ephemeral regions, or are the result of convection acting upon ARs,
tearing flux away and recycling it over time \citep[][]{2001ASPC..236..363P}.
That basically implies that they are composed of
flux produced by the global dynamo being one possibility and
magnetic flux produced by a local dynamo being another \citep[][and
for a review see \cite{2013SSRv..178..141M}]{2007A&A...465L..43V,
2008A&A...481L...5S,2013A&A...555A..33B,2012A&A...547A..93S}.
It is still an open question whether the quiet Sun's magnetic field is created
mainly by the global dynamo or a local turbulent dynamo.
One possibility
to investigate this is the latitude dependence, where the global dynamo
would likely lead to a significantly different distribution of quiet-Sun areas
as a function of latitude, while the action of a local dynamo would not.
Another approach is to trace quiet-Sun
regions in time over a solar cycle. While a global dynamo would lead to a
significant change as the cycle progresses, a local dynamo would not.
This approach was applied by \cite{2013A&A...555A..33B}, based on circular
and linear polarization signals measured with \hinode/SOT-SP
during about half a solar cycle (during the years 2006 to 2012).
No significant changes, in both linear and circular polarization, were found,
in particular for magnetic features with a LOS magnetic flux of less than
$10^{19}$~Mx. Thus, their results are
favoring a local turbulent dynamo, at least for the creation of weak
internetwork fields, and supporting what has been suspected in
earlier studies \citep[][and references therein and
see also section~\ref{PCH:QS} for the importance of a local dynamo in
the Sun's polar regions]{2003A&A...411..615S}.

% \begin{figure}
%       \centering
%       \includegraphics[width=\textwidth,draft]{\fpath bfly_hathaway}
%       \put(-340,165){\bf\sf(a)}
%       \put(-340,80){\bf\sf(b)}
%       \caption{(a) Daily observations of sunspot area since 1874, averaged over
%       individual solar rotations. The sunspot area in equal-area latitude strips
%       (vertical axis) is shown as a function of time (horizontal axis). The color
%       (code indicates the area covered by a sunspot or a group of sunspots
%       located within a certain latitude strip. (b) Average area of the visible solar
%       surface covered by sunspots (\% of the visible hemisphere; vertical axis)
%       as a function of time (horizontal axis). White numbers above the time axis
%       indicate the number of the solar cycle.
%       (Courtesy of D.~Hathaway/NASA/ARC. Reproduced by permission.)
%       }
%       \label{fig:hathaway}
% \end{figure}

\subsubsection{Temporal evolution of the magnetic field}
\label{sss:evolution_photo}

The emergence of magnetic flux ropes from below the surface within
ARs is usually followed by the growth and separation of the opposite polarity
patches. Most commonly, loop footpoints move apart almost linearly
with time \citep[][]{cen_soc_07}. But also more complex motions
such as a circular ones are possible, although only if the emerging loop
possesses a writhe or a twist \citep[][]{bug_mar_12}. Then, physical
long-term \citep[][]{2003A&A...397..305L} and apparent
short-term rotational motions \citep[][]{lop_dem_00,luo_dem_11}
of the opposite polarity patches are usually observed.
And also apparent shearing and rotational motions have been noted
\citep[][]{gib_fan_04,liu_zha_06}.
(Note that whenever we speak of shearing without any specification,
a horizontal motion, \ie, parallel to the solar surface, is referred to.)
Sunspots also can show an apparent rotational motion around
their center shortly after emergence. The related coronal
magnetic loops (which magnetically connect the rotating sunspots) are often
twisted and visible as sigmoid structures in coronal images \citep{bro_nig_03}.
Coronal structures above rotating sunspots are also prone to cause flaring activity.

Once the ${\rm\Omega}$-loops have emerged, the enhanced magnetic
field at their footpoints (the magnetic patches) interact with the convection
in different ways \citep[][]{schr_tit_97}.
At the beginning, the magnetic field is generally roughly in
equipartition with the flows (typically granular flows).
At the solar surface this corresponds to field strengths of 300~G to 500~G.
Once the field has emerged, it gets concentrated to form kG (kilogauss)
features by its interaction with convective
flows \citep{par_78a,2008ApJ...677L.145N,2010A&A...513A...1D}.
Recent studies suggest that the concentration of the field can be
followed by a weakening of the field and that this can cycle multiple times
\citep[][]{2014ApJ...789....6R}.
The flows also move the magnetic features around, causing each to carry
out a random walk, although the exact nature of the motion can differ,
depending on the location of the magnetic feature
\citep[][]{abr_car_11,jaf_cam_14}.

\begin{figure}
      \centering
      \includegraphics[width=0.9\textwidth]{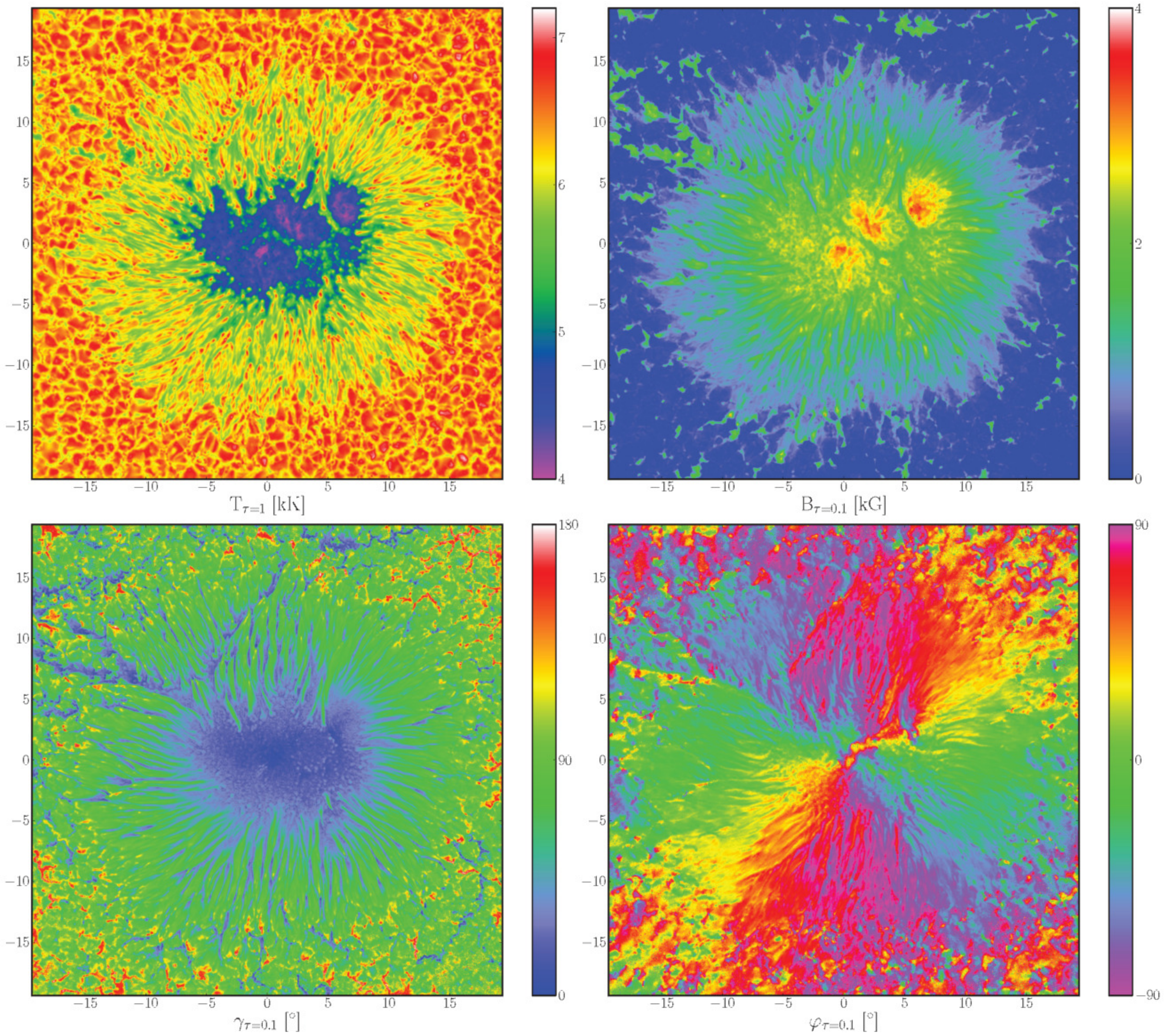}
      \put(-307,260){\bf\sf(a)}
      \put(-307,125){\bf\sf(b)}
      \put(-157,260){\bf\sf(c)}
      \put(-157,125){\bf\sf(d)}
      \caption{Structure of the magnetic field and the temperature of a relatively symmetric
      sunspot recorded by the \hinode/SOT-SP on January 2007 which scanned the
      sunspot area from 12:36~UT to 13:00~UT. Plotted are maps of the (a)
      temperature, (b) field strength, (c) field inclination and (d) azimuth. The fine-scale
      structures of this sunspot have been analysed by \cite{2013A&A...557A..25T} and
      \cite{2013A&A...557A..24V}.
      }
      \label{fig:sunspot}
\end{figure}
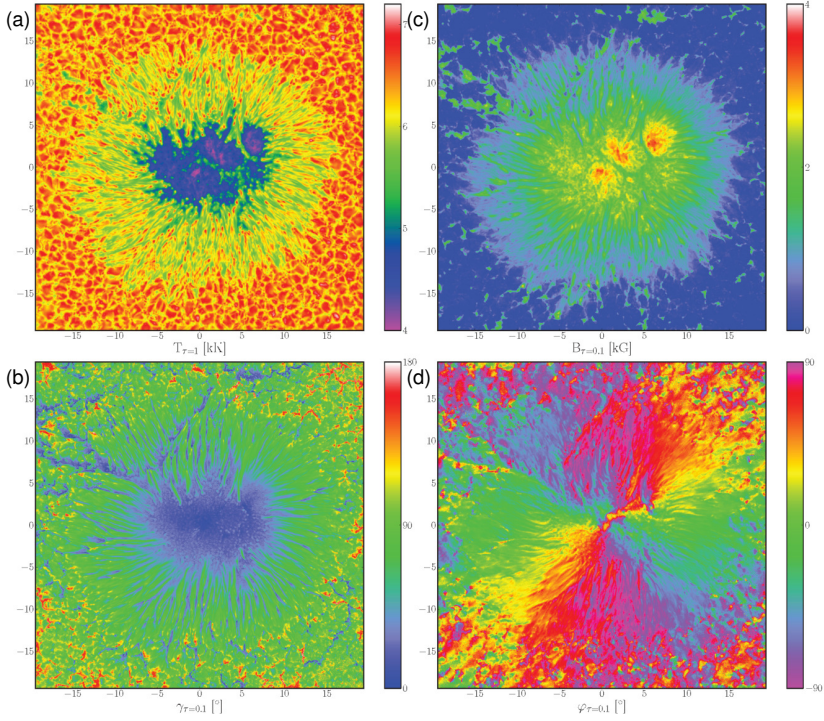

The random walk of the magnetic patches, imposed by the convective motions,
necessarily leads to the encounter of opposite-polarity fields.
These, in the case of smaller flux tubes,
often do not correspond to the other footpoint of the original
${\rm\Omega}$-loop, so that in larger ARs a fair amount of cancellation
takes place \citep{liv_wan_85}. When fields of same polarity meet, larger
flux concentrations are (sometimes only temporarily) formed
\citep{1984ssdp.conf...30M}.
Only if there is enough magnetic flux of a single polarity present,
then part of it coalesces into a sunspot.
Proper sunspots consist of a central umbra and a surrounding penumbra
(see Figure~\ref{fig:sunspot}).
The latter is a filamentary structure of weaker, more
horizontal magnetic field which surrounds the more vertically oriented
stronger umbral magnetic field \citep[for reviews see][]{sol_03,bor_ich_11}.
Typically, magnetic field strengths of about 1~kG
are found in penumbrae while the maximum umbral field strengths
usually range between 2~kG and 4~kG \citep{tit_fra_93,lit_elm_93,scha_13}.
In extreme cases, values as large as 6~kG have been reported
\citep[][]{liv_har_06}. Only recently, \cite{2013A&A...557A..24V} reported
$\approx$\,7~kG in a sunspot, although surprisingly not in the umbral area but
near the outer edge of the penumbra, in a strong downflow region.
Sunspots have diameters between 3~Mm (megameter) and 60~Mm and live
for a few hours to months \citep[][and see also the
review by \cite{sol_03}]{bra_lou_64}, with the lifetime
being linearly correlated to the maximum area covered \citep{wal_55}.
Sunspots grow rapidly after their emergence, soon reach a maximum
size and decay slowly afterward.

Sunspots are often preceded, accompanied and followed by
``faculae'' (called ``plage'' at chromospheric
levels) which have a spatially averaged field strength of typically between
100~G and 500~G \citep[][]{tit_top_92} and are composed of
magnetic elements of a range of sizes, with comparatively
field-free or weak-field gas in between. Faculae tend to surround and generally
outlive the sunspots by a significant amount. Consequently, an old AR is generally
composed of faculae only which then decay and disperse to form enhanced
network fields. The flux in ARs that is not in the form of sunspots is
concentrated in either pores or magnetic
elements. The properties of pores include diameters of some Mm
and field strengths of 1~kG to 2~kG
\citep[][]{tho_wei_04,sob_dmo_12}. Magnetic elements
have diameters smaller than $\approx$\,350~km and exhibit field strengths of
1~kG to 2~kG \citep[][]{ste_73,1992ApJ...391..832R,1992A&A...263..323R}.
They are the chief magnetic constituents of faculae, are bright
(\ie, hotter than their surroundings) particularly in the
mid-photosphere and above \citep[see reviews by][]{sol_93,sol_inh_06} and are present even in the internetwork
\citep[][]{lag_sol_10}.

Sooner or later, larger magnetic features (\eg, sunspots) break up and
dissolve, their fragments becoming subject to transport and distortion by
the convective flows.
The smallest and most dynamic convective elements in the QS are granules.
Granules have typically diameters of
500~km to 1.5~Mm, a single turnover time of a few minutes and lifetimes
of minutes \citep[][]{nor_ste_09,zho_wan_10}.
Roughly, the turnover time is the time it takes for hot matter to be transported
up through the solar surface, cooled there and transported down again in an
intergranular lane, while the lifetime is the time over which a given granule
maintains its identity (\eg, in a series of images of the solar surface).
In the QS, the magnetic field is additionally swept to the edges of
supergranular cells \citep[for a review see][]{rie_rin_10} with
typical diameters of 20~Mm to 30~Mm. This happens on a timescale of several
hours and leads to the formation of a patchy magnetic network outlining the
boundaries of the supergranular cells.

The transport of magnetic flux to the edges of the granular and supergranular
cells leads to an enhancement of magnetic flux if the accumulated flux is
of the same polarity. Only when magnetic elements of opposite
polarity meet, do they (partially) cancel. In fact, the most significant
process of disappearance of magnetic flux appeared to be the
cancellation of magnetic elements of opposite polarity
\citep{liv_wan_85}.
\cite{wan_shi_88} concluded that the flux cancellation occurs as a
consequence of magnetic reconnection in or above the photosphere, which
is likely due to the expansion of the field, so that the opposite polarities
meet mainly in the upper photosphere \citep[][]{2011A&A...533A..86C}.
However, a recent study of \cite{2013ApJ...774..127L} suggests that at least in the
QS, flux dispersal is the more common route by which magnetic elements
are destroyed, although the exact physical process of flux removal could not
be studied (dissipation at small spatial scales is likely to play a role).
Another explanation for the apparent disappearance of magnetic flux
is that the continuous buffeting of the magnetic flux concentrations
leads to the fragmentation of some of the flux into entities whose lesser
magnetic flux may then be below the detection threshold of
a particular instrument \citep[][]{1996ApJ...463..365B}.

\subsubsection{Relative importance of magnetic forces}
\label{sss:forces_photo}

Typical values for the particle number density in the solar
photosphere are on the order of $n\approx10^{23}$~m$^{-3}$
\citep[at the temperature minimum;][]{fou_04}.
Typical quiet-Sun and active-region
 magnetic field strengths cover the range
100~G to 2~kG. As a consequence, the ratio of the plasma pressure to the
magnetic pressure (usually referred to as plasma-$\beta$, or simply denoted
by $\beta$)  is on the order of 1 to 10
\citep[when averaged over larger regions,][]{gar_01}.
Note that a value of $\beta\gtrsim1$ implies that the
pressure exerted by the plasma is higher than that exerted by the magnetic
field, \ie, that the plasma motion controls the dynamics (and the
photosphere is therefore generally said to be ``non force-free'').
Locally, however, due to the evacuation of magnetic features values
of $\beta<1$ are often found.
Consequently plasma pressure forces might not be dominant everywhere
in the photosphere. Sunspots and kG magnetic
elements (for instance at supergranular boundaries) likely represent such
exceptions \citep[][]{pri_82}.

\begin{figure}
      \centering
      \includegraphics[width=\textwidth]{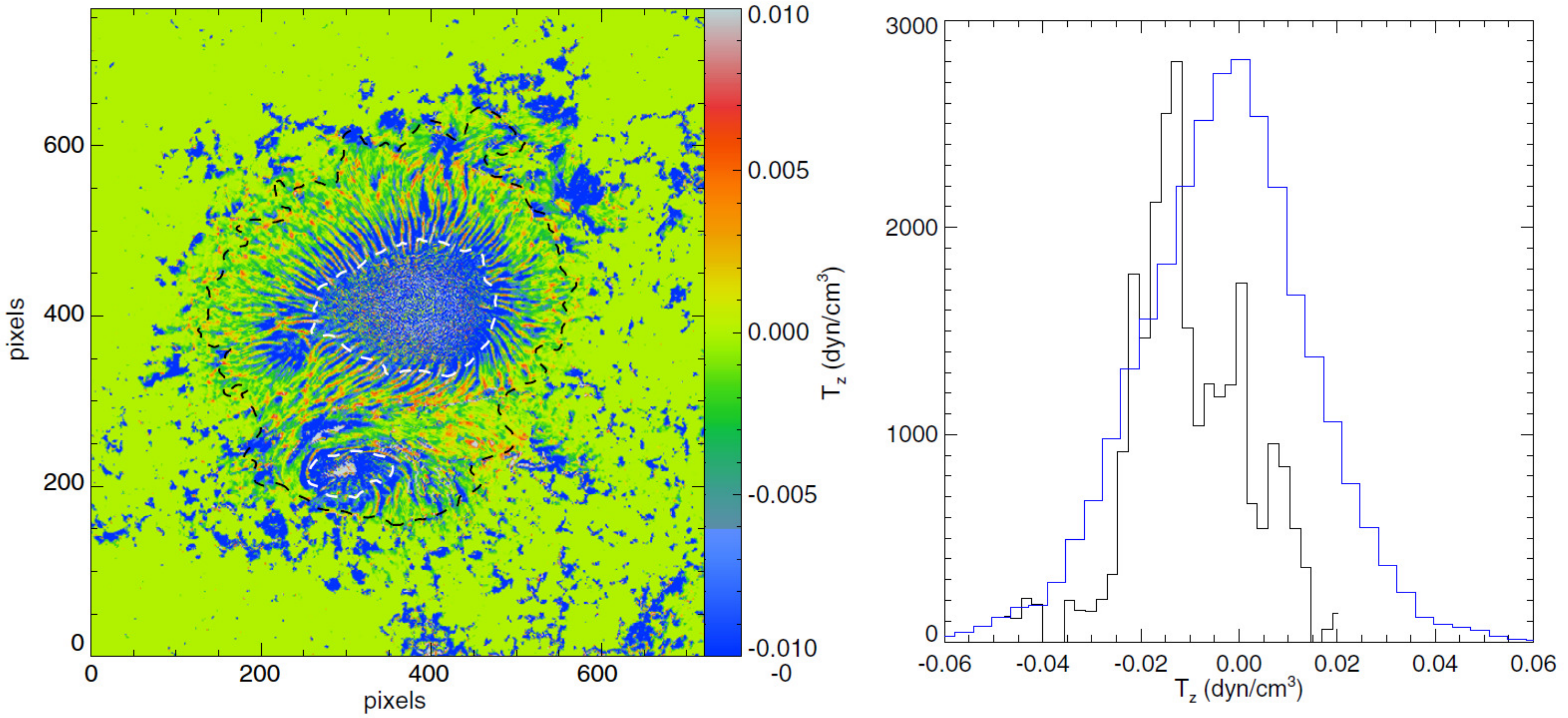}
      \put(-315,140){\bf\sf(a)}
      \put(-130,140){\bf\sf(b)}
      \caption{(a) Map of the vertical tension force ($T_z$) of an
      active-region sunspot. Gray and black dashed lines outline the
      boundaries of the umbra and penumbra, respectively. $T_z$
      has high negative values at most places over
      the sunspot. (b) Histograms of $T_z$ in the umbra (black) and
      penumbra (blue). The histogram peak for the umbral field is shifted
      towards higher negative values, \ie, the umbral field is more force-free than
      that in the penumbra.
      (Adapted from Figure 2 of \cite{tiw_12}. \copyright~AAS. Reproduced with permission.) }
      \label{fig:tiwari}
\end{figure}

\cite{sol_wal_93} found that in the layers of sunspots
near the bottom of the photosphere, $\beta$ is likely
above unity everywhere. It was found to drop from higher values in the
umbral center and to reach $\beta\approx1$ at the umbral boundary,
followed by another increase towards the outer penumbral boundary.
In contrast, \cite{mat_sol_04}, who used the same spectral lines as
\cite{sol_wal_93}, presented a case where both, the entire umbra as
well as the inner penumbra of a sunspot had a $\beta$ slightly
below unity.
More recently, \cite{tiw_12} statistically addressed this topic using
high-resolution magnetic field information for 19 sunspots. He
found that in mid-photospheric layers
most of the fine structures over most of the sunspot areas were
nearly force-free with the tendency that umbral fields were less
forced, while penumbral fields were more (see Fig.~\ref{fig:tiwari}).
This combination of large plasma-$\beta$ in a spatially averaged
sense and small values of $\beta$ locally has important consequences.
A comparatively low $\beta$ inside strong-field features helps to explain
why they maintain their identity for often considerable lengths of time.
The high average $\beta$ implies that magnetic features as a whole,
more or less passively, follow convective motions. That in turn explains
that the magnetic field in the corona can become tangled and complex
(see section~\ref{s:active}).

The relative importance of magnetic forces in
entire ARs were also estimated.
\cite{met_95} found a value of $\simeq0.4$ for the net Lorentz force
(\ie, the ratio of the total vertical Lorentz force and magnetic pressure
force, integrated over the area of the considered AR)
and concluded that the
analysed AR cannot be validly considered as to be force-free at a
photospheric level. In contrast, \cite{1987ApJ...314..782G} found that
another analysed AR was indeed force-free, except for some, localized
areas (areas for which flaring activity was noticed).
\cite{moo_cho_02} analysed the forces within three flare-productive ARs
and found a median of $\approx 0.1$ for the net vertical Lorentz force
and argued that the magnetic field at photospheric levels may not be as
far from being force-free as commonly assumed.
(See also section~\ref{qs_forcefree}, for a discussion on the force-freeness
of quiet-Sun regions.)

The above compilation shows that the findings, so far, are not entirely
conclusive regarding how close to being force-free the photospheric
magnetic field really is, they rather show that the amount of forcing (by the
gas) depends on the situation being considered.
Therefore, special care is required when using
photospheric vector magnetic field data as input for, \eg, coronal magnetic
field models. Such modelling often relies on routine
measurements of the magnetic field, which are to date predominantly
performed at photospheric levels (see section~\ref{s:modeling}).

\subsection{Chromosphere}
\label{ss:chromo}

The chromosphere lies on top of the photosphere with a thickness of
about 1~Mm to 2~Mm, starting from the temperature minimum
in traditional one-dimensional model atmospheres. In reality, the
chromosphere is far more complex and its thickness is likely to vary
strongly from one horizontal location to another. Importantly, it should
be thought of more as a temperature rather than a static height regime,
with the temperatures increasing from the temperature minimum to
$\approx$\,$10^4$~K \citep[][]{sti_02}.
Sketches indicating the
rich variety of phenomena in the chromosphere and its complexity have
been presented in reviews b \cite{wed_lag_09} and
\cite{2012RSPTA.370.3129R}.
Just as the small-scale dynamics of the photosphere are dominated by granular
convection, those of the chromosphere are dominated by waves.
In internetwork regions these are mainly acoustic waves with a three-minute period,
produced in
the convection zone
\citep[for reviews see][]{1991SoPh..134...15R,1997ApJ...481..500C,2004A&A...414.1121W}.
But there is also mounting evidence of MHD waves in the chromospheric layers of
magnetic structures \citep[][]{2006ApJ...647L..73H,2007Sci...318.1574D}.

\subsubsection{Characteristic chromospheric magnetic structures}
\label{sss:prop_chromo}

The enhanced magnetic flux concentrations outlining the supergranular cells
(the magnetic network) in the photosphere coincide with the bright network
seen in chromospheric spectral lines (\eg, of Ca\,{\sc ii}).
The spatial agreement results from the fact that the magnetic features are
nearly vertical \citep[]{mar_lit_97,2014arXiv1408.2443J},
which is the result of the large field strength of the photospheric
($\approx$\,kG) flux tubes. Strong fields produce nearly evacuated structures
which result in the flux tubes being buoyant \citep[][]{par_55}
and therefore cause a radial (\ie, vertical)
orientation of the field. The vertical orientation is maintained also in
the presence of horizontal granular flows \citep[][]{schu_84} which bend
the magnetic elements \citep[][]{ste_gro_96}. The magnetic elements appear
bright in chromospheric radiation and larger in size than in the photosphere
\citep[][]{gai_85}.
Smaller magnetic features are brighter than their surroundings in
photospheric radiation due to the vertical,
evacuated structures being less opaque than their surroundings. As a
consequence, the radiation from the flux tube's walls may penetrate
deep into the thin flux tube's interior which then appears bright
\citep[][and for reviews see
\cite{sol_93} and \cite{2007AIPC..919...74S}]{1976SoPh...50..269S}.
To explain the enhanced brightness in the chromosphere, however,
additional sources of heating, such as the dissipation of waves propagating
along the field lines \citep{1997LNP...489...75R} are necessary.

Both in active-region plage and in the network, the kG
magnetic field structures appear more diffuse
in the chromosphere than in the photosphere \citep[][]{jon_85,pet_pat_09}.
While the photospheric field is
mainly radially oriented, the chromospheric field expands in all directions,
forming a magnetic canopy \citep[][]{gio_80,jon_gio_82},
which is likely to be a natural consequence of the excess heating inside
magnetic elements \citep{sol_ste_90}.
\cite{cho_sak_01} compared LOS chromospheric magnetic field as observed
in the Ca\,{\sc{ii}} 8542~\AA\ spectral line with a current-free magnetic
field model. The latter was based on photospheric LOS magnetic field
observations in the Fe\,{\sc{i}} 8686~\AA\ spectral line. Analysing 137~ARs,
they found that the chromospheric observations were reproduced best by
a current-free model field at a height of $\approx$\,800~km above the
photosphere, in agreement with the expected formation height of the
Ca\,{\sc{ii}} 8542~\AA\ line.
Their results also suggested a decreasing correlation between the
observed and modelled LOS magnetic field with increasing field strength,
which they attributed to change of the spectral line's formation height in
strong-field regions (although a real deviation from a potential configuration
remains a possibility).

On larger (active-region) scales, often observed as dark
elongated features in H$\alpha$ 6563~\AA\ and He\,{\sc i} 10830~\AA\
images are filaments \citep[``prominences'' when observed above
the limb; for reviews see][]{lab_hei_10,mac_kar_10}.
Filaments straddle polarity inversion lines and typically exhibit heights of
$\approx$\,50~Mm, lengths of
$\approx$\,200~Mm and a thickness of a few Mm \citep[][]{sti_02}.
They are involved in many eruptive processes (``eruptive'' filaments),
but outside of ARs often persist for a long time in the QS (``quiescent'' filaments).
As suggested by the name, active-region
filaments concentrate around the activity belt, while quiescent filaments
can be located everywhere on the Sun.
In principle, they are thought to be comparatively cool ($T\lesssim10^4$~K)
chromospheric material suspended in the corona, sustained by the geometry
of the magnetic field.
Early investigations of large samples of polar prominences
(quiescent as well as eruptive) mainly based on Hanle effect measurements,
revealed characteristic longitudinal field strengths on the order of 1~G to 10~G
\citep[][]{1977A&A....60...79L,1983SoPh...83..135L,1983SoPh...89....3A}.
For active-region filaments, the interpretation of the Zeeman effect
revealed strengths of some 100~G to 1~kG for the vertical as well
as horizontal field \citep[][]{2005ApJ...622.1275L,kuk_mar_12,xu_lag_12}.
\cite{xu_lag_12} were furthermore able to trace the photospheric
and chromospheric signatures of the same active-region filament,
and detected differing morphologies.
This led them to suggest that an emerging magnetic flux rope
may, besides sustaining filament material at low atmospheric heights
(upper photosphere to low chromosphere),
at the same time be able to store plasma at the top part of the
flux rope, \ie, at greater (mid chromospheric) heights.

On smaller scales, around sunspots, a radially outward directed filamentary
pattern is observed, persisting for hours to days, the sum of which is
called the (chromospheric) ``super-penumbra'' \citep[][]{hal_08b}. These
structures are seen in almost all chromospheric spectral lines, including
H$\alpha$ 6563~\AA, Ca\,{\sc{ii}} 8542~\AA\ and Ly$\alpha$ 1216~\AA\
and, though more rarely, also in Ca\,{\sc{ii}}~H and K \citep[see][and
references therein]{pie_hir_09}.
A common assumption is that these chromospheric
``fibrils'' outline the direction of closed magnetic field structures
in the upper photosphere and chromosphere, linking the
spot with the surrounding flux of opposite polarity
\citep[][]{nak_raa_71,woo_cha_99},
allowing a mass flow
away from the spot \citep{eve_09} or into the spot
\citep[``inverse Evershed effect'';][and for a review see \cite{sol_03}]{1913ApJ....37..322S}.
In a similar fashion, fibrils (seen, \eg, in H$\alpha$) are thought
to connect opposite polarity magnetic flux elements in the QS
\citep[][]{rea_wan_11,bec_cho_14},
although some fibrils may follow the chromospheric part of magnetic
field lines that continue into the corona (see also section~\ref{s:quiet}).
The fibril pattern around
sunspots is often observed to be oriented radially outwards and forming
whirls which exhibit rotation patterns specific to the hemisphere where
they are observed \citep[][]{1908ApJ....28..100H,ric_41,1996ApL&C..34...77P}.
\cite{vec_cau_07} underlined the likeliness of fibril-like structures seen in
Ca\,{\sc ii}~8542~\AA\ images to outline the canopy at chromospheric levels.
The fibrils are thought to follow the canopy magnetic field
of sunspot super-penumbrae, whose base rises slowly from the edge of the
spots as one moves radially outward alongside a decreasing magnetic
field strength \citep[][and see also section~\ref{intro_canopy}]{gio_80,gio_jon_82,1994A&A...283..221S}.

\subsubsection{Indirect tracing of chromospheric fields}
\label{sss:obs_chromo}

\begin{figure}
      \centering
      \includegraphics[width=0.41\textwidth]{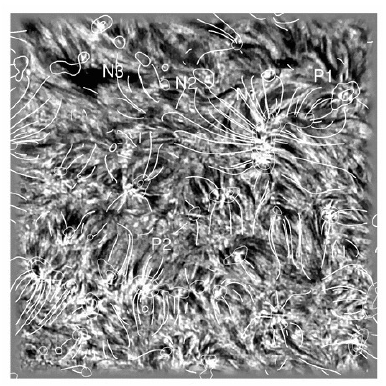}
      \includegraphics[width=0.5\textwidth]{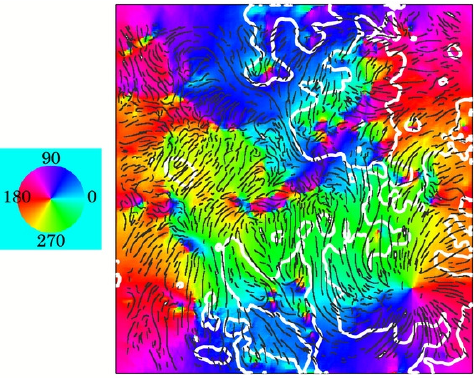}
      \put(-320,125){\bf\sf(a)}
      \put(-140,125){\bf\sf(b)}
      \caption{ (a) Comparison of H$\alpha$ fibrils (observed at 6563~\AA; gray-scale
      background) with projections of potential magnetic field lines over-plotted.
      The FOV spans roughly $300\arcs\times300\arcs$.
      Except for some local areas denoted as P1 and P2 (upper right and lower
      middle part of the image, respectively), poor agreement
      between filed and the direction of fibrils is recognized in particular
      around N1, N2, and N3 in the upper mid to left part of the frame.
      (Figure 2 of \cite{woo_cha_99}. With kind permission from Springer Science and Business Media.)
      (b) Comparison of H$\alpha$ fibrils (at 6563~\AA; black curves)
      with the chromospheric potential magnetic field azimuth,
      counted counter-clockwise from 0$^\circ$ at solar west. The white
      contours outline magnetic PILs.
      The FOV is roughly $254\arcs\times264\arcs$.
      Good agreement is found between
      the potential field azimuth and the fibril orientation in some places,
      while a clear deviation of the two direction is seen in others.
      (Adapted from Figure 5 of \cite{jin_yua_11}. \copyright~AAS. Reproduced with permission.)
      }
      \label{fig:fibrils}
\end{figure}

\cite{woo_cha_99} investigated the non-potentiality
of fibril structures in the QS. They performed a comparison of field lines from a
potential field model with fibrils observed in H$\alpha$ 6563~\AA.
They found, under the assumption that the fibrils trace magnetic field lines,
that the observed fibril structure aligns
well with the magnetic field model in some places, but
in others not (see Fig.~\ref{fig:fibrils}a).
They concluded from this finding that the quiet-Sun's chromospheric magnetic
field is far from a
potential state (\ie, it carries currents on small scales). This interpretation
has been tested for active-region fibrils by \cite{jin_yua_11} who based
their study on a
potential magnetic field model starting from chromospheric magnetograms.
 Again it was found that in some places the modelled horizontal field
agrees well with the segmented fibril orientation but in other places not
(Fig.~\ref{fig:fibrils}b).
It appeared that there is a link between the horizontal shear of
the involved field and the mismatch between model and observation:
the higher the shear of the observed chromospheric magnetic field structures,
the lower the agreement with a potential magnetic field model.
Consequently, potential field models, either based on photospheric or
chromospheric magnetic field data, can in general not be assumed to
adequately reproduce the (chromospheric) magnetic field, assuming that fibrils
indeed outline the orientation of the chromospheric magnetic field.

\begin{figure}
      \centering
      \includegraphics[width=0.6\textwidth]{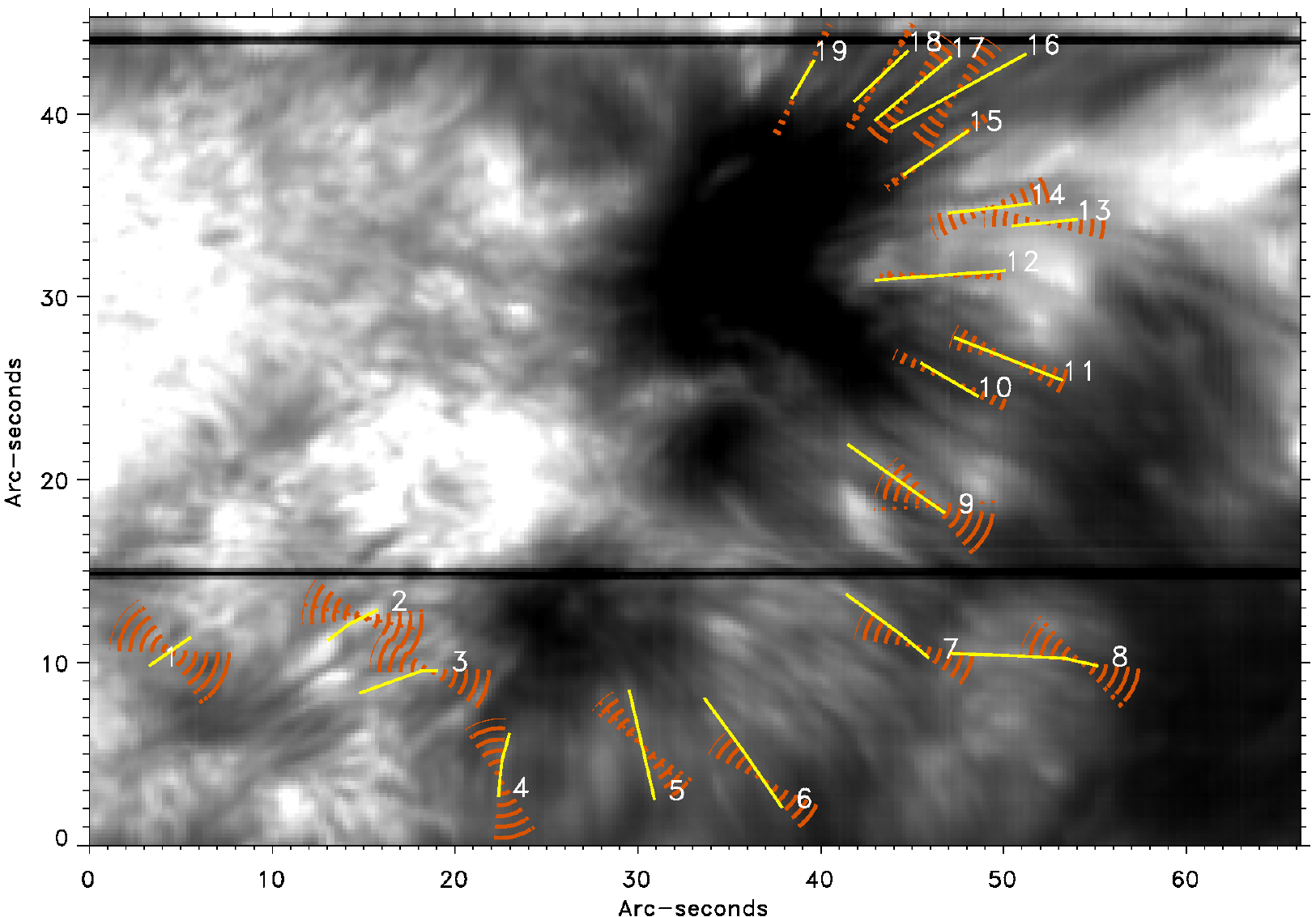}
      \includegraphics[width=0.35\textwidth]{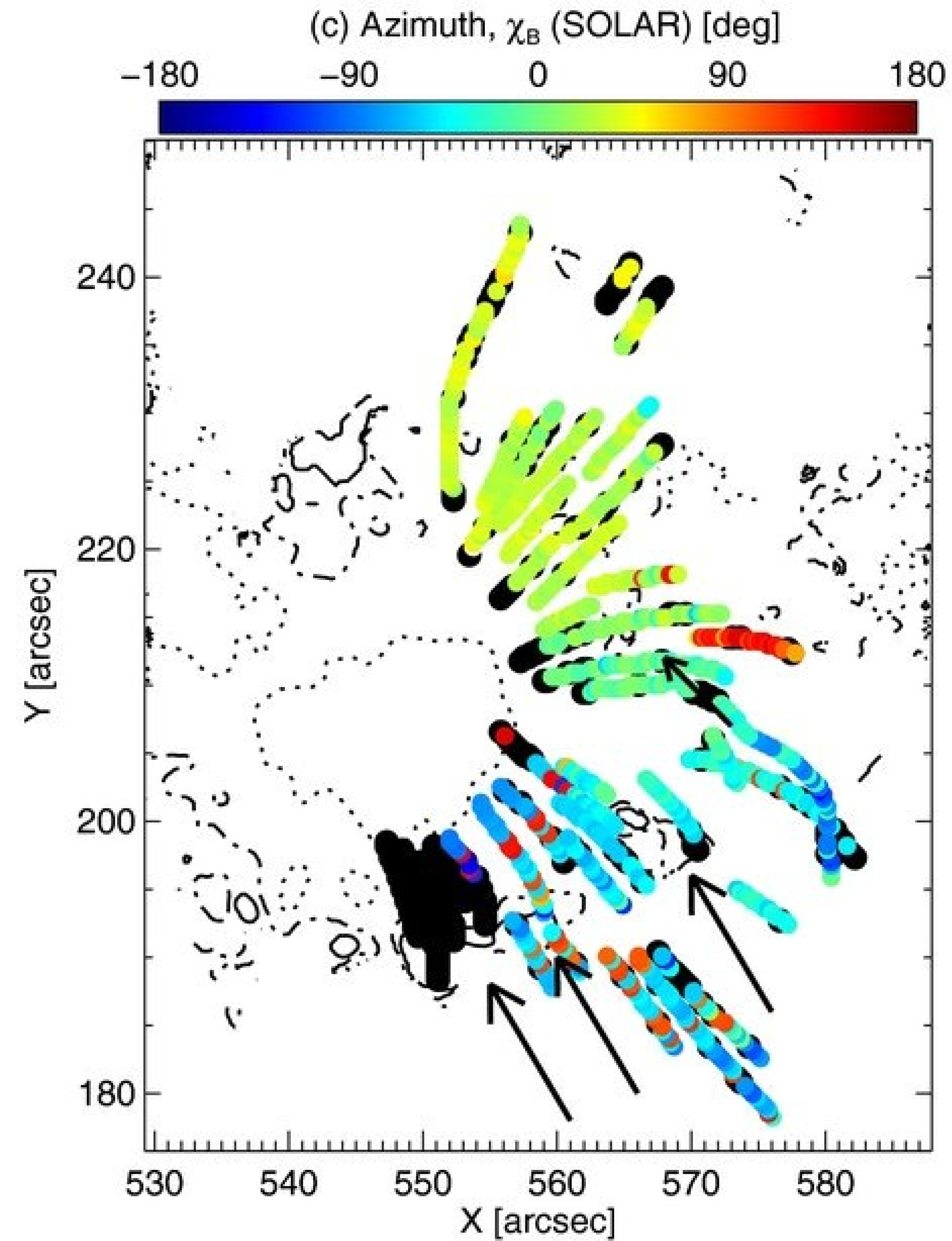}
      \put(-335,135){\bf\sf(a)}
      \put(-120,135){\bf\sf(b)}
      \caption{ (a) Comparison of visually determined directions (yellow lines)
      of Ca\,{\sc ii} fibrils (observed at 8542~\AA; gray-scale background) to the magnetic field
      azimuth compatible with linear polarization signals (orange cones). While the
      fibril orientation is picked up by the reconstructed horizontal magnetic field
      orientation for a fair number of fibrils (\eg, fibrils 9 to 19),
      it is only poorly recovered for others (\eg, fibrils 1 to 5).
      (Figure 1 of \cite{dlcr_soc_11}. Reproduced with permission from Astronomy \& Astrophysics, \copyright~ESO.)
      (b) Spatial map of the magnetic field azimuth along selected super-penumbral
      fibrils, inferred from He\,{\sc i} at 10830~\AA\ observations.
      The dotted and dot-dashed contours
      indicate a photospheric magnetic field inclination of 135$^\circ$ and 90$^\circ$,
      respectively. Black dots mark severe deviations between the inferred and
      observed fibril orientation. They are restricted
      to where  fibrils turn to their photospheric rooting points.
      (Adapted from Figure 10 of \cite{scha_pen_13}. \copyright~AAS. Reproduced with permission.)
      }
      \label{fig:fibrils2}
\end{figure}

An alternative interpretation of the results of \cite{woo_cha_99} and \cite{jin_yua_11} is
that fibrils do not outline the orientation of the chromospheric magnetic field.
This, however, is in direct contrast with the results of recent
numerical simulations which suggested that H$\alpha$ fibrils are
visible manifestations of high-density ridges aligned with the magnetic field
\citep{lee_car_12}, thus serving as an indirect tracer of the vertical-to-horizontal
transition of the magnetic field orientation around magnetic flux concentrations.
This was addressed by \cite{dlcr_soc_11} who compared the observed
orientation of fibrils in Ca\,{\sc{ii}} 8542~\AA\ images to the chromospheric
magnetic field vector, inferred from observed polarization signals originating
from the same spectral line. They found that most of the fibrils in the
surrounding
of a penumbral boundary nicely followed the magnetic field direction but also
recorded a significant mismatch for a considerable number of fibrils
(Fig.~\ref{fig:fibrils2}a). They
also noted a too rapid decrease of the linear polarization signal when moving
out of the penumbral area, if the fibril pattern indeed were to outline the
super-penumbral field direction. The rapid decrease of the linear
polarization signal, however, may be attributed to the height of
the canopy base relative to the formation height of the spectral line.
This was re-addressed recently by \cite{scha_pen_13} who, in contrast to
\cite{dlcr_soc_11}, found a clear coincidence of the projected direction of
super-penumbral fibrils and the inferred magnetic field (to within $\pm10^\circ$)
using He\,{\sc i} 10830~\AA\ observations. They detected a notable change of the
inclination only close to where the fibrils turn towards their rooting point in the
sunspot (Fig.~\ref{fig:fibrils2}b).
Moreover, based on their findings, they explicitly support
schemes which propose the inclusion of the spatial information delivered by
chromospheric fibril observations to increase the success of force-free coronal
magnetic field models. Such proposed schemes use the fibril information to
increase the match between the modelled and observed horizontal field at
chromospheric heights \citep[where the magnetic field vector is not routinely
measured; see][]{2005SoPh..228...67W,wie_tha_08,yam_kus_12}.

\subsubsection{Plasma-$\beta$ in the chromosphere}
\label{sss:dynamics_chromo}

\begin{figure}
      \centering
      \includegraphics[width=0.75\textwidth]{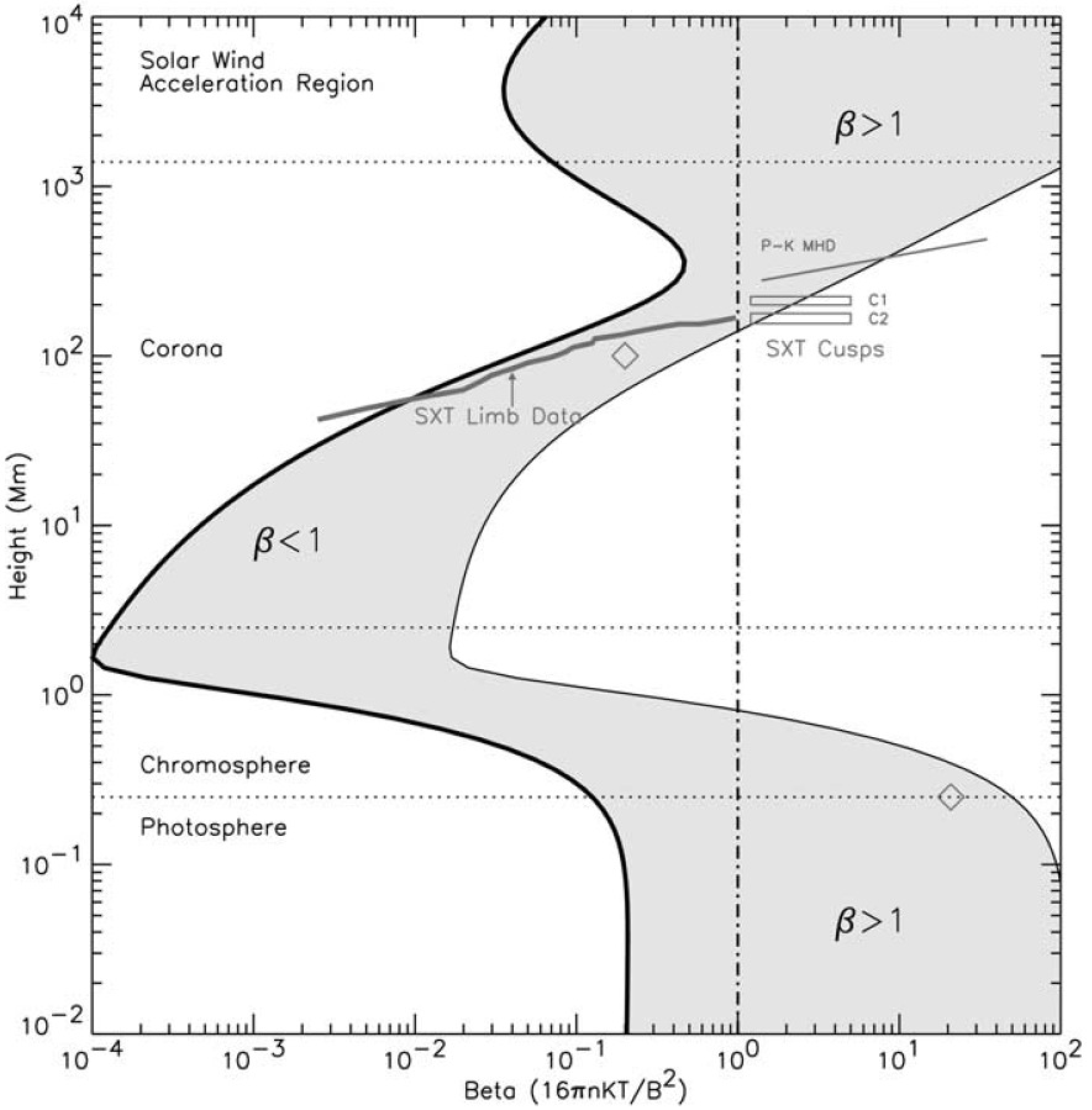}
      \caption{Distribution of the plasma-$\beta$ as a function of height
      above an AR. The shaded area represents the vertical run of $\beta$
      for open and closed fields originating between a sunspot (represented
      by the thin solid line) and a plage region (represented by the thick solid line).
      $\beta$ becomes only small ($\simeq10^{-2}$) at heights above the upper
      chromosphere and mid corona. Note that this low-$\beta$ region
      is sandwiched between high-$\beta$ regions (the photosphere and low to
      mid chromosphere below as well as the upper corona and solar wind
      acceleration region above).
      (Figure 3 from \cite{gar_01}. With kind permission from Springer Science and Business Media.)}
      \label{fig:gar_01_fig3}
\end{figure}

Density and temperature are heavily
structured in the highly dynamic chromospheric environment so that the
relative strength of the plasma pressure and magnetic forces also varies
strongly with position, at a
given height. The height at which the magnetic forces start to
dominate over others (\ie, where $\beta<<1$) is expected to
be strongly corrugated relative to the solar surface.
In the QS, that height is expected to vary between
$\approx$\,800~km and 1.6~Mm above the
photosphere \citep[][]{ros_bog_02}. In ARs, this height is likely to be
lower, as shown by \cite{met_95}. They used chromospheric vector
magnetic field measurements inferred from
observations in the Na\,{\sc{i}} 5896~\AA\ spectral line to test the
relative contribution of the plasma pressure and magnetic forces in
an AR. They found that the atmosphere above that AR could be
considered to be force-free from $\approx$\,400~km above the solar
surface. \cite{gar_01} was able to confirm that finding by combining
a plasma pressure and magnetic field model to estimate the pattern of
interchanging dominance of plasma and magnetic pressure with height in
the solar atmosphere (see Figure~\ref{fig:gar_01_fig3}).
He concluded that the magnetic forces above
sunspots should start to dominate from relatively low heights
($\gtrsim$\,400~km above the photosphere). Above plage regions, the model
results suggest this to be true from $\gtrsim$\,800~km above a photospheric
level upwards. In summary, ARs can be considered to be
force-free in most of the chromosphere (in contrast to quiet-Sun areas;
see section \ref{qs_forcefree}).

\subsection{Magnetic coupling from the lower solar atmosphere to the corona}
\label{ss:coupling}

\subsubsection{Magnetic canopy}
\label{intro_canopy}

\begin{figure}
   \centering
   \includegraphics[width=\textwidth]{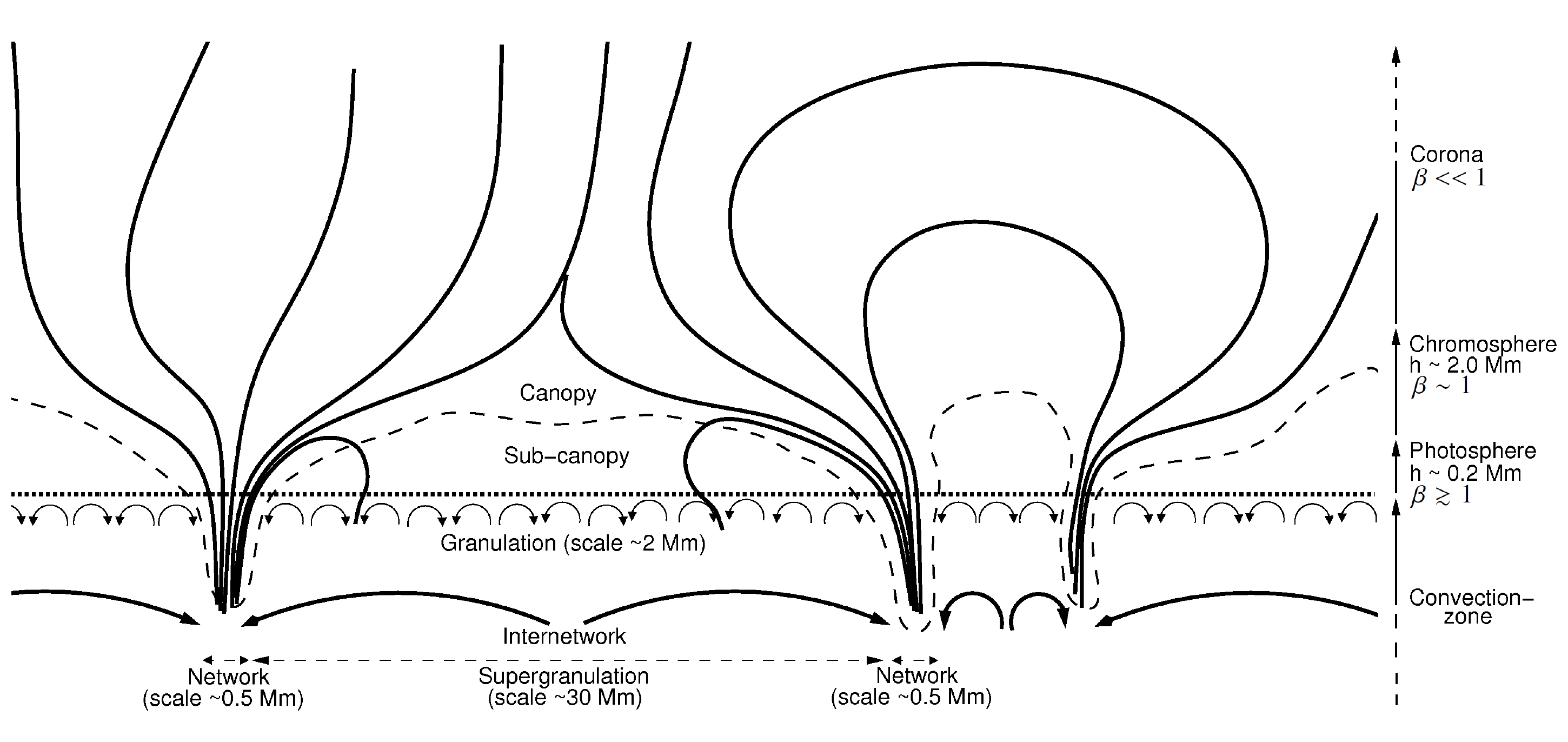}
   \caption{Sketch of the quiet-Sun magnetic field structure in a vertical
   cross section through the atmospheric layers of the Sun. Swept to the
   edges of supergranular cells by large-scale convective flows (thick, large
   arrows at bottom), intense magnetic network elements or sunspots form.
   Small-scale convective flows (thin, small arrows below dotted horizontal
   line representing the photosphere) result in the photospheric
   granular pattern. The magnetic field lines (solid lines) expand at
   chromospheric heights and form the nearly horizontal magnetic canopy
   (dashed line). (Adapted from \cite{jud_06} and \cite{wed_lag_09}.)}
   \label{fig:canopy}
\end{figure}

At photospheric levels only a
small fraction of the solar surface is occupied by strong magnetic field
($\lesssim5$\%). In contrast to that, the
coronal magnetic field fills the entire coronal volume and is distributed
relatively uniformly in strength (although not in orientation). Consequently, the
photospheric field must spread out with increasing height in the solar
atmosphere.
The magnetic field expands until it either turns over and returns to connect
back to the photosphere or it meets the expanding field of the neighbouring
flux tubes. It then forms a ``magnetic canopy'', \ie, a base almost parallel
to the solar surface and overlying a nearly field-free atmosphere (see
Fig.~\ref{fig:canopy}). For a comprehensive review of the current picture
of the magnetic coupling of the photosphere, chromosphere and transition
region to the corona we refer to \cite{wed_lag_09}
and restrict ourselves to a brief summary here.
Estimates for the merging height of photospheric flux tubes range from some
100~km for active-region to $\approx$\,1~Mm for quiet-Sun
magnetic fields \citep[][]{spr_81,gio_jon_82,1990GMS....58..113R}.
(Note that these estimates essentially depend on the filling factor, \ie,
whether the considered region exhibits a high or low mean magnetic field
strength.)

The expansion of the magnetic field with height is a consequence of the
small gas-pressure scale height
\citep[$\approx$\,100~km in non-magnetic regions;][]{dur_88}.
From the lateral pressure
balance follows that the field strength must rapidly decrease with height.
(Remember that lateral pressure balance requires the gas pressure inside
a flux tube to be lower than outside it.)
With increasing height,
the magnetic field strength drops due to the fall-off of the gas
pressure and flux conservation implies that the
magnetic field must spread out, \ie, the extension of the magnetic
structures must increase rapidly.
Since magnetic features are hotter than their surroundings in the
middle/upper photosphere and chromosphere,
the internal gas pressure
drops more slowly with height than the external gas pressure.
As a consequence, at certain heights, the internal pressure force exceeds
the external. This removes the lateral confinement of the
magnetic structures and allows the structures to expand unhindered,
until it hits field from another photospheric source. This implies a
significant horizontal component of the field over a large part of the
volume \citep[the canopy;][]{sol_ste_90,bra_cra_91}.
The different merging heights thus depend, besides on the distance
between neighbouring magnetic features, on the temperature difference
between the magnetic field structures and their surroundings, causing
successively lower canopy heights for increasing temperature differences
(see chapter~5 of \cite{asch_05}).
Above this merging height the magnetic field becomes increasingly
homogeneous. Generally, the field of a magnetic element is seen to be
shaped roughly like a wine glass.
The direction of the field then depends on the structure of the magnetic
field in its surroundings and the connectivity of the field lines on a larger
scale (\ie, whether they are closed or open and where they return to the
solar surface).

In plage regions, the flux tubes merge already in the mid to upper
photosphere, so that the atmosphere above is almost fully magnetic
\citep{bue_ste_93}.
Model results suggested that, in the QS and in CHs where magnetic
features are further apart, this base is located somewhere in the lower
chromosphere \citep{gab_76,jon_gio_82,sol_ste_90,sol_ste_91}.
Quite some time after the first speculations on the height of canopy-type
magnetic fields, observational evidence for the merging heights in plage
of on the order of several hundreds of km have been delivered
\citep[][]{ste_piz_89,gue_mat_91,1995A&A...293..240B}.
\cite{ros_bog_02} performed numerical simulations of the propagation
of waves through a model atmosphere, resembling properties of the
chromospheric
network and internetwork, and found the canopy height to vary between
$\approx$\,800~km and 1.6~Mm above the base of the photosphere.
However, a considerable number of findings, especially in the QS,
led to serious doubts upon the reality of a large-scale, undisturbed
magnetic canopy there (for details see section~\ref{ss:canopy_qs}).

Even though some aspects of the magnetic canopy, especially in the QS are
still to be elaborated further, its basic nature seems clear: it is not a
simple, rigid structure and also not at a constant height in the solar atmosphere.
Instead, its shape and
height is different for regions on the Sun with different amounts of magnetic flux
and it also varies with time. Above the canopy, the coronal volume is filled more
or less uniformly with magnetic field.

\subsection{Corona}
\label{ss:corona}

\subsubsection{Transition region and coronal base}

The corona is to be thought of as a temperature regime, covering a few
times $10^5$~K in open field regions (such as CHs; see section~\ref{s:chs}),
$\approx$\,1~MK to 2~MK (megakelvin)
in the predominantly closed field of the quiet-Sun corona,
and up to 2~MK to 6~MK in
ARs \citep[see chapter~1 of][]{asch_05}. It even can briefly reach values of
10~MK to 20~MK during strong flares.
It spans the atmospheric layers between the transition region
(within which the temperatures increase from $\approx$\,$10^4$~K to
$\approx$\,1~MK) and the height where the solar wind is accelerated,
\ie, spanning several hundreds of Mm in height \citep[][]{gar_01}.
The very narrow transition region not only bridges a large difference in temperature,
but also separates the dilute coronal plasma (with number
densities of $n\lesssim10^{12}$~m$^{-3}$) from the dense
($n\gtrsim10^{16}$~m$^{-3}$)
chromosphere (see chapter~1 of \cite{asch_05}). The base
of the corona is not to be thought of as a horizontal layer somewhere
above the solar surface. As the thickness of the chromosphere beneath
varies, so does the height of the coronal base above the solar surface
(see section~\ref{ss:coupling} for details).

\subsubsection{Morphology of coronal magnetic fields}
\label{sss:morphology_corona}

Two very distinct magnetic configurations are present in the corona. The field
is either arranged in the form of closed loops of enhanced emission, or in
the form of open field lines seemingly not connecting back to the solar
surface \citep[][]{schr_tit_99,sol_inh_06}.
Arcades (ensembles) of bright coronal loops connect regions of opposite
magnetic polarity on the solar surface and are often, but not necessarily
always, rooted in an AR.
Large-scale loop systems (sometimes exhibiting sigmoidal shapes)
are often found to connect neighbouring ARs and/or ARs with their
quiet-Sun surrounding \citep[][]{stro_94}.
Following \cite{rea_10},
the observed coronal loop systems span a wide range of length scales,
from a few Mm (bright points) up to giant arches which may
span 1~Gm (gigameter).
Several loop arcades neighbouring each other
are often found in magnetically complex ARs and often host
eruptive processes such as flares or CMEs (see section~\ref{s:active}).
Therefore, in the majority of cases, bright coronal loops
(see section~\ref{sss:coronal_loops} for more details) are
concentrated around the activity belts.

Most of the quiet-Sun magnetic field (see section~\ref{s:quiet}) that
reaches the corona is rooted in the magnetic network.
At greater heights, they fan out to form funnels and to fill the coronal volume
above \citep[][]{gab_76,dow_rab_86}.
Along the open field structures, plasma is efficiently transported
outwards, which allows charged particles to escape from the solar
atmosphere. Especially during solar activity minimum, open magnetic flux is
concentrated around the poles, causing depleted regions which emit less than their
surrounding temperatures above 1~MK and consequently appear dark in coronal
images (therefore termed ``coronal holes''; see section~\ref{s:chs}).
At lower latitudes the coronal structure is dominated by
``helmet streamers'' and ``pseudo streamers'',
extending out to several solar radii in height \citep[][]{schw_06}.
Helmet and pseudo streamers are visible as enhanced
emission in the form of a cusp
above the limb, bridging the space between open fields of opposite and
same polarity, respectively (see \cite{1971SoPh...18..258P} and
\cite{wan_she_07}, respectively.

\section{Magnetic field modelling}
\label{s:modeling}

The solar magnetic field is routinely measured mainly
in the photosphere, whereas direct measurements in the
higher solar atmosphere are available for individual cases.
If the 3D magnetic field vector in the chromosphere and corona were to
be measured routinely with high accuracy, cadence and resolution,
indirect modelling approaches (as discussed in the following sections) would
not be required. Since this is not yet the case (see section~\ref{model:direct}),
modelling approaches of different sophistication have been developed
with the aim of computing the magnetic field in the upper solar atmosphere,
generally starting from measurements made in the lower atmosphere.

One possibility is to use the longitudinal photospheric magnetic
field component, or the measured full magnetic field vector
(if available) as boundary condition for force-free magnetic
field reconstruction techniques. This is possible since the
solar corona is almost force-free, because
the magnetic pressure is several orders of magnitude higher than
the plasma pressure. That allows neglecting nonmagnetic
forces to lowest order and applying such methods
(see section~\ref{model:extrapol}). Because these models are snapshots
and assume stationarity and stability
of the coronal magnetic field configuration,
they are not to be used for modelling of dynamic features
(such as CMEs, flares or eruptive prominences). Moreover, these
models do not provide a self-consistent description of the coronal plasma.
Time-dependent simulations are required for these aims, usually within
the MHD approach (see section~\ref{model:mhd}). Full MHD
models (see section~\ref{model:full_mhd}) are both theoretically and
observationally very challenging
because plasma and magnetic field have to be modelled self-consistently.

Complementary to these
numerical approaches one can use the fact that
the emitting coronal plasma (as visible in coronal images;
see section~\ref{sss:coronal_loops}) is frozen into the magnetic
field and consequently the coronal loops visible in the images
outline magnetic field structures.
Therefore coronal images can be used to identify the 3D shape of the
magnetic field structures when images from multiple viewpoints exist (\eg, from the
\stereo-spacecrafts, \soho\ or \sdo). A 3D reconstruction of
structures seen above the solar limb
can be performed by stereoscopic and tomographic methods
(see section~\ref{sss:stereoscopy} and \ref{sss:tomography}, respectively).
Coronal images are also frequently used to validate
the results of coronal magnetic field models. In some cases, time sequences
of coronal images show oscillating coronal loops, which allow
estimating the coronal magnetic field strength by coronal seismology
(see section~\ref{sss:seismology}).

The main aim of this section is to give a short overview of the methods
for deriving the 3D magnetic field structure of the
upper solar atmosphere (although we start this section with a short review
of direct measurements of chromospheric and coronal magnetic fields).
We refer to specialized reviews
and the primary literature for further details. Outside the scope of the present
review are methods of the interaction of the convection zone
with the solar atmosphere by flux emergence \citep[the interested reader can
find a recent review on the theory of flux emergence in][]{2014LRSP...11....3C}.
Methods to analyse the 3D coronal magnetic field topology are described in
section~\ref{ss:local_topology} and also in a review by \cite{lon_05}.

\subsection{Direct coronal magnetic field measurements}
\label{model:direct}

Direct measurements of the solar magnetic field are an important tool
for understanding the magnetic field in the upper solar atmosphere.
Here, we briefly introduce the most important methods for measuring the
chromospheric and coronal magnetic field directly. The difficulties of
performing such measurements are only briefly touched
upon here \citep[for details see][]{rao_05,whi_05,vdg_cul_09,car_09}.
Thanks to instrumentation, \eg, the ground-based
\nso/DKIST (planned to become operational in 2019), together with
powerful inversion techniques, coronal field measurements might become
a prosperous method in future.

\subsubsection{Chromospheric magnetic field measurements in the infrared}

Infrared lines have been used to derive the magnetic
field vector near the coronal base in the upper chromosphere.
Initial measurements of the LOS magnetic field were performed by
\cite{1971IAUS...43..279H}, \cite{1995A&A...293..252R} and \cite{1995ApJ...441L..51P}
and the first vector magnetic field measurement by \cite{1996SoPh..164..265R}.
\cite{2003Natur.425..692S} applied the same method using
the He\,{\sc I} 10830~\AA\ line, which
is optically thin. Consequently, the measurements
are related to different formation heights,
following the fluctuating height of the coronal base.
The authors managed the 3D structure of the chromospheric
loops to be reconstructed by applying the following criteria. If a randomly selected
pixel matches in field strength and direction the two neighbouring pixels,
then the radiation is assumed to originate from the same loop. Because
the full magnetic field vector is inferred, this allows to reconstruct the
loop in 3D, with the additional constraint that the field strength decreases
with height.
% A thin current sheet in the chromosphere was detected as well.
The 3D structure deduced for the emerging loops was questioned by
\cite{2009A&A...493.1121J} but it was later shown by
\cite{2011A&A...532A..63M} that the proposed geometry provided
a better representation to the data than the flat alternative proposed by
\cite{2009A&A...493.1121J}.
Simultaneously with these chromospheric measurements, the photospheric
field vector was measured as well, and extrapolated into the
chromosphere using force-free modelling
techniques (see section~\ref{model:extrapol}), where the
NLFF model was found to agree best with the chromospheric
observations \citep[for details see][]{2005A&A...433..701W}.

\subsubsection{Coronal magnetic field measurements in infrared}

Coronal measurements in the infrared are possible
from the ground with a coronagraph, or with instruments from space.
An overview on some aspects of the usage of infrared lines to measure the
coronal magnetic field can be found in \cite{2014LRSP...11....2P}.
That review also gives a detailed discussion of advantages and disadvantages
of using infrared lines in general (not restricted to coronal magnetic fields).
More than ten coronal lines in the infrared have been identified and
some of them are magnetically sensitive. The Fe\,{\sc xiii} 10750~\AA\ line,
for instance, has
been used to measure the Stokes vector in the corona, which in
principle would allow determining the magnetic field vector by an
inversion. A general problem with coronal observations is, however,
that due to the optically
thin coronal plasma, any recorded radiation form the corona is integrated over the
LOS. This naturally complicates the interpretation
of the measurements, so that to
derive the spatially resolved coronal magnetic field vector in 3D,
measurements from multiple viewpoints are necessary. The situation
has some similarities with deriving the coronal density by
a tomographic inversion (see section~\ref{model:stereo}).

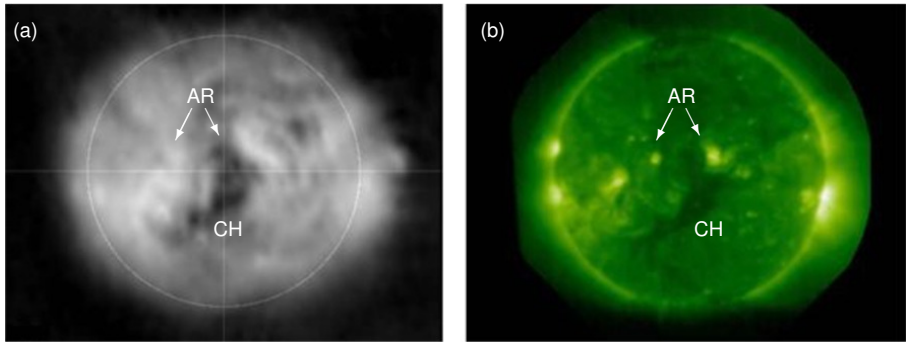
\begin{figure}
   \centering
   \includegraphics[width=\textwidth]{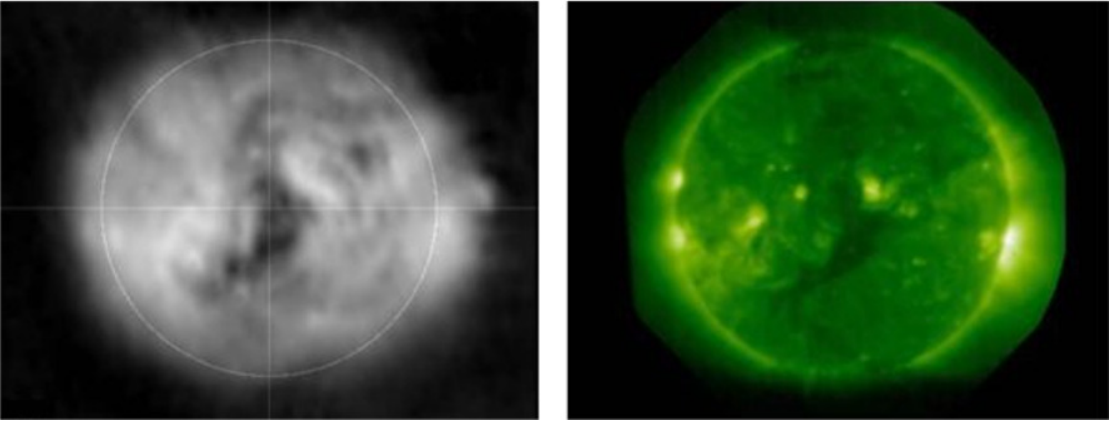}
%   \put(-346,105){\bf\sf\white(a)}
   \put(-335,115){\bf\sf\white(a)}
   \put(-160,115){\bf\sf\white(b)}
   \put(-260,40){\bf\sf\white CH}
   \put(-80,40){\bf\sf\white CH}
   \put(-270,90){\bf\sf\white AR}
   \put(-90,90){\bf\sf\white AR}
   \begin{picture}(60,40)
      \put(-88,88){\white\vector(-1,-2){6}}
      \put(-83,88){\white\vector(1,-2){5}}
      \put(-268,88){\white\vector(-1,-2){6}}
      \put(-263,88){\white\vector(1,-2){5}}
   \end{picture}
   \caption{
%    {\it Top:}
      (a) Full-disk NRH radio emission observed on
   June 27, 2004 at 432~MHz and (b) \goes/SXI X-ray emission.
   The close relationship between the radio and X-ray emission can be seen
   in the form of bright emission around ARs and minimal
   emission in CHs.
   (Adapted from Figure 5 of \cite{mer_cha_12}. Reproduced with permission from Astronomy \& Astrophysics, \copyright~ESO.)
}
   \label{fig:radio_corona}
\end{figure}

\subsubsection{Coronal magnetic field measurements at radio wavelengths}
\label{model:radio}

Radio signatures emitted from the active-region corona, currently
represent the most widely used direct measure of the magnetic field.
Because they are produced only in specific circumstances when
electrons are guided by a magnetic field, they allow the reconstruction of
the magnetic field strength in the corona
\citep[][]{1991ApJ...366L..43W,1994ApJ...434..786S,whi_kun_97,
1997ApJ...488..488B,1998ApJ...501..853L,1999ApJ...510..413L}.
Note that hard X-ray emission often goes hand in hand with radio
emission since the efficient emission of both requires electron energies
of $\gtrsim10$~keV \citep[see
Figure~\ref{fig:radio_corona} and the review by][]{whi_ben_11}.
Information on the height of the on-disk radio source in the corona is
not accessible through such measurements, except occasionally
for coronal structures at different heights above the solar limb using
near infrared wavelengths \citep[][]{1982A&A...112..350A,1982A&A...116..248A,
lin_pen_00,2004ApJ...613L.177L}
or radio observations \citep[][]{2006ApJ...641L..69B}.
For on-disk measurements, the lacking height information may be
compensated by (force-free) magnetic field modelling of the coronal
structure, starting from photospheric magnetograms \citep[][]
{2008ApJ...680.1496L,2012SoPh..276...61B,2012ARep...56..790K}.
Furthermore, radio maps can be used also for a stereoscopic 3D
reconstruction (see section~\ref{sss:stereoscopy}).

\subsection{Force-free modelling from photospheric measurements}
\label{model:extrapol}

The solar magnetic field vector is measured routinely with high accuracy
only in the photosphere, \eg, by \sdo/HMI at a constant resolution of
$1\arcs$ over the whole solar disk.
Under reasonable assumptions we can extrapolate these photospheric
measurements
into the higher solar atmosphere, where direct magnetic field measurements
are more challenging (see section~\ref{model:direct}). So, which assumptions
are reasonable in the solar atmosphere? A key to answering this question is the
comparison of magnetic and non-magnetic forces and in particular the
plasma-$\beta$. While the plasma-$\beta$ is around
unity in the photosphere it becomes very small (about $10^{-4}$ to $10^{-2}$)
in the corona (at least in ARs; see
Figure~\ref{fig:gar_01_fig3} and section~\ref{sss:dynamics_chromo} for details).
Consequently non-magnetic
forces can be neglected in the low $\beta$ corona, and the coronal
magnetic field can be modelled as a force-free field (the Lorentz-force vanishes).
The electric current density,
$$
{\bf j} =\frac{1}{\mu_0}\, \nabla \times {\bf B},
$$
has either to be parallel or anti-parallel to the magnetic field, leading to
$$
\nabla \times {\bf B} = \alpha\, {\bf B},
$$
where a positive (negative) value of $\alpha$ means that the electric current
flows parallel (anti-parallel) to the magnetic field.

While a low plasma-$\beta$ is a reasonable justification
for using force-free models, the opposite is not true.
A high (of order one or more) plasma-$\beta$ does not exclude force-free
magnetic fields. If the non-magnetic forces compensate each other
(\eg,the plasma pressure gradient is compensated by the gravity force
in a magneto-static equilibrium)
then the Lorentz force can still vanish, even if $\beta$ is not small.
In the general high-$\beta$ case, however, non-magnetic forces have to be
considered self-consistently, \eg, in a magneto-static or stationary
MHD model. We will only summarize some
basics about the possibilities and problems of force-free models and avoid
mathematical and computational details.
For a more detailed overview on the methods used to
compute solar force-free fields see \cite{2012LRSP....9....5W}.
Depending on the force-free parameter (or function), $\alpha$,
one distinguishes between potential (current-free) fields $(\alpha =0)$,
linear force-free fields (LFF; $\alpha$ is globally constant) and the general case
that $\alpha$ changes in space, \ie, the nonlinear force-free (NLFF) approach.

\subsubsection{Potential and linear force-free fields}
\label{pfss}

The simplest case, a potential
field, requires only the LOS photospheric magnetic field component as boundary
condition. Current-free equilibria are mathematically simple and
represent the lowest possible energy state of a coronal magnetic field.
For computations on a global scale (PFSS models),
one assumes that all field lines become radial at the ``source surface''
\citep[at about 2.5~solar radii; see][for details]{schatten69}.
Potential field models are popular because they are easy to compute
and are capable of reproducing the basic coronal magnetic field structure.
More sophisticated methods (as discussed in the following) are numerically
expensive and often use a current-free field solution as initial guess
for an iteratively sought, non-potential solution.

In order to employ a LFF magnetic field model, only the photospheric LOS
magnetic field component is required as well, but such models contain one
additional free parameter $(\alpha)$.
The value of $\alpha$ (constant in space) can be inferred from additional
observations, \eg, in the form of an average value of the entire photospheric
distribution of $\alpha=(\nabla \times {\bf B})_{z}/B_{z}$.
Note that $\alpha$ is the ratio of the vertical (LOS) current density
and the vertical (LOS) magnetic field magnitude, and that the vertical (LOS)
current density can be derived from the horizontal (transverse) magnetic field.
(In that case, the knowledge of all three vector components of the magnetic
field is required.) Alternatively, $\alpha$ can be
deduced and/or optimized by the comparison of model magnetic field lines
and coronal observations (either directly with coronal loops seen in EUV
images, or coronal loops extracted from such images; and see also
section~\ref{sss:stereoscopy}).
% , or as part of magnetic stereoscopy; see also section~\ref{model:stereo}).

On global scales, LFF models are mathematically and computationally
possible, but are not frequently employed, mainly for two
reasons. Firstly, the maximum allowed value of $\alpha$ scales with
the inverse of the length scale of the computational domain.
Consequently very small values of $\alpha$ are possible but they are
so small that they have no significant effect
(\ie, the resulting magnetic field is almost similar to a potential field configurations).
Secondly, observations show that both signs of $\alpha$ can be
present in different regions on the Sun, at the same time.
That is a contradiction to the LFF assumption, namely
that $\alpha$ is constant (\ie, has the same value for different regions on
the Sun).

On smaller scales (in particular to analyze ARs), however, LFF
models were used, though more frequently before the time when vector
magnetograms started to became routinely available
(as provided to date by, \eg, \sdo/HMI).
On these smaller scales, the maximum value
of $\alpha$ can be significantly larger than on global scales and
consequently active-region LFF fields can be very
different from potential ones, \eg, the associated field lines can be
sheared.
Also for LFF models employed on active-region scales, however,
the observation of different values of $\alpha$ in different portions of the
same AR contradicts the basic assumption of a single value of
$\alpha$ being representative for the entire AR under consideration.

\subsubsection{Nonlinear force-free fields}

Given the limitations of potential and LFF approaches (as discussed above)
for a meaningful and self-consistent modelling of coronal
magnetic fields, one has to take into account that $\alpha$ is
a function of position. This spatial dependence is accounted for
in NLFF models, which
are much more challenging, both mathematically (one has to solve
nonlinear equations) and observationally (mostly
photospheric vector magnetograms are required as input,
instead of just the longitudinal (vertical) field component).
Measurement inaccuracies
in photospheric vector magnetograms
(\eg, due to noisy Stokes profiles and instrumental effects)
affect the quality of NLFF coronal magnetic field models.
The modelled coronal field, however, is less sensitive to these
measurement errors than the photospheric field vector itself
\citep[][]{2010A&A...511A...4W}.
A review on methods for computing NLFF fields has been given by
\cite{2008JGRA..113.3S02W}. The corresponding numerical
implementations have been intensively reviewed,
and repeatedly evaluated and improved within
the last decade
% by an international collaboration (NonLinear Force-Free Field (NLFFF) consortium)
\citep[see][]{2006SoPh..235..161S,2008SoPh..247..269M,
2008ApJ...675.1637S,2009ApJ...696.1780D}.
% for joint publications of this team.
The numerical schemes have been implemented
in cartesian and spherical geometry to perform active-region and
global magnetic field modelling, respectively.
As boundary condition, either the magnetic field vector at the
bottom boundary of the computational domain or, alternatively,
the vertical magnetic field and vertical electric current density is
usually required.

A difficulty arises from the fact that the plasma in the
corona is a low-$\beta$ plasma, but that of the photosphere is not.
In the photosphere, $\beta$ is on average of
the order of unity or more \citep[][]{gar_01},
although locally considerably smaller values may be found
\citep[\eg, in the interiors of magnetic
elements; see][]{1990A&A...239..356Z,1992A&A...263..323R}.
Note that a non-vanishing plasma-$\beta$ does not exclude
the existence of a force-free field,
but one has to be careful when using photospheric measurements
as boundary condition for NLFF computations. Because then it cannot be
guaranteed that the photospheric magnetic field vector is consistent with
the assumption of a force-free field in the corona. One can
find out whether the vector magnetic field measurements are consistent by
writing the force-free equations as the divergence of the Maxwell stress
tensor, integration over the entire computational volume and applying
Gauss' law. For force-free consistency,
the value of the resulting surface integrals have to vanish
\citep[see][for details]{aly_89}, or in practice must then be sufficiently small.
Theoretically, the surface integrals
need to be evaluated over the entire boundary of the computational domain,
but in practice this is only possible for the
the bottom (photospheric) boundary, where the field is measured. This is justified
for ARs that are surrounded by weak (quiet-Sun) fields where the gross part
of the magnetic flux closes within the AR (\ie, on the bottom boundary) and the
contribution of the other boundaries can be neglected.

Only exceptionally, however, active-region
vector magnetograms fulfill the force-free criteria
\citep[for such an example see][]{2012SoPh..281...37W}.
In the majority of cases, they are not force-free, simply because
the photosphere is a non-force-free environment. Additionally,
polarization signals are often affected by the temperature in
the sampled magnetic features and introduce biases between,
\eg, sunspots and magnetic elements forming plage regions
\citep[][]{1987A&A...176..139G,sol_93}.
To circumvent this problem a procedure dubbed
``preprocessing'' has been developed. The method uses
(force-free inconsistent) photospheric vector field measurements as input
and provides a force-free consistent vector field as output
\citep[see][for details]{2006SoPh..233..215W}. An alternative is
to measure the magnetic field vector higher in the solar atmosphere, \eg,
in the low-$\beta$ chromosphere (exclusively, or in addition to
photospheric measurements).

To our knowledge, the first and so far only
NLFF extrapolation from vector magnetograms observed
simultaneously at multiple heights (at a photospheric and chromospheric level)
have been performed by \cite{2012ApJ...748...23Y}, in order to study the
structure of an AR filament.
One difficulty in combining and comparing two such
data sets is that the exact height in the atmosphere of the chromospheric
measurement is unknown.
As a reasonable approximation the authors
assumed that the chromospheric measurements refer to the height
of best agreement with the NLFF reconstruction based on the photospheric
vector field (about 2~Mm above the solar surface surface).

Despite the difficulties discussed above, NLFF
extrapolations are a powerful tool for deriving the 3D coronal magnetic field
above ARs. On the other hand the applicability of force-free models
to quiet-Sun magnetic fields is questionable because it is very likely neither
force-free nor quasi-steady \citep[see][and
section~\ref{qs_forcefree} for details]{schr_vba_05}.

\subsection{MHD models}
\label{model:mhd}

\subsubsection{MHD models of the coronal magnetic field}
\label{model:full_mhd}

A full understanding of the physical processes in
the upper solar atmosphere requires the knowledge of the plasma
that populates the investigated magnetic structures. Deriving these
properties in the outer solar atmosphere, however, remains a challenging task.
Most commonly used models for a self-consistent
description of the plasma and magnetic field are based
on the MHD approximation.
Interestingly, even though
the MHD approximation is strictly valid only in collisional plasmas,
the collision-free coronal
plasma is often modelled using such an approach.
More sophisticated, collisionless kinetic models cannot be applied to model
large-scale structures in the solar corona since the considered scales are
several orders of magnitude larger than the relevant (microscopic) scales
which have to be resolved in kinetic simulations
(\eg, the gyro-radius or Debye-length).
This approach, however, is frequently applied to model
the solar wind plasma \citep[see review by][]{mar_06}.

One approach to derive plasma quantities, which can then be compared
to observations, is forward modelling aided by time-dependent MHD simulations
\citep[see][]{2006ApJ...638.1086P}. As an initial state,
a potential field is computed from the measured photospheric
(LOS or vertical) magnetic field component. (Note that for
MHD simulations the magnetic field data has usually to be
scaled to a lower spatial resolution.)
A strength of the forward MHD modelling technique is that the resulting
plasma quantities can be used to
compute synthetic spectra, which can be
compared with observed chromospheric and coronal images/spectra
 \citep[\eg][using \soho/SUMER EUV data]{2006ApJ...638.1086P}.

\subsubsection{MHS models}
A simpler approach, when refraining from performing
numerically expensive time-dependent MHD simulations
is to use a reduced set of equations, \eg, MHS or stationary MHD.
This allows a self-consistent modelling of magnetic field and plasma
\eg, in the high-$\beta$ regimes containing the photosphere and lower chromosphere,
and beyond the source surface in global simulations.
Generally, these equilibria require the computation of nonlinear equations, which are
numerically even more challenging (and slower converging)
than the set of NLFF equations, in particular in a mixed-$\beta$ plasma
\citep[see][for an implementation in cartesian and spherical geometry,
respectively]{2006A&A...457.1053W,2007A&A...475..701W}.

Mathematically simpler, and
computationally much faster, is the subclass of MS models,
which are based on the assumption that electric currents flow on
spherical shells perpendicular to gravity (resulting in horizontal,\ie,
parallel to the lower boundary, currents in cartesian geometry).
This approach allows linearizing the MS equations and solving them
with a separation-ansatz
\citep[see][for a cartesian and spherical approach,
respectively]{1991ApJ...370..427L,1986ApJ...306..271B,1995A&A...301..628N}.
Because of the linearity of the underlying equations, a field-parallel
electric current can be superposed (for a constant value of $\alpha$).
The final current distribution consists of two parts: a LFF one
and another one that compensates non-magnetic forces such as
pressure gradients and gravity. These classes of MS equilibria
require only LOS photospheric magnetograms as boundary conditions,
are relative easy to implement and allow the specification of two free parameters
(the force-free parameter $\alpha$ and additionally a parameter
which controls the non-magnetic forces). The limitations on $\alpha$ are
similar to those discussed for LFF modelling approaches (see section~\ref{model:extrapol}).
In these models, plasma pressure and density are computed self-consistently
in order to compensate the Lorentz-force. Above a certain height the corresponding
configurations become almost force-free, which in principle
allows it to model a forced photosphere and chromosphere, together
with a force-free corona above. A limitation of MS
equilibria is that the two free parameters are globally
constant and the method does not guarantee a positive plasma
pressure and density. To ensure positive values of these quantities,
one either has to add a sufficiently large
background atmosphere (which may lead to unrealistically high values of the
plasma-$\beta$), or is limited to small values of the parameter controlling the
non-magnetic forces. Note that, as force-free approaches, MS
models are only snapshots of the coronal field
and the temporal evolution of such configurations
can only occur as a series of equilibria,
in response to temporally changing boundary conditions.

\subsubsection{Flux transport models}
\label{model:mhd:flux}

So far (for the aim of coronal magnetic field modelling)
we have discussed only the coronal response to photospheric changes, but did
not try to understand the evolution of the photospheric field itself.
This can be done on a large (global) scale with the help of flux transport
models \citep[][and for recent
reviews see chapter~2 in \cite{2012LRSP....9....6M} as well as
\cite{2014SSRv..tmp...43J}]{1964ApJ...140.1547L}.
The aim of magnetic flux transport models is to simulate how
(newly emerged) flux is transported
horizontally on the solar surface, \ie, in the photosphere.
The magnetic field is assumed to be radially oriented.
The main contributing flows and velocities on large
scales are differential rotation and meridional flows.
On smaller scales, convective processes on granular and
supergranular scales become important too, where
the granular scales are generally ignored.

A natural application of flux transport models is to
compute the evolution of active-region or global coronal fields
\citep[see][]{1987ApJ...319..481S,
2004A&A...426.1075B}, as well as to investigate
the development, structure and decay of polar
CHs \citep[see][]{1989SoPh..119..323S}, and
to estimate the Sun's open magnetic flux.
Flux transport computations performed in
recent times often start from observed magnetograms, \eg, full-disk
\citep[][]{schr_der_03} or synoptic \citep[][]{2004SoPh..219...55D}
LOS magnetograms from \soho/MDI. As a welcome side product,
fluxes are obtained also for regions where no LOS measurements
can be performed or where they are not reliable (\ie, at the far side
and around the poles of the Sun, respectively).
Additionally, such computations can be used to compensate gaps in
the original full-disk or synoptic LOS data.
%  Umformuliert:
% As a result one gets a data set which contains
% observed magnetograms whenever available with data gaps
% (in time and also for the rear side and poles of the Sun) filled
% with the output of the flux-transport model.
To our knowledge, current flux transport models provide only
the radial component of the photospheric field (\ie, not the full field vector),
however.

For the aim of coronal
magnetic field modelling, the resulting (synthetic) magnetic flux maps
can be used in a similar fashion as LOS magnetogram data.
Most popular for combined models of photospheric flux transport and
coronal field models are global potential field models.
A more sophisticated approach is to combine the flux transport model
with a NLFF approach, based on a magneto-frictional MHD relaxation code
(see \cite{2000ApJ...539..983V,2006ApJ...641..577M,2012LRSP....9....6M}).
In contrast to the NLFF extrapolation technique
based on vector magnetograms, this evolutionary method
requires only the radial photospheric field component.
Both, the photospheric and coronal magnetic field is
evolved in time by a combined approach: the photospheric
field by the flux transport model and the coronal field
by the magneto-frictional code.

\subsection{Coronal stereoscopy, tomography and seismology}
\label{model:stereo}

Rather than measuring or modelling the magnetic field itself,
we can get insights into the structure and shape
(but not the field strength) of magnetic
field lines by analysing images of the emitting coronal
plasma. This is possible because of, owing to the high conductivity,
the coronal plasma is frozen into the field and thus serves as tracer
of it.  Special techniques (coronal stereoscopy and tomography) have been
developed to reconstruct the 3D coronal structure from sets of
simultaneously observed 2D images
\citep[see][for a recent review]{2011LRSP....8....5A}.
Here we briefly summarize the techniques relevant for
magnetic field structures.

\subsubsection{Stereoscopy and magnetic stereoscopy}
\label{sss:stereoscopy}

Stereoscopy is classically carried out with
two (or more) images obtained from different vantage
points. It is preferably done with clear solid edges, which are, however,
not available for optically thin coronal structures
(such as loops or plumes). While some early work on solar stereoscopy
has been done from a single viewing direction
\citep[and using the rotation of the Sun to mimic multiple vantage points;
see][for details]{1985SoPh...96...93B}, the
application of both techniques got a big boost with the launch of
the \stereo\ spacecrafts.

A natural approach is to compare and combine the results of
coronal stereoscopy and magnetic field extrapolations from
the photosphere (called magnetic stereoscopy; \cite{2006SoPh..236...25W}
and for a review see \cite{2009AnGeo..27.2925W}).
In early applications, before vector magnetograms from \sdo/HMI became
routinely available, magnetic stereoscopy has been mainly performed
with the help of LFF fields (based on LOS magnetograms). The method
was designed to automatically find the optimal force-free parameter
$\alpha$ of the LFF model \citep[see][for the first application of this method to
\stereo/SECCHI images and \soho/MDI magnetograms]{2007ApJ...671L.205F}.
Stereoscopy and magnetic field extrapolations have complementary
strengths and weaknesses and it is by no means clear whether the
reconstructed 3D loops validly represent the actual coronal loops
\citep[see][for a comparison of force-free field modeling
and stereoscopy]{2009ApJ...696.1780D}. Nevertheless, a comparison of
the result of both approaches at least serves as a consistency check and
allows to approximate related uncertainties.
Recently, some attempts have been made to use coronal information
(either stereoscopically reconstructed 3D loops or 2D projections of loops
extracted from coronal images) to constrain NLFF fields in addition
to photospheric measurements (\cite{2014ApJ...783..102M},
\cite{2014ApJ...785...34A} and Chifu \etal, 2014, \aap, submitted).

Maps of optically thin radio emission (see section~\ref{model:radio}.)
can be treated basically similarly to EUV and SXR images.
This is different for observations of optically thick sources, which have a
similar opacity as a solid 3D body. Consequently for a given 3D magnetic
field structure, one finds different
\citep[see section~3.5 in][for details]{2011LRSP....8....5A}
gyroresonance layers that are visible
as equi-contours in 2D images, dependent on frequency and harmonic.
For slowly evolving magnetic fields, which remain almost static for a few days,
the solar rotation can be used for a stereoscopic 3D reconstruction of
the magnetic field structure. Here, the structures have a high opacity,
making stereoscopy more straightforward compared to using images of
optically thin sources. A comparison of this method
with force-free magnetic field reconstruction methods based on
photospheric data revealed that a potential field model failed to reconstruct
a corresponding structure, whereas a NLFF approach showed
a reasonable agreement \citep[see][for details]{1999ApJ...510..413L}.

\subsubsection{Tomography and vector tomography}
\label{sss:tomography}

A complementary approach, which is specifically
tuned to optically thin structures, is solar tomography.
To our knowledge, this was first proposed by
\cite{1994ApJ...423..871D}. This method uses LOS-integrated
coronal images, preferably from multiple viewpoints, as input.
Unfortunately, a large number of viewpoints
is not available for solar observations and we are currently limited
to a maximum of {\mod three} viewpoints
(\stereo-A and B, plus either \sdo\, or \soho).
In future, \so\ will provide a fourth viewpoint.
In principle, one can extend the number of viewpoints by taking images
sequentially one after the other, while the Sun rotates.
Because the vector tomographic inversion requires data from
multiple viewpoints, one would need to observe the
rotating Sun for several days if only one viewpoint, \eg~from Earth,
is available. Then, the analysis is limited to static or slowly
evolving structures.

Fortunately for the
aim of coronal magnetic field investigation, the large-scale
magnetic field structure changes more slowly compared with the plasma
(which exhibits flows and reacts to, \eg, heating and cooling).
Sources for a tomographic inversion are EUV images,
white light images in which the radiation is dominated by
Thompson scattering, and radio maps.
Consequently, the physical conditions for both stereoscopy and
tomography of the Sun are not ideal as compared with the stereoscopy of solid
objects on Earth and
% the use of tomography in medicine and
one has to find suitable ways of combining different techniques
to obtain the best scientific insight from available observations.

The inversion of magnetic field related polarization signals is more challenging
than inferring the plasma density, because the magnetic field is a vector.
The corresponding vector tomography methods
\citep[][]{2006A&A...456..665K,2013ApJ...775...25K} require,
as in ordinary scalar tomography, an assumption for a regularization
(in addition to LOS-integrated magnetic field measurements of magnetically
sensitive coronal lines).
This may be, \eg, the assumption of a solenoidal magnetic field.
The corresponding
regularization integrals require a boundary condition, which can be
derived from photospheric magnetic field measurements, be they
LOS or radial components. As a consequence, the vector tomographic inversion
is not independent from photospheric measurements.
In principle, there is a large potential of combining methods of
NLFF coronal magnetic field models with vector tomographic inversions.
Models for both approaches can be
derived from optimization principles, which makes a combination
mathematically straightforward. The computational
implementation, however, remains challenging.

\subsubsection{Coronal seismology}
\label{sss:seismology}

While the stereoscopic reconstruction from coronal images provides
only the 3D shape of coronal loops, but not their field strength,
we can get insights into the coronal magnetic field strength
by analysing loop oscillations, which are often visible
in time series of coronal images or spectra.
The principle has been well known for several
decades: the properties of waves travelling through a magnetized
medium react to the magnetic field strength
(for global seismology see \citealt{1970PASJ...22..341U}, and
for local coronal seismology see \citealt{1984ApJ...279..857R}).
It is outside the scope
 of our article to review the rich history of coronal seismology, but
 the reader can find an overview of solar coronal waves and oscillations,
 including an introduction to coronal seismology in
\cite{2005LRSP....2....3N}.
The basics for seismology are the analysis of waves, here within
the limit of an MHD approach. MHD waves (slow and fast magneto-acoustic waves,
and Alfv{\'e}n waves) are sensitive to the magnetic field of the medium
through which they travel. In principle, the method has similarities with using
acoustic waves for helioseismology of the Sun's interior. Coronal
images and spectra of high spatial resolution and time cadence
(for example from TRACE, \sdo/AIA and \soho/SUMER,
\hinode/EIS, respectively),  allow a reliable application of coronal seismology to
compute the magnetic field strength \citep[see][]{2008A&A...487L..17V}.

\section{Global coronal magnetic fields}
\label{s:global}

\subsection{Magnetic field topology}
\label{global_top}

The structure of the solar corona is dominated
by magnetic fields that emerge from below the solar surface
and expand into the atmosphere. The plasma confined in these fields
is visible in coronal images and outlines
open and closed coronal magnetic
fields. Most of today's knowledge on the nature of the
coronal magnetic field, however, was gained from
coronal magnetic field models (see section~\ref{s:modeling}) and their
comparison with observations of the radiation emitted by coronal plasma.
Such a combination of modelling and observations is necessary
due to the lack of routine direct measurements of the 3D
magnetic field vector in the upper solar atmosphere
\citep[see review by][]{car_09}.

\begin{figure}
  \centering
  \includegraphics[width=\textwidth]{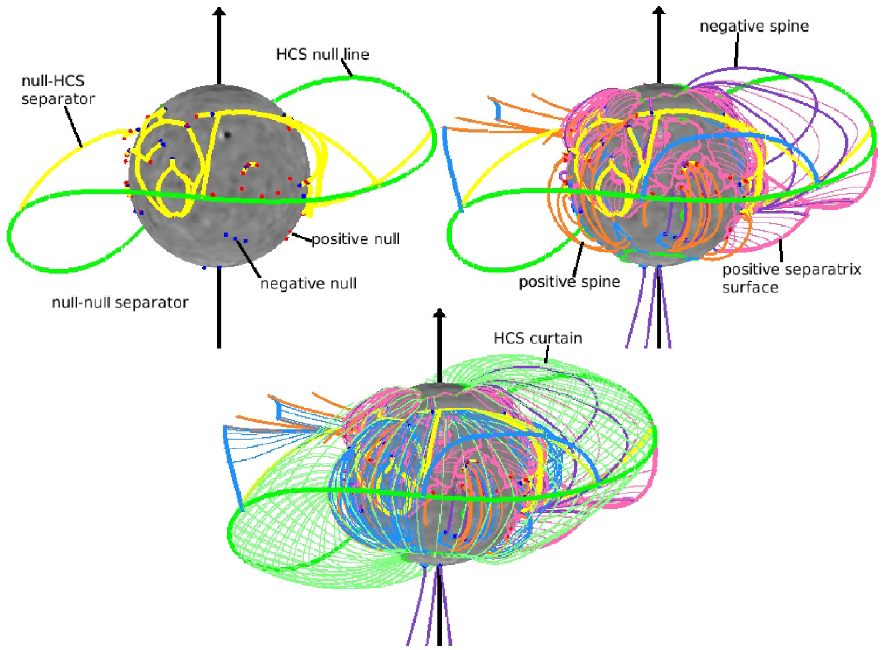}
  \put(-285,240){\bf\sf(a)}
  \put(-115,240){\bf\sf(b)}
  \put(-200,125){\bf\sf(c)}
  \caption{Illustration of the basic elements of the global coronal
  magnetic-field topology, which is characteristic for the solar
  minimum period and was calculated from a PFSS model.
  The gray-scale sphere reflects the \solis/VSM radial
  component of the synoptic photospheric magnetic field.
  (a) The thick green line marks the neutral line on the source surface,
  which is the base of the heliospheric current sheet.
  Blue and red dots mark null points in the corona, \ie, places
  where the magnetic field vanishes. Yellow thick
  lines represent separators, \ie, field lines that mark the intersection of
  two separatrix surfaces from opposite-polarity null points.
  (b) Separatrix surfaces (pink and blue lines) and spine field lines
  associated with the null points (orange and purple lines)
  are plotted on top of the features shown in (a).
  (c) Superimposed on the features depicted in (b) is the ``heliospheric
  current sheet curtain'' (green thin lines) which separates closed from open fields.
  (Adapted from Figure 6 of \cite{pla_par_14}. Reproduced with permission from Astronomy \& Astrophysics, \copyright~ESO.)
  }
  \label{fig:pla_par_14_fig6}
\end{figure}

\subsubsection{Performance of PFSS models}

\cite{pla_par_14} recently presented a detailed topological picture
of the global solar corona based on
PFSS modelling (see section \ref{model:extrapol}).
Even during times of minimal solar activity,
a quite complex picture of the coronal magnetic field is revealed (see
Figure~\ref{fig:pla_par_14_fig6}).
These authors summarized the building blocks of
the coronal topology to include the neutral line at the source surface,
separatrix surfaces which physically separate closed and open fields
(``separatrix curtains''), and various types of smaller separatrix surfaces
closing below the source surface. The coronal neutral line separates the
large-scale opposite polarity regimes of the coronal magnetic field.
Above the source surface (a regime not modelled within the PFSS approach)
the neutral line is used as a proxy for the base of the
heliospheric current sheet \citep[for details on topological considerations
of solar magnetic fields and the associated terminology see][and see
section~\ref{ss:local_topology} for its representation on active-region
scales]{lon_05}.

PFSS models are widely used to picture the structure of the Sun's global
magnetic field, basically owing to their mathematical simplicity
and because only LOS photospheric magnetograms are required as
boundary condition.
 The ability of such models to adequately reflect some of the
observed structures, however, seems to be limited or at least dependent on
the case studied and/or the specific analysis that is carried out.
Thus, \cite{wan_she_07} used a PFSS model to show that the magnetic structure
of pseudo-streamers, as seen in \soho/LASCO white light images,
is rooted between
open fields emanating from photospheric regions of the same polarity. Using
the same model approach, \cite{zhu_sae_08} investigated mid-latitude
coronal streamers.
During periods of high solar activity, however, they found no satisfactoring
agreement of number and positions of the streamers, especially of
those originating from polar regions.

One has to keep in mind that during times of high solar activity large parts of
the solar atmosphere are filled with non-potential magnetic
fields, in particular in ARs.
A PFSS approach will then be limited in its success of
reproducing the corresponding coronal magnetic field structure
\citep[][]{nit_der_08}. \cite{rus_hag_08} estimated that, at best, in only
about 50\% of the cases, a PFSS model might be capable of
reproducing the locations of open magnetic field structures associated to
flaring ARs. More recently, \cite{kra_air_14} highlighted the limited success
of PFSS models in reproducing coronal streamers even during periods of
minimal solar activity. Depending on the height of the source surface,
they found that the PFSS models could not or could only partly render the
positions of coronal streamers and CHs seen in \stereo/EUVI
195~\AA\ images.

One reason for the partial mismatch
between PFSS model results and observed coronal structures is
their current-free nature. A very likely other reason is the use
of synoptic maps as
photospheric boundary condition. Synoptic maps are usually
created by combining the data near the central meridian of
full-disk, LOS magnetograms that are acquired daily over one Carrington
rotation. They are thus representative for the activity near disk center
in the course of one Carrington rotation, with the extreme longitudes
on the map having been recorded 27 days apart. To model the locations
of streamers and CH boundaries at a certain instance, however,
real-time knowledge of the far-side magnetic field configuration
needs to be available as well.
\cite{schr_der_03} have shown that the lack of knowing the instantaneous
magnetic field distribution on the far side of the Sun can be compensated.
They used acoustic far-side imaging to forecast strong magnetic field
concentrations before they appeared on the east solar limb.
After including this information in the PFSS
model, the CH boundaries as observed in SXR and He\,{\sc i} synoptic
maps matched quite well. This, and the above compilation
of case studies involving PFSS modelling,
implies that one needs to judge, by comparing with
observations, the quality of PFSS model results in each case.

\subsubsection{Achievements of global MHD models}

In PFSS models the effect of the solar wind is taken into account only
by the assumption that all field lines become radial at the source surface.
This is not sufficient to investigate wind properties
themselves, because they
require the application of models that include plasma flows,
such as global MHD models.
Such approaches are mathematically more complex and
incorporate more physics than PFSS
models. They consequently suffer from requiring also
longer computation times (\cite{ril_lin_06}; for a review see
\cite{2012LRSP....9....6M}, see also section \ref{model:mhd}).

\begin{figure}
  \centering
  \includegraphics[width=\textwidth]{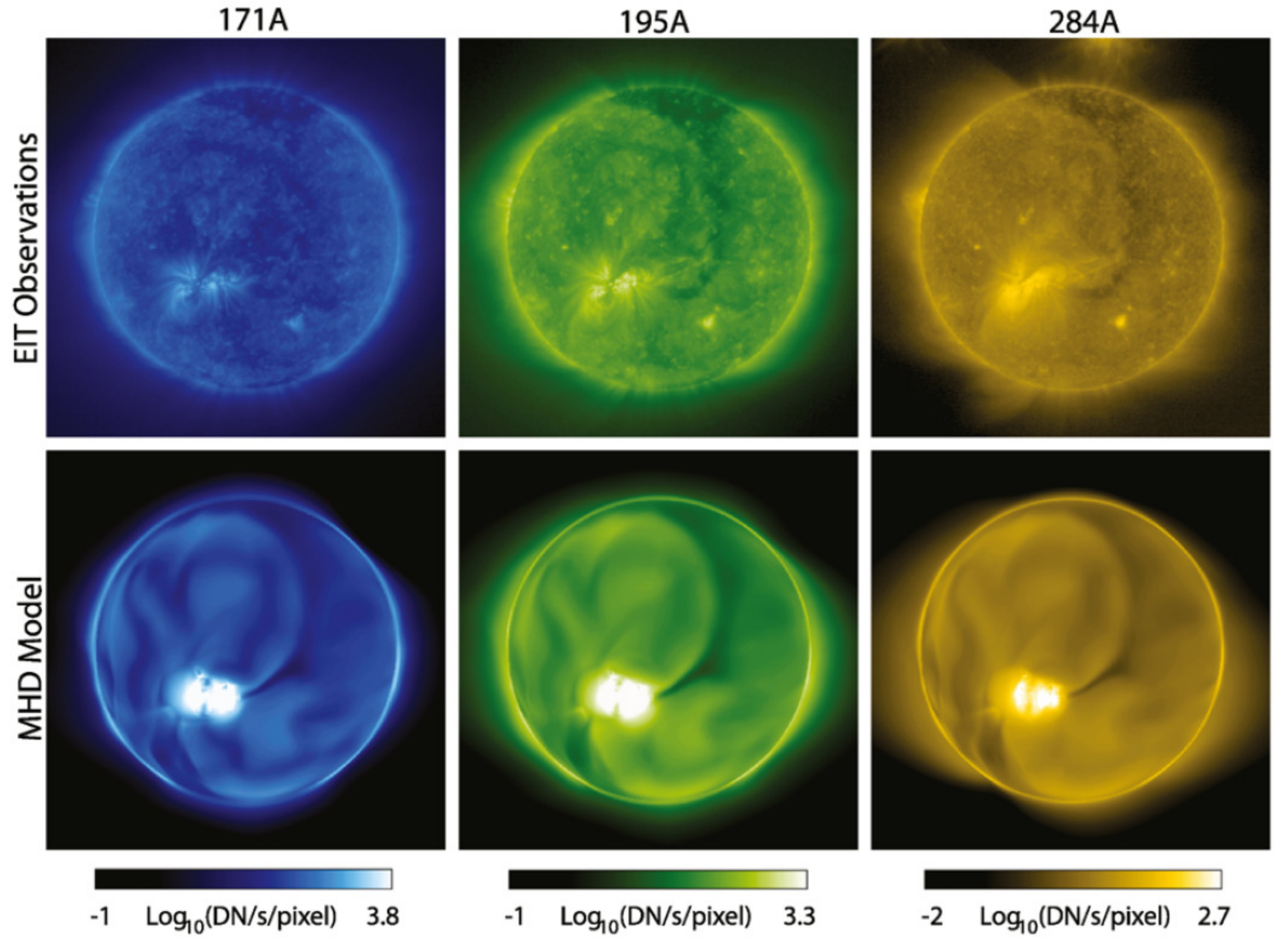}
  \put(-322,230){\bf\sf\white(a)}
  \put(-215,230){\bf\sf\white(b)}
  \put(-105,230){\bf\sf\white(c)}
  \put(-322,120){\bf\sf\white(d)}
  \put(-215,120){\bf\sf\white(e)}
  \put(-105,120){\bf\sf\white(f)}
  \caption{Synoptic maps of the coronal EUV emission at (a) 171~\AA,
  (b) 195~\AA\ and (c) 284~\AA\ for Carrington rotation 1913
  (covering the period August 22 to September 18, 1996 at zero longitude),
  recorded by \soho/EIT.
  A large CH (seen dark with respect to the
  quiet-Sun emission) extends from the north pole into the southern hemisphere.
  Synthetic (d) 171~\AA, (e) 195~\AA\ and (f) 284~\AA\
  spectroheliograms,
  computed from the densities and temperatures obtained with a global MHD
  model are shown below. Importantly, the overall brightness of
  the images compares. In
  addition, several features, including the cross-equatorial extension of the
  northern polar CH and the the position of the AR south of the equator are
  well recovered.
  (Adapted from Figure 6 of \cite{ril_lio_11}. With kind permission from Springer Science and Business Media.)
  }
  \label{fig:riley11}
\end{figure}
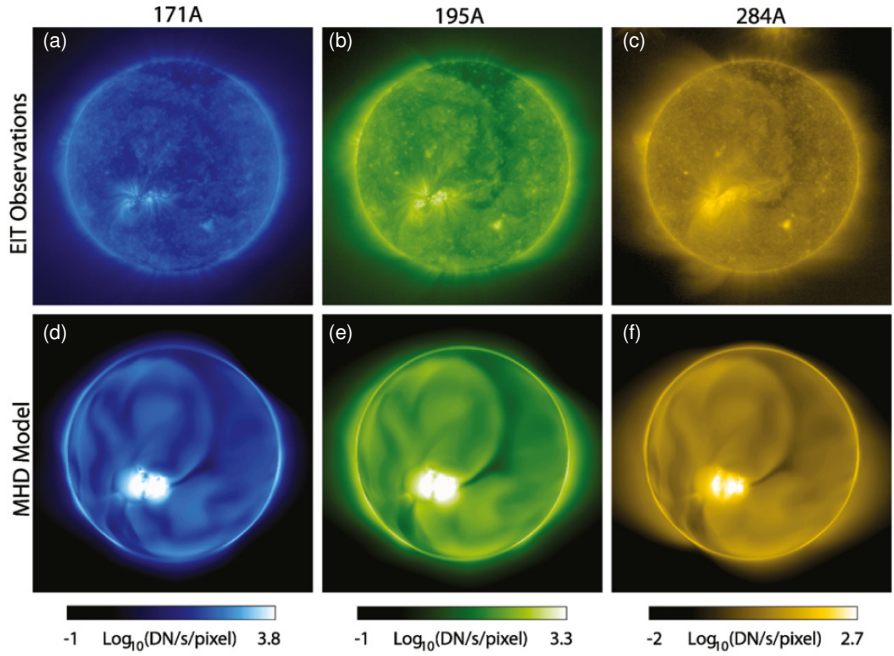

If employed, they deliver properties beyond the
coronal magnetic field, normally the density and temperature
of the coronal structures. \cite{ril_lio_11} showed that such models have the
ability to quantitatively reproduce the signatures seen in coronal images,
provided additional assumptions on the coronal heating function
are included (see Figure~\ref{fig:riley11}).
The north polar CH extends well down into the
southern hemisphere and its shape is well recovered within the MHD
model result. The location of the AR south of the solar equator is
reproduced as well, but appears too bright compared to the observed
EUV emission. Small-scale features as well as plasma
emission from open field lines anchored in polar regions
have not been recovered.
Advanced MHD techniques have been used to
successfully model the variation of the solar wind speed
\citep{hu_fen_08,nak_tan_09,yan_fen_12}. This includes the fast
and slow solar wind, the sector structure of the interplanetary magnetic
field as well as the shape and location of CHs (see section~\ref{s:chs}).

\subsection{Cross-equatorial fields}

Global observations often
show loop structures extending across the solar equator, thereby connecting
the two hemispheres of the Sun (``trans-equatorial loops''; TELs).
These are systems of magnetic field lines
bridging the solar equator that connect active and/or
quiet-Sun regions.
The TELs become sheared above the solar surface due
to their line-tied footpoints being subject to differential rotation
(in the long term), or vortex motions and/or
the rotation of sunspots (on shorter time scales).
This may also cause the constituent field lines of the TELs to become twisted
around a common axis \citep[][]{bao_sak_02}, which may then be observed
in the form of a sigmoid in coronal images.
Consequently, an observed sigmoidal structure does not necessarily imply
that the underlying non-potential field geometry was already present when
the fields emerged \citep{pev_can_97}.

\subsubsection{Creation of transequatorial loops}

The creation of TELs is still far from being understood, mainly
because the most frequently employed 2D dynamo models cannot account for
their intrinsic 3D nature and formation process \citep[][]{jia_cho_07}.
3D models aspiring to resolve the problem are still in their infancy
\citep[see][for recent applications]{2013MNRAS.436.3366Y,2014ApJ...785L...8M}.

Active-region magnetic fields that
connect across the solar equator and form sigmoidal loop
systems have been described by, \cite{sve_kri_77} and
\cite{tsu_96}. They observed newly created loops,
connecting across the Sun's equator in \yohkoh/SXT images and
thus that the number of such newly brightened connections increased with time.
\cite{yok_mas_10} addressed the question how TELs, appearing bright in
\yohkoh/SXT images, can connect very distant regions on the solar surface.
In their picture, a series of reconnection processes between equator-bridging
weak magnetic fields and their neighbouring strong active-region fields
reconfigures the magnetic connectivity such that strong equator-bridging
fields develop. Besides, they pictured how
this process may result in simultaneous chromospheric evaporation
signatures at both footpoints of the TEL system \citep[which had been
reported by observationally by][]{har_mat_03}.
\cite{2010ApJ...723L..28L,liu_wan_11} analysed a coronal
current sheet which appeared bright in \soho/EIT 195~\AA\ images
above a cusp-shaped flaring loop that connected locations on either
side of the solar equator. They presented the first comprehensive
set of observations, providing
support for the standard picture of flare/CME events
\citep[usually referred to as the ``CHSKP model''][and see also
section~\ref{sss:evolution_corona}]{car_64,hir_74,stu_96,kop_pne_76}.
These included the convergence in the legs of the TEL
system in the area where the formation of a cusp-shaped flare loop
was formed later on, co-temporal radio signatures,
as well as an expanding post-flare arcade and coronal dimming in the
atmosphere above it.

\subsubsection{Properties of transequatorial loops}

\begin{figure}
      \centering
      \includegraphics[width=0.75\textwidth]{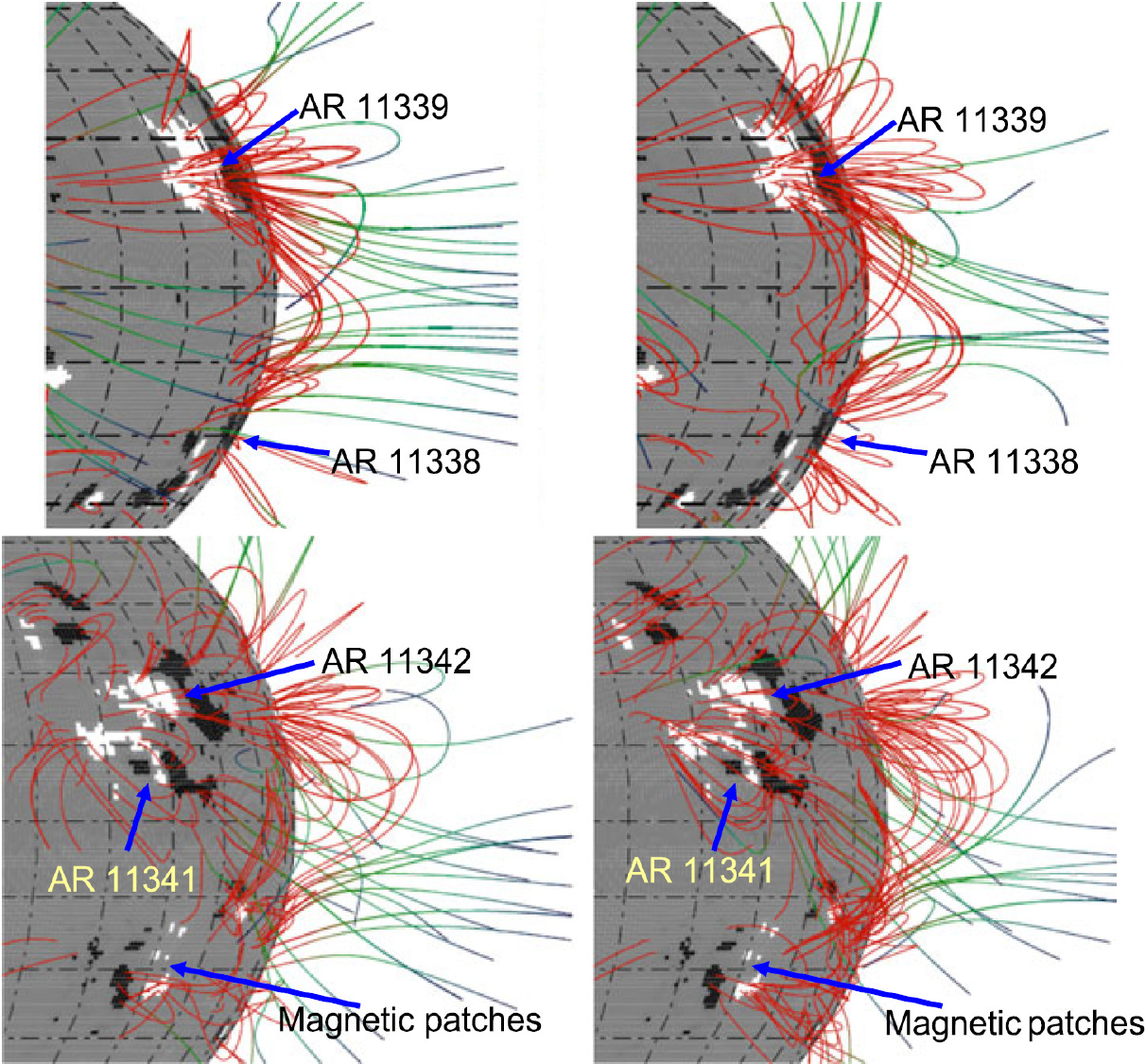}
      \put(-260,225){\bf\sf(a)}
      \put(-130,225){\bf\sf(b)}
      \put(-270,112){\bf\sf(c)}
      \put(-140,112){\bf\sf(d)}
      \caption{Model magnetic field configurations connecting strong magnetic fields
      on either side of the solar equator. Selected field lines are calculated from a global
      (a) potential field and (b) a NLFFF model connecting AR 11339 and AR 11338. Potential
      and NLFFF model field lines connecting two ARs (11342 and 11341) to
      strong magnetic patches are shown in (c) and (d), respectively. Red and green line
      colors indicate closed and open magnetic field lines, respectively. The gray-scale
      background reflects the measured radial \sdo/HMI magnetic field. It can be seen
      that the non-potential TELs deviate only little from a potential configuration.
      (Adapted from Figure~5 of \citealt{tad_wie_14}. With kind permission from Springer Science and Business Media.)}
      \label{fig:tad_wie_14_fig5}
\end{figure}
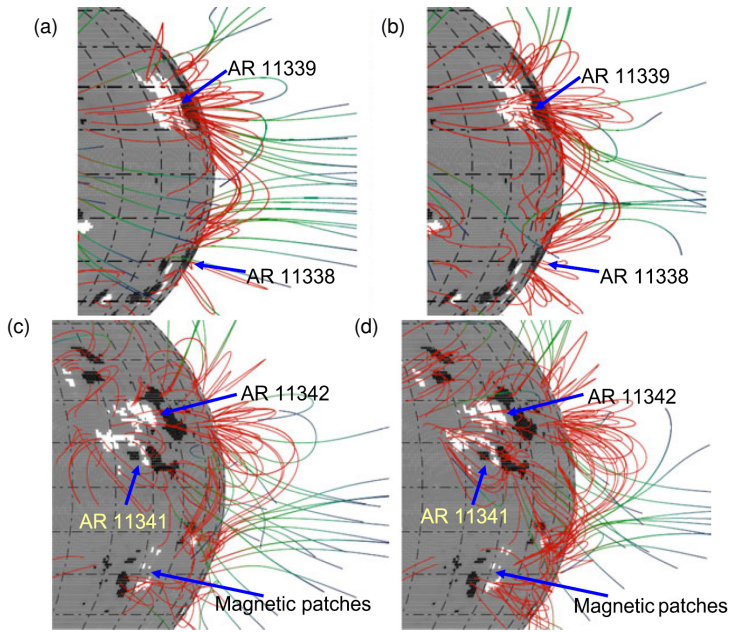

More frequently global NLFF modelling techniques are being used to
investigate the magnetic connectivity between
ARs located on either side of the solar equator
(see Figure~\ref{fig:tad_wie_14_fig5}).
For instance, \cite{tad_wie_14} found that TELs
carry only weak electric currents.
The departure from a potential state, however, stresses the
importance of using magnetic field
models that allow departures from a current-free state.
At the same time, global force-free magnetic field modelling
can strongly depend on the boundary conditions supplied.
Measurement errors are particularly
 high in the transverse component of weak quiet-Sun fields.
 Photospheric vector field measurements
 with different instrumentation can differ from each other and
 influence the corresponding coronal field models.
\cite{tad_wie_13} performed a comparison of global NLFF
models based on full disk vector magnetograms from
 \solis/VSM and \sdo/HMI. In this work some of the
 TELs clearly observed in coronal images where reproduced
 with the help of NLFF modelling based on \solis/VSM data, but
 not from the modelling based on \sdo/HMI data.

\cite{pev_00} tested the importance of the chirality (handedness) of
active-region magnetic fields
 for the formation of TEL systems. The results
suggested that in roughly two thirds of the cases the connected
active-region fields were of the same handedness. Recently, \cite{che_lun_10}
examined the twist of a larger number of TELs (a subset of the samples
analysed by \cite{che_bao_06}). They found that the ones that linked ARs
displayed an obvious sigmoidal shape and were related to a flaring
activity stronger than C-class (\ie, peak SXR fluxes of ${>}10^{6}$~W\,m$^{-2}$).
They calculated the ratio $\tilde{L}/D$, where $\tilde{L}$ is the apparent
length of the TEL system and $D$ is the apparent distance
between the locations where the TEL system seems rooted at the solar
surface. $\tilde{L}$ was measured by tracing the length of the coronal
loops at the outer edges of the sigmoidal loop system, where they are
well distinguished from the faint emission from the (quiet-Sun) background.
Higher values of that ratio $\tilde{L}/D$ indicate a more pronounced sigmoidal
shape and thus imply a stronger twisting of the associated field lines.
They found that most of the TELs possess only weak sigmoidal shapes,
indicating a low degree of non-potentiality. It appears that flares above
C-class preferentially originate from structures of a specific amount of twist
($\tilde{L}/D\approx1.4$). It is an important future task to model the
associated cross-equatorial 3D coronal magnetic field and its evolution
with the help of global force-free and time-dependent MHD models
in order to reveal the importance for eruptions to occur (see also
section~\ref{ss:eruptive_conditions}).

\subsection{Spatio-temporal aspects of activity}
\label{ss:global_cyclic}

\subsubsection{Cyclic changes of the coronal magnetic field}

As the coronal field responds to photospheric changes,
it also changes on global scales with an approximately
decade-long periodicity. Bright loop systems appear at higher
latitudes ($\gtrsim30^\circ$) at the beginning of a solar cycle and
the activity belts
progressively move closer to the equator as the cycle progresses.

Coronal holes vary in shape and position with the solar cycle as well
\citep[see review by][and section~\ref{s:chs}]{cra_09}. While polar CHs,
if present at all, are found to cover only small areas around the poles
during solar maximum, they tend to reach their largest extents around solar
minimum and then occasionally extend to latitudes as low as
$\approx60^\circ$. Low-latitude CHs emerge
preferentially around solar maximum and near the activity belts
due to the occasional accumulation of unipolar flux during the emergence
of ARs. If a sufficient amount of unipolar flux emerges and clusters together
to form extended patches, the associated fields may be open. As a
consequence, the open flux on global scales varies in the course of the
solar cycle \citep[its long-term variation was computed by][]{2000Natur.408..445S,2002A&A...383..706S,
2002ApJ...577.1006S,2011A&A...528A..83J}.
Since the interplanetary magnetic field is fed by the magnetically
open regions on the Sun, a similar modulation over the course of a solar cycle
was expected too. Thus, it was surprising that the magnitude of the radial
interplanetary magnetic field strength does not show a strong dependence
on the activity level.
\cite{wan_she_00} and \cite{wan_lea_00} investigated why this is so and
argued that the reduced area occupied by open fields around solar
maximum is compensated by their, on average, higher field strengths.
As a consequence, the open solar flux is nearly maintained throughout
the solar cycle.

During solar activity maximum strongly emitting (bright) loop systems
occupy a considerably larger volume within the corona.
Associated active features, including flares \citep[][]{bai_03,zha_mur_11}
and coronal streamers \citep[][]{li_11} as well as super-ARs \citep[\ie, ARs
associated to repeated flaring and mass ejections, cf.,][]{tia_liu_02}
were found to be distributed inhomogeneously in solar longitude.
They seemingly relate to ``active nests'' or
``active longitudes'', which had originally been postulated based on
similar trends seen in surface magnetic field observations.
\cite{chi_32}, for instance,
found sunspots to preferentially emerge at particular longitudes on
the solar disk, which has been almost immediately attributed to projection
effects and/or selection criteria by \cite{car_33}.
Follow-up studies concerning the preferred locations (longitudes) of
sunspot formation delivered inconsistent and partly contradictory results.
Even today, the possible number, migration, life-times, long-term
behaviour and the particular method to track them remains a subject of
debate \citep[see][]{ber_uso_03,uso_ber_05,pel_bro_06,uso_ber_07,
2013ApJ...770..149W,2014SoPh..289..579G}.

\subsubsection{Association to dynamic events}

\begin{figure}
  \centering
  \includegraphics[width=\textwidth]{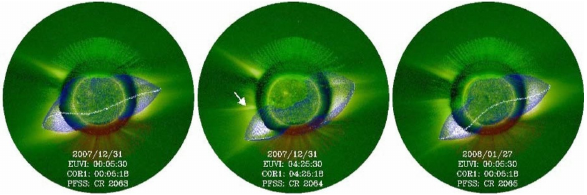}
  \put(-340,100){\bf\sf(a)}
  \put(-230,100){\bf\sf(b)}
  \put(-115,100){\bf\sf(c)}
  \caption{Composite \stereo/SECCHI EUVI 195~\AA\ (Sun at the center
  of each image) and COR1
  (surrounding disk) images (a) before, (b) immediately after and (c) 27 days
  after a CME. The white arrow in (b) indicates a streamer-like structure behind
  the CME which was interpreted as the current sheet in its wake,
  also because it did not exist before (compare (a)), nor about a month
  after the ejection (compare with (c)).
  The blue lines represent magnetic field lines calculated from a global PFSS
  model and projected onto the \stereo/SECCHI images. In a similar manner,
  the white line displays the source surface neutral line. It can be seen that, as
  a consequence of the CME, the coronal streamer migrated southwards
  (compare the position of the coronal streamer above the south-east limb of
  the Sun in (a) and (b)) but survived and persisted for more than one
  additional solar
  rotation (note the location of the streamer in (c)).
  (Figure 3 of \cite{liu_luh_09}. \copyright~AAS. Reproduced with permission.)
  }
  \label{fig:liu_luh_09_fig3}
\end{figure}

Mass ejecta cause severe changes of the coronal magnetic field, leaving
it behind in a massively reconfigured configuration.
Changes in the coronal magnetic field
configuration, owing to magnetic reconnection of
non-potential closed magnetic fields below coronal streamers,
for example, are in
fact thought to be the building blocks of the basic mechanism of
coronal mass ejecta \citep[for reviews see][]{for_00,che_11}.
\cite{liu_luh_09}, for instance, presented observations
of short-lived as well as of lasting deformations in association with an
observed CME. The transient modifications were seen in the form
of structures, neighbouring the ejection site, pushed aside
and bouncing back. Lasting distortions were observed in the form of
displacements of the associated helmet streamer and the shrinkage
of coronal holes (see Figure~\ref{fig:liu_luh_09_fig3}).

On the other hand, it has also been reported that,
in a considerable number of cases, the coronal environment
barely responded to or appeared insensitive to the occurance of mass
ejections. This viewpoint has been championed by \cite{sime_89},
who argued that the global evolution of the coronal
magnetic field gives rise to CMEs but is not really influenced by them.
In other words, the ejecta may be associated to the streamer belt but may
not have a lasting effect on it.
This was supported by the observation that
the streamer belt appeared to reestablish itself within a few
days after ejections \citep{zha_hoe_96}. Although, it is certain
that CMEs are associated with the belt of coronal streamers
\citep{hun_93}, the disruption
(or disappearance) of an associated streamer was observed
in only $\approx15\%$ of the cases observed
\citep[][]{sub_der_99,flo_lam_14}.
In summary, whether mass ejections are only a response to or a contributing
factor to the coronal restructuring is not yet clear \citep{liu_luh_09}.

It is generally agreed,
however, that only mass ejections (which inevitably drag the embedded
magnetic field along) are capable of physically reducing the coronal magnetic
helicity, which is tightly related to the structural properties
of the magnetic field \citep{mof_69}. We discuss this in the following.

\subsection{Magnetic helicity budget}
\label{ss:hbudget_global}

\subsubsection{Helicity dissipation and helicity transport}

Magnetic helicity is dissipated on significantly longer scales
than the magnetic field and consequently the
dissipation time in the corona is too long to relevantly reduce the
helicity \citep{ber_84}.
For the helicity budget of ARs and quiet-Sun regions
see sections \ref{ss:helicity_ar} and \ref{QS:helicity}, respectively.
Since magnetic helicity cannot be
efficiently dissipated, it is approximately conserved in the absence of
ejecta, \ie~the magnetic helicity in the solar corona will continuously
build up.
That has important consequences for the magnetic field relaxation towards
a lower energy state. It implies that in the course of a
flare, but lacking a mass expulsion, a non-potential field
can only relax to another, lower-energy
configuration of the same helicity content
\citep[\ie, a constant-$\alpha$ field; see][]{tay_74,hey_pri_84,tay_86}.
However, the coronal magnetic field may never entirely relax to this
constant-$\alpha$ (conserved-helicity) state, because the required
``complete'' reconnection is not expected to occur due to the line-tying of
the coronal magnetic field structures at photospheric levels.
This also inhibits the simultaneous formation of numerous current-sheets
\citep[see][and references therein]{ant_dev_99b}.

The majority of ejecta, including CMEs associated with disappearing
filaments \citep[][]{yur_wan_01,cho_par_13} or eruptive X-ray loops
\citep[][]{man_poh_05,zhe_jia_11}, propagate away from the Sun in the
form of magnetic clouds (MCs) and hence actively carry helicity away.
MCs are force-free regions of enhanced magnetic
field strength, with the field vector monotonically rotating as they journey with
the solar wind. There are the moving large-scale helical structures
described by
\cite{bur_sit_81,bur_88}. They are a sub-set of interplanetary CMEs that
consist of plasma and magnetic field, expanding behind a shock wave into
interplanetary space. \cite{gop_han_98}
suggested that MCs originate from the structure overlying an
eruptive prominence and its associated CME and include the coronal cavity and a
bright frontal structure.
A typical MC may carry a magnetic helicity of some
$10^{41}$~Mx$^2$ to $10^{43}$~Mx$^2$ \citep{cho_par_13}.
This, and considerations based on surface magnetic field measurements,
permit to estimate the amount of helicity transported away on a global scale
during one solar cycle as $\approx10^{45}$~Mx$^2$ to $10^{46}$~Mx$^2$
\citep{bie_rus_95,rus_97,dev_00,ber_ruz_00,geo_rus_09,zha_yan_13}.

\subsubsection{Hemispheric trends}

When estimating the global
helicity budget, it is of importance to
take the Sun's differential rotation into account.
\cite{dev_00}
estimated the effect which the Sun's differential rotation has on the shearing
of an active-region magnetic field. He estimated an accumulated helicity as
$\approx10^{43}$~Mx$^2$ during a characteristic active-region
lifetime of $\approx120$~days. However, the differential
rotation cannot represent the only source of helicity supply to the corona
\cite{dem_man_02b}. \cite{geo_rus_09} suspected that the contribution
of differential rotation to the total amount of injected magnetic helicity
amounts only to about 20\%. The dominant source of injection, they argued,
must be due to the plasma flows within ARs (see
section~\ref{ss:hstorage_ar}). This agrees with earlier results
which stated that the amount of helicity injected by differential rotation in
ARs may comprise roughly 10\% to 50\% of that injected by motions within
the ARs themselves \citep[][and references therein]{dem_par_09}.

The injection of magnetic helicity does not appear to display any periodicity,
indicating it to be a rather unforeseeable process \citep{geo_rus_09}.
Although temporal periodic patterns have not been found, some systematics
regarding the spatial distribution of helical features on a global scale are
known. Several observational features indicating the handedness of
structures, such as sunspot whorls \citep[][]{hal_25,ric_41}, chirality
of filaments \citep[][]{rus_67,pev_bal_03,ber_rus_05} and S-shaped
coronal X-ray brightening \citep{rus_kum_96,can_hud_99} revealed
a dominant positive helicity pattern in the northern solar hemisphere and a
dominant negative one south of the solar equator.
The associated magnetic fields in the northern and southern hemisphere are
dominated by right- and left-handedness, respectively
\citep[][]{see_90,pev_can_95}.
Interestingly, these patterns do not change from one activity cycle to the next.

This tendency, however, was found to be less pronounced for
active-region filaments than for filaments in the QS \cite[][]{pev_bal_03}.
\cite{bao_sak_02} gave a hint why there is a weaker hemispheric
trend of active-region magnetic fields. They argued that the chirality
introduced by the same mechanism acting on a rising flux tube throughout
the convection zone, or acting on the flux tube after emergence through the
photosphere may be opposite. The dominant contribution would then
determine the chirality of an AR which would not necessarily follow the
hemispheric trend. Using a magneto-frictional model, \cite{yea_mac_08}
were able to reproduce the skew of more than 90\% of considered
filaments (observed in H$\alpha$ images) correctly, even including
exceptions from the hemispheric trend. \cite{yea_mac_12} recently
presented the first long-term simulation of the chirality of high-latitude
filaments and were able to recover the hemispheric trend. According to
them, the apparent handedness depends on which of the two effects,
creation of helicity by differential rotation or its transport from active
latitudes, is stronger. They found that the latter is generally stronger,
except during the early years of a
solar cycle. Further evidence for the dependence of the chirality of features
in the solar atmosphere on the phase of the solar cycle
comes from SXR loops \citep{2010ApJ...719.1955Z} and
active-region magnetic fields \citep{2010MNRAS.402L..30Z}.

Having started on magnetic fields on active-region scales, we
continue to discuss their magnetic and helicity properties in the next section.

\section{Coronal active-region magnetic fields}
\label{s:active}

\subsection{Coronal loops}
\label{sss:coronal_loops}

The frozen-in condition for the coronal plasma and
magnetic field is valid in most of the coronal environment.
 The sole
exceptions are the small diffusion regions in strong
current concentrations,
the so-called ``current sheets'' (see section~\ref{ss:local_topology}).
Much of the plasma in the corona is confined
by the magnetic field in the form of thin closed flux tubes. The
loop plasma may be heated by a number of
possible mechanisms \citep[for a review see][]{rea_10}.
As a consequence,  pressure and density of the gas within the
magnetic loops are enhanced compared to the values reached by the
surrounding gas and the
coronal loops appear bright in coronal radiation.
This property is of great advantage for the
investigation of the dynamics in the corona: the bright structures serve
as an indirect tracer of the coronal magnetic field.

Owing to its high temperature the coronal plasma emits
radiation predominantly at X-ray, EUV and radio
wavelengths. The bulk of the coronal loops hosts plasma of temperatures
of $\approx$\,$10^5$~K to $\approx$\,10~MK. The lower and upper
limits are representative of cool and flaring loops,
respectively \citep{rea_10}.
Coronal loops often fan out from relatively compact footpoints at the
solar surface, arranging themselves in dome-like structures with multiple
layers. The individual loops are thought
to have a very narrow distribution of temperatures within the loops
themselves -- neglecting their footpoints --
though the loops within an AR may have very different temperatures
\citep[][]{del_mas_03,asch_boe_11}.
In the corona, where the field fans out with height, thin constituent
strands are discernible.
At present, several spectral EUV channels are used to observe and study
coronal loop fine structures since they appear well defined
around the wavelengths  171~\AA\ and 193~\AA.
At the highest spatial resolution presently achievable,
in the form of Hi-C data, the density and temperature
structure across the observed thin strands varies on a spatial scale
of roughly $0\farcs1$; alternatively, one may state that a diameter of $0\farcs1$
represents an upper limit for the strands which make up an observed
loop \citep{cir_gol_13,pet_bin_13}.

\begin{figure}
   \centering
   \includegraphics[width=0.85\textwidth]{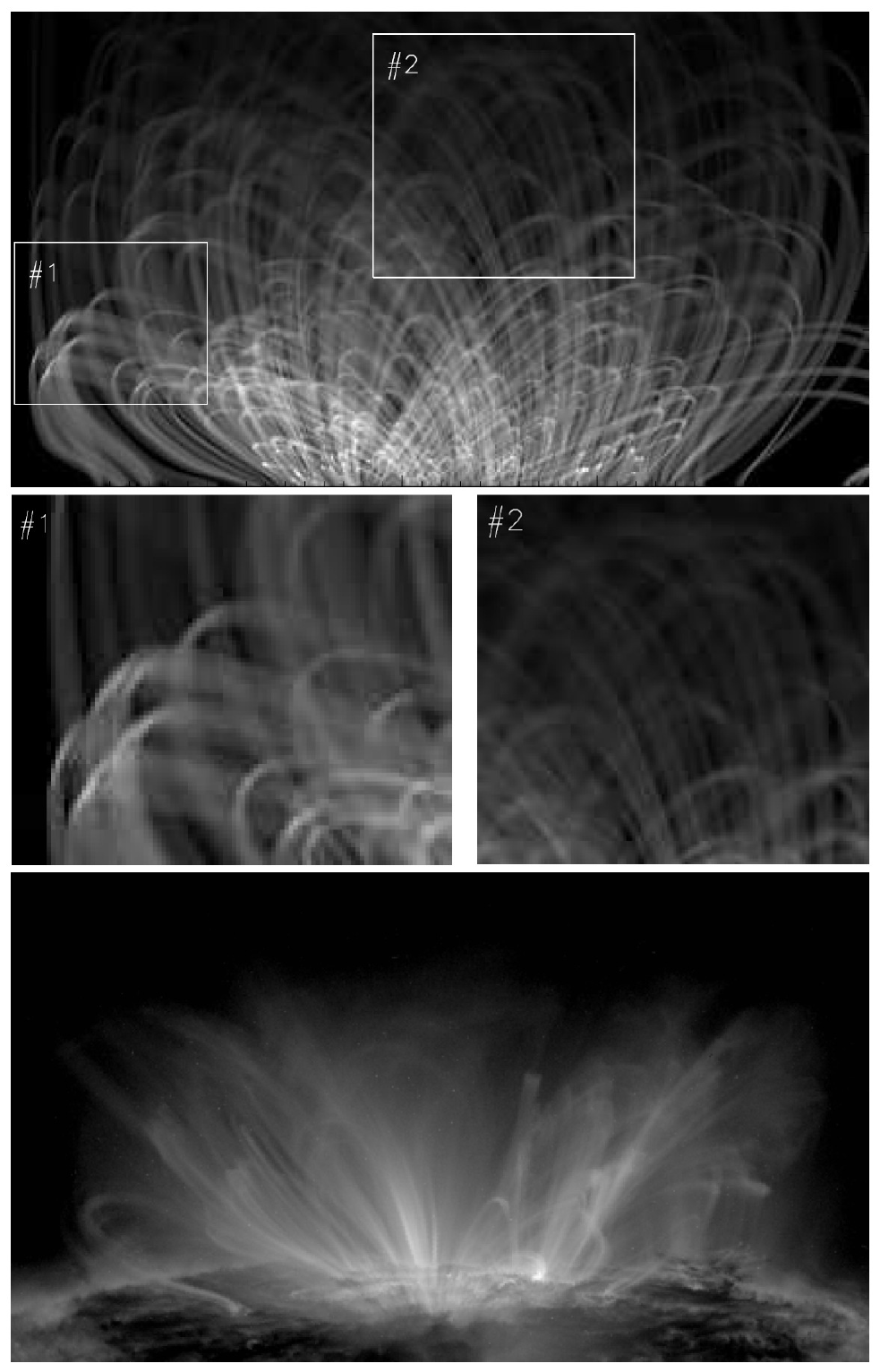}
   \put(-280,435){\bf\sf\white(a)}
   \put(-270,275){\bf\sf\white(b)}
   \put(-113,275){\bf\sf\white(c)}
   \put(-280,154){\bf\sf\white(d)}
   \caption{(a) Coronal emission simulated from flux tubes (calculated from
   a potential magnetic field model) filled with hydrostatic, isothermal plasma.
   The gray-scale reflects the squared column emission in arbitrary units.
   Rectangular areas labeled as \#1 and \#2
   highlight areas where loops do not notably expand with respect to their
   neighbours and where loops turn towards the observer, respectively.
   Close-ups of these regions are shown in panels (b) and (c), respectively.
   (d) EUV emission from active-region loops as seen above the solar
   limb in an \sdo/AIA 171~\AA\ image. The numerically synthesized brightness
   variations are well visible in the observed corona:
   the emission of loops with less varying apparent cross section is enhanced
   (compare panel (b)) and horizontally elongated where loops turn towards
   the LOS (compare panel (c)).
   (Adapted from Figures 17 and 18 of \cite{2013ApJ...775..120M}. \copyright~AAS. Reproduced with permission.)
   }
   \label{fig:mal_schr_13_fig17_fig18}
\end{figure}
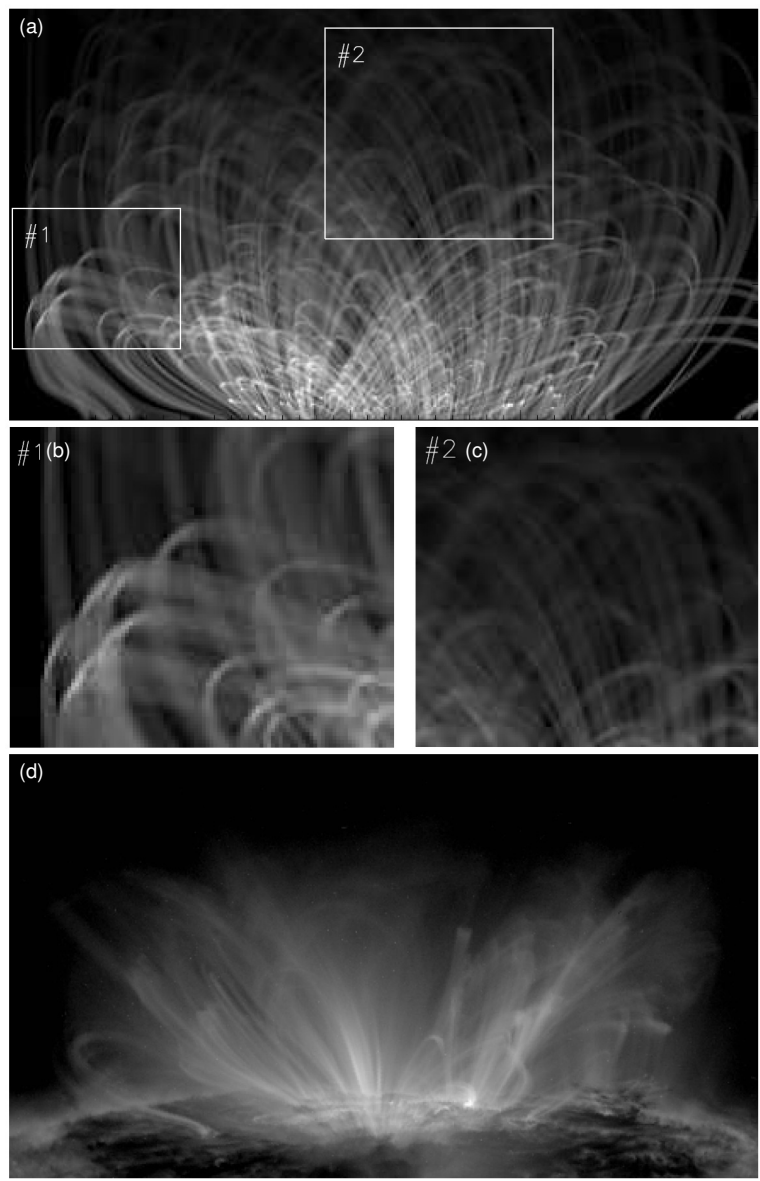

It has long been puzzling why especially the loops seen in coronal images
do not show a significant variation of their width with height in the
atmosphere. Given magnetic flux tubes expand with height in
the solar atmosphere, one would naturally expect this to be reflected in form
of a clear height-dependence of the emission observed from the thin threads
which compose the flux tubes \citep[][]{def_07}.
Instead, an apparently constant cross section and more or less constant
brightness along the loops, but
no significant expansion was observed at the two wavelengths
mentioned above.
This is the result of the analysis of coronal loops seen
in \yohkoh/SXT images \citep[][]{kli_lem_92}, and EUV observations with
\trace\ \citep[][]{def_07} as well as with \sdo/AIA \citep[][]{asch_boe_11}.
It has been argued that this might
just reflect the fact that the coronal loops are entities of a constant diameter
\citep[][]{2000SoPh..193...53K}, although force-free magnetic field
models do not support such an interpretation.

Physical reasons for the
geometric distribution of the observed coronal loop emission
is now beginning to come from numerical experiments.
\cite{pet_bin_12} used an MHD model to investigate the temporal evolution of
the corresponding synthesized coronal emission. They were able to show
how emission of seemingly constant width may arise from an expanding flux
tube.
They argued that the plasma indeed fills the fanning-out magnetic field
structure, but that it does not  equally contribute to the
emission perpendicular to the plasma loops' axis. The radiation from the
plasma at the outer edges, \ie, the
 ``envelope'' of the expanding magnetic field
is emitted from plasma at a lower temperature. Consequently, this radiation is
missed in images showing the bulk of the emission at, say, 171~\AA\ and
resulting in coronal images showing loops of nearly constant cross section.
\cite{pet_bin_12}
 also pointed out that the appearance of the coronal loops
may even be caused by the specific perspective at which the coronal
loops are seen \citep[see also][]{2008ApJ...679L.161M}.
Similar results have been found recently by \cite{2013ApJ...775..120M}, who
investigated a large sample of flux tubes that
were based on a potential field model. They were able to assign well known
observed features to both, projection effects as well as the deviation of the
flux tubes' cross section from a circle along its length.
These include the enhanced brightness of loops
which seemingly do not expand much in the plane of sky, compared to
neighbouring loops which seem to do so, and the characteristic elongated
bright emission where loops turn towards the LOS (see
Figure~\ref{fig:mal_schr_13_fig17_fig18}).

\subsection{Local field topology}
\label{ss:local_topology}

The bright coronal loops within ARs connect photospheric locations of enhanced
magnetic flux of opposite polarity.
On a global scale, as discussed in section~\ref{s:global}, such loops
are observed only within certain latitude bands, the activity belts.
The active-region fields are often composed of strong sunspot fields
surrounded by plage regions of weaker field strengths. Prominent
ensembles of active-region loops preferentially connect locations of strong
magnetic flux inside an AR (\ie, sunspots and/or plage regions), or
they have one footpoint located in an AR and the other in the enhanced
flux of the surrounding network \citep[][]{schr_tit_99}.

\subsubsection{Magnetic skeleton}

\begin{figure}
   \centering
   \includegraphics[width=0.75\textwidth]{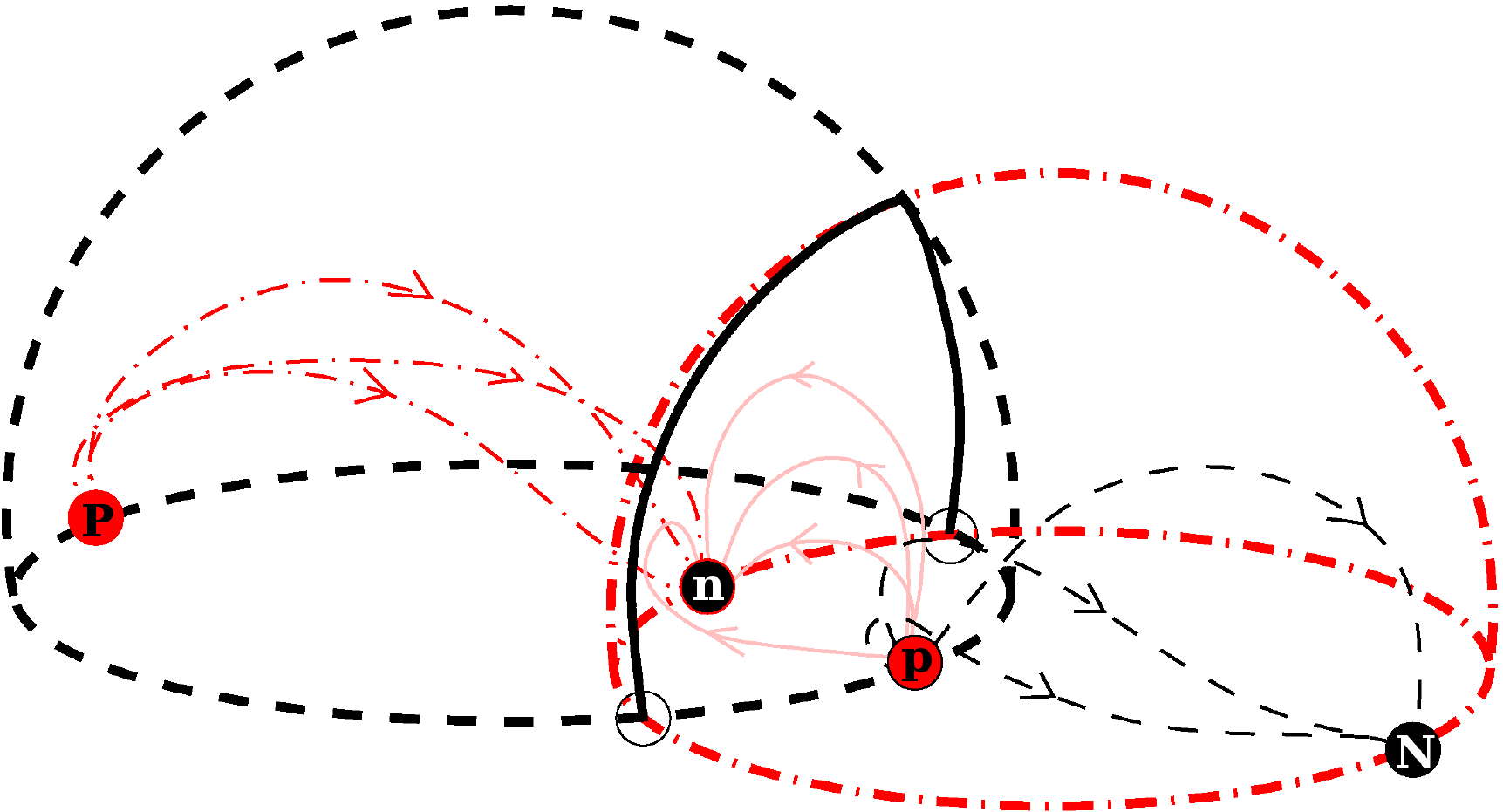}
   \caption{Sketch of a 3D magnetic skeleton. Circles with red and black
   filling represent magnetic sources ($P$, $p$) and sinks ($N$, $n$),
   respectively, \ie, positive and negative flux concentrations.
   Black (dashed) and red (dash-dotted) curves represent sample field lines
   which close within the domain bounded by the positive and negative
   separatrix surfaces (thick red dash-dotted and thick black dashed outlined
   domes, respectively). The separator field line (black solid curve) runs
   where the positive and negative separatrix surfaces intersect, connecting
   two null points (black circles). It also marks the intersection of at least four
   distinct flux domains. Pink solid curves represent field lines that connect
   $p$ and $n$, which must close below the separator. Note that
   $P$ and $N$ can only be connected by field lines running outside of both
   separatrix surfaces, \ie~ they have to bridge the separator.
   (Adapted from \cite{pri_for_02}.)}
   \label{fig:topology}
\end{figure}

The building blocks of the coronal magnetic field
include the locations of a vanishing magnetic field
(``null points''), field lines which separate topologically distinct regions
of space (forming ``separatrix surfaces''), as well as special field lines
at the intersection of such surfaces which connect two null points
(``separators'').
The sum of all these special locations, field lines and surfaces is
called the ``magnetic skeleton'' \citep[see][and
Figure~\ref{fig:topology}]{lon_05,pri_07}.
Separatrix surfaces and separators are places where the magnetic
field is discontinuous, \ie, the field lines with footpoints on either side
connect to rather different positions on the Sun. Consider, for instance,
a field line which originates from the vicinity of the positive polarity
$P$ (indicated by the red, dashed-dotted lines) in Figure~\ref{fig:topology}.
If the positive-polarity footpoint of the field line is located within the
separatrix surface associated to $P$ (black dashed dome), the field
line can only close down in the vicinity of the neighbouring negative
polarity $n$. If the footpoint, however,
is shifted to a location just outside of the associated separatrix surface,
the field line can no longer connect to $n$, because field lines do not
cross separatrix surfaces or separators.
Instead, it may connect to another neighbouring negative polarity $N$ by
running all along outside of the separatrix surfaces associated to $P$
and $N$ (red dash-dotted dome).

\subsubsection{Association to current-sheets and magnetic reconnection}
\label{sss:cs_mr}

When a magnetic configuration evolves quasi-statically or dynamically,
different flux domains develop as pictured above.
It is possible to trace the different magnetic flux domains in the corona and
their footprint on the solar surface, given that the 3D magnetic field
configuration is known \citep{pri_dem_95,dem_bag_97,dem_06, tit_hor_02}.
One can determine the position of the conjugate-polarity footpoint of a
certain field line. Applying a horizontal shift to the location of
the footpoint from which the field line calculation is started, the places
of strongest variations regarding the location of the opposite-polarity
footpoint are detected, indicating the location of
``quasi-separatrix layers'' \citep[QSLs;][]{pri_dem_95,dem_bag_97,tit_hor_02}.
There, the magnetic connectivity is not discontinuous but has steep
gradients. This leads to the formation of current sheets \citep[][]{tit_07}
in which electric currents
may be efficiently dissipated \citep{aul_par_05,bel_06,par_mas_09}.
Note that this implies that current sheets preferentially form where QSLs
border each other \citep{schr_der_10}.

Current sheets are regions in space where the magnetic field strength,
and thus the magnetic energy density, is locally enhanced.
In the collisionless environment of the solar corona, current sheets
are transient features which diffuse away but,  as long as they exist, they
have important consequences for the field and plasma they contain.
Most importantly, the resistivity within a current sheet is locally enhanced
and as a consequence, unlike in the coronal surroundings of the current sheet,
the plasma is
-- in contrary to the surrounding corona --
not frozen into the field. Thus, they are
favourable locations for changes of the magnetic topology to happen,
\ie, where the magnetic field may change direction and/or magnitude
(see chapter~2 in \cite{pri_82}, \cite{2007mare.book.....P},
chapter~6 in \cite{2014masu.book.....P}, and also section~\ref{ss:erelease}).

As pointed out by \cite{2008ApJ...675.1656P}, knowledge
about the magnetic skeleton is necessary if one aims to determine the location,
type, rate and frequency of reconnection events.
Therefore, the magnetic skeleton has been investigated for complex magnetic
field configurations also with the help of MHD experiments satisfying solar-like
parameters.
\cite{2009A&A...501..321M} performed MHD simulations, using a potential
field based on \soho/MDI observations as an initial equilibrium. They investigated
the evolution of the magnetic field around an EUV bright point observed in
\trace\ 171~\AA\ images. They simulated the effect of the observed rotation of one
of the main photospheric magnetic sources of a bright point. They could
also show that the resulting build-up of electric currents may have enabled
magnetic reconnection at the separatrix surfaces associated with the rotated
magnetic source.
They commented that more research must be undertaken in order to
determine which parts of the separatrix surfaces host strong electric currents.
This was partly addressed by \cite{2010ApJ...725L.214P}, who studied the
magnetic skeleton in the course of magnetic flux emergence into a pre-existing
magnetic field. Their results indicated that locations along separators
are favourable locations for magnetic reconnection to occur.

\subsubsection{Relation to dynamic phenomena}
\label{sss:dyn_phen}

Because the trigger for magnetic reconnection has often been suspected
around an existing coronal null point or within a QSL,
studies have been undertaken to relate flare-associated features to the
magnetic topology of the corona
\citep[][]{luo_man_07,bak_vdg_09,vdg_cul_12,2012ApJ...757..149S}.
\cite{aul_del_00} inferred the location
of a coronal null point and separatrix surfaces from a potential field
reconstruction and highlighted that, a current-free model cannot
be expected to reproduce the observations in detail -- especially in the
presence of strong shear. Nevertheless, it can be expected that it should be
capable of recovering the basic, underlying coronal magnetic field topology.
A similar finding was presented by \cite{su_vba_09}, who employed,
besides a potential field model, also a NLFF field model with an artificially
inserted flux rope, forced to emerge into the force-free model corona.
Their work revealed a line of coronal null points not only in the
NLFF but also in the potential field representation. This supported the
idea that a potential field model may indeed be sufficient to gain information
on the basic coronal magnetic field topology.
A less supportive conclusion was reached by
\cite{2012ApJ...757..149S}, who found that the number and location of
coronal null points were not the same for NLFF and current-free field
models. Specifically, the position of a coronal null,
as inferred from an NLFF field solution was clearly displaced from the
location calculated from the associated potential field model. This
mismatch, they argued, might have been caused by the high non-potentiality
of the investigated active-region field which made the current-free model
fail to realistically account for its structure.

\begin{figure}
   \centering
   \includegraphics[width=\textwidth]{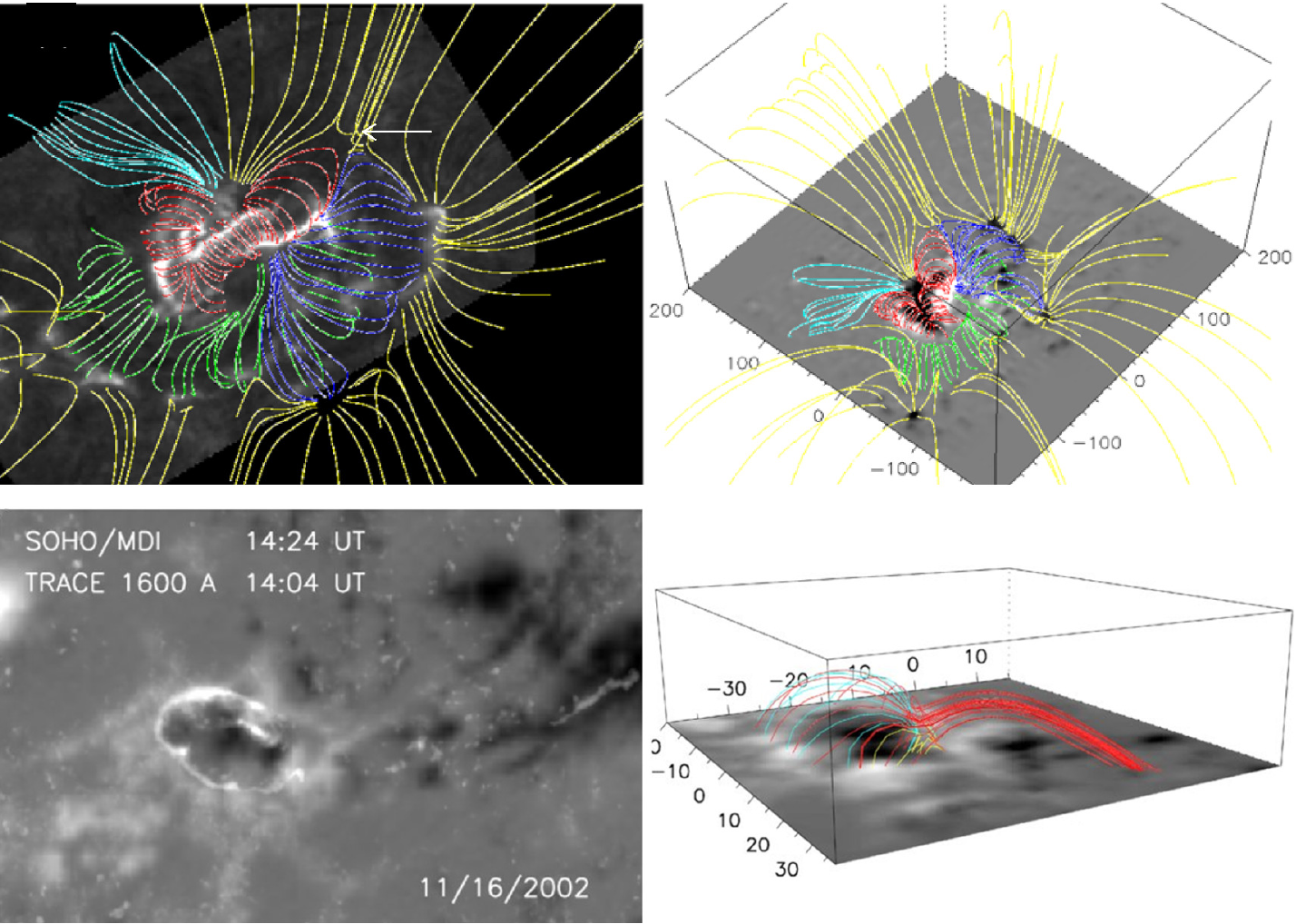}
   \put(-335,230){\bf\sf\white(a)}
   \put(-165,230){\bf\sf(b)}
   \put(-190,100){\bf\sf\white(c)}
   \put(-165,100){\bf\sf(d)}
   \caption{
   (a) Magnetic topology of AR 10365 on 27 May 2003,
   estimated from a current-free field model. The white arrow in (a)
   indicates the possible location of a coronal null point, based on
   the associated current-free field configuration. Selected
   field lines that outline the basic magnetic field configuration are
   displayed on top of a near-in-time H$\alpha$ image taken at the
   Solar Observatory Tower Meudon. Strongest
   emission in (a) outlines the location of flare ribbons (that is, where
   newly reconnected field lines are line-tied to the low atmosphere).
   (b) The same model configuration, shown on top of the \soho/MDI
   LOS magnetogram (gray-scale, where positive/negative
   polarity corresponds to white/black areas). Sample
   magnetic field lines, connecting the AR and its periphery are shown
   in yellow. Green, light and dark blue, as well as red field lines
   connect the observed flare ribbons to other regions within the AR.
   (Adapted from Figure 8 of \cite{2009SoPh..258...53C}. With
    kind permission from Springer Science and Business Media.)
   (c) \trace\ 1600~\AA\ emission (bright ribbon and kernels) associated
   to a C-class flare on November 11, 2002, overlaid on a \soho/MDI
   photospheric LOS magnetic field (gray-scale background; white/black
   represents positive/negative polarity) of AR 10191.
   (d) Sample field lines outlining the potential field reconstruction
   of the associated coronal field. The red, yellow and blue lines indicate
   a coronal null-point topology.
   (Adapted from Figures 2 and 3 of \cite{mas_par_09}.
   \copyright~AAS. Reproduced with permission.)
   }
   \label{fig:skeleton}
\end{figure}
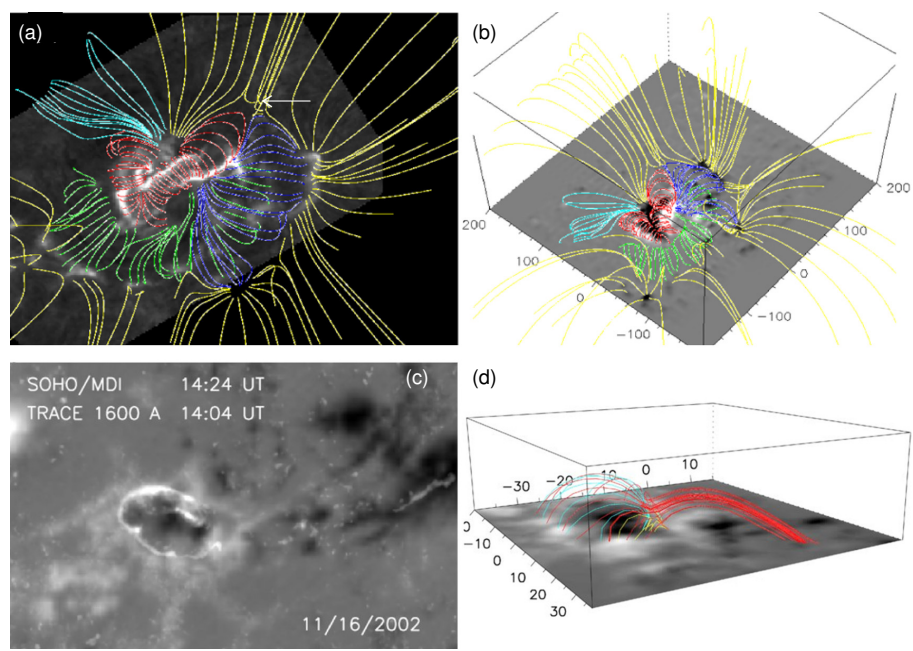

\cite{aul_del_00} and \cite{su_vba_09} found the
shape of observed flare ribbons to be closely associated to the
intersection of separatrix surfaces with the lower atmosphere.
The spatial proximity between flare ribbons and QSLs agrees with the
picture of confined flares
\citep[for reviews see][]{pri_for_02,shi_mag_11}, which relates flare
ribbon emissions to particles that are accelerated at the coronal
reconnection site and that follow the separatrix field lines downward.
Eventually encountering the denser layers of the lower atmosphere, they
lose their energy due to collisions, but increase the radiation emitted at
these locations by heating the surrounding plasma. Also using a potential
field model, \cite{mas_par_09} explained the flare ribbons observed in
association with a small flare by the photospheric intersections of field
lines that passed the vicinity of a coronal null point \citep[similar to what was
found by][see
Figure~\ref{fig:skeleton}]{2009SoPh..258...53C}. They found that these field
lines formed a surface which enclosed the flux domain of a ``parasitic
polarity'' (that is a patch of a certain magnetic polarity surrounded by
fields of opposite polarity) and that the geometry agreed closely with the
theory of confined flares.

Note that magnetic field lines are fictitious constructs,
meant to outline the direction of forcing by the magnetic field. Thus,
they are often related to physical coronal structures, since charged
particles in the corona feel the force exerted
by the magnetic field (Lorentz force). As a consequence, for instance, heat
conduction is most efficient along the magnetic field. Similarly, density
enhancements and small localized disturbances
propagate along field lines, and plasma flows are mechanically confined
by the field \citep[][]{lon_05}. Such
flows persist on time scales similar to that of the slow quasi-static evolution
of the magnetic field, \ie, of the order of hours, and have indeed been found near
QSLs. Early on,
\cite{2004A&A...428..629M} were able to show the quasi-static nature of
active-region flows and their close connection to the coronal magnetic field
structure. Using a LFF field model, they were able to
associate sharp changes in the Doppler velocity (inferred from
\soho/SUMER Dopplergrams) to the border between topologically different
field configurations (open and closed). Furthermore, there seems to be a
connection
between the strength of the flows and the underlying magnetic field: the
strongest flows seem to be associated to that portion of QSLs which are
situated above strong magnetic fields \citep{bak_vdg_09}.

\subsection{Temporal evolution of active-region magnetic fields}
\label{ss:ar_evol}
\label{sss:evolution_corona}

The active-region corona that spans from the coronal base
until about 100~Mm above the solar surface, is a
low-$\beta$ environment \citep[see][and
section~\ref{sss:dynamics_chromo}]{gar_01}. That implies that the
evolution of the magnetic field dictates the mass, momentum and energy
flow. This is because the dynamic pressure exerted by the solar wind within
the Alfv{\'e}n radius, \ie,
the range within which any escaping material is forced to rotate rigidly with
the Sun, is small compared to the magnetic pressure, the latter also well
exceeding the gas pressure there (see chapter~9 in \cite{sti_02}).
The shuffling around of magnetic field lines
at photospheric levels by the convective motions at (sub-)photospheric
levels determines the temporal evolution of the coronal field above.
Thanks to the
high Alfv{\'e}n velocity ($\approx10^3$~km\,s$^{-1}$), the coronal plasma
can quasi-statically adjust to the driving (sub-)surface convective motions
\citep[with
characteristic speeds of a few km\,s$^{-1}$; see chapter~6 in][]{sti_02}.
This, together
with the underlying field topology determines the fate of coronal active-region
magnetic fields. In most cases the evolution of the corona is
 a slow transition between neighbouring equilibrium states.

A slow, quasi-static evolution means that the coronal magnetic fields adjust
to the random or systematic motion of their line-tied photospheric footpoints
without a sudden energy release, \ie, without a major eruption.
Note that the condition of the magnetic field being frozen in the
plasma and its more or less passive advection by photospheric
flows is referred to as line-tying \citep[see][]{lon_05}. That does,
however, not preclude the presence of dynamic phenomena.
For instance, at the edges of ARs, outflows have often
been observed with
velocities of tens of km\,s$^{-1}$.
This has actually been proposed to be
important for the solar wind \cite{1997SoPh..170..163B,2001ApJ...553L..81W,
2007Sci...318.1585S}. Such flows might arise from the reconnection between
small-scale emerging fields and larger-scale open field lines of ARs
\citep[][]{2014Ap&SS.351..417L}. Also SXR jets are sometimes observed
at the boundaries of ARs \citep{1994ApJ...422..906S}, often associated
with parasitic polarities
\citep{schm_guo_13}. But despite such small-scale events,
for most of the time, the magnetic field of an AR evolves slowly and can be
considered to be in equilibrium.
At typical coronal conditions, the magnetic energy density -- which is a
measure for the magnitude of the associated pressure -- is at least three
orders of magnitude larger than the gravitational, thermal or kinetic energy
density \citep{for_00}. Under equilibrium conditions, this implies that
the Lorentz force vanishes and that the field can be considered to be force-free
(see section~\ref{model:extrapol}). Once equilibrium is lost, however, this
might no longer be true and dynamic forces come into play that
compensate the non-vanishing magnetic forces.

\subsubsection{Dynamic evolution: eruptive phenomena}

Forced away from its equilibrium position, or, more precisely, from a quasi-static
oscillation around its equilibrium position, a configuration of the coronal field
may develop in two ways. Either it reaches a metastable state or a
non-equilibrium state \citep{pri_for_02}. A metastable configuration is stable
with respect to small perturbations because it involves a stabilizing element.
A corresponding coronal configuration would be in the form of, \eg, a twisted
flux rope bridged by a loop arcade, where the latter holds the configuration
steady \citep{stu_web_01}. A non-equilibrium state, on the other hand,
is instable against any further perturbation and its evolution is then
determined by the evolutionary path of that initially small perturbation.
We are facing different types of evolution, in the form of a growth of the perturbation,
an oscillation, or eventually a damping out. Given the appropriate trigger,
a meta- or non-stable state may evolve further to a state where mass and/or
energy is released due to an associated change of the field topology.
Examples of force-free equilibria with stable and instable configurations
(dependent on model parameters) have been constructed by
\cite{1999A&A...351..707T}. Instable
configurations have been investigated with numerical MHD simulations by
 \cite{2005ApJ...630L..97T}. Importantly, the morphology and dynamic
evolution of a modelled erupting flux rope (an eruption triggered by an ideal
kink instability) was found to be in agreement with an observed eruptive flare.

Associate releases of mass and/or energy are called flares, eruptive prominences
and CMEs, depending on the emission and dynamics observed \citep[for
reviews see,][]{ben_08,fle_den_11,mac_kar_10,2014LRSP...11....1P,
hud_bou_06,web_how_12}.
Flares involve sudden enhancements of electromagnetic radiation,
caused by accelerated particles and excessive heating of the coronal plasma.
During an eruptive prominence the material initially trapped inside is
partially or completely expelled from the Sun.
CMEs involve the expulsion of huge masses of coronal plasma into
interplanetary space, spanning the range of roughly $\approx$\,$10^{12}$~kg
to $\approx$\,$10^{16}$~kg
\citep[see][for a comprehensive analysis of CME masses during
a full solar cycle]{2010ApJ...722.1522V,2011ApJ...730...59V}.
Many attempts have been made to develop models to describe the features
associated with a loss of equilibrium, including the pre-eruptive coronal
magnetic field structure, the triggering mechanism and the temporal evolution
of the eruption itself. It is outside the scope of this review to summarize
all existing models of eruptive processes in the solar atmosphere. Instead,
we refer here to the comprehensive reviews on flare \citep{for_lin_06,che_11}
and CME models \citep{for_00,shi_mag_11}.

\subsection{Favourable conditions for eruptions}
\label{ss:eruptive_conditions}

To find the conditions under which the magnetic field of a (part of an) ARs is likely to erupt,
one has to analyse the corona as well as the underlying parts of the atmosphere
before an eruption occurs
\citep[for reviews see][]{schr_09,che_11}.
Favourable conditions for eruptions include strongly twisted or sheared magnetic structures
embedded in a less-sheared magnetic system \citep[][]{hao_guo_12},
the rapid emergence/evolution of strong magnetic flux concentrations
\citep[][]{2004ApJ...605..931W, 2012ApJ...748...77S},
and/or a complex magnetic field topology \citep[][]{2012ApJ...757..149S}.

\subsubsection{Magnetic flux emergence and complex magnetic field structure}

A complex magnetic field structure may
include parasitic polarities, highly sheared fields, strong gradients and long
PILs, not necessarily all at the same time
or in the same event \citep{vdg_cul_09}.
For magnetically complex ARs a tendency to be more CME-productive
has indeed been found \citep{che_wan_11}. Large flares mainly originate
from ARs with a complex photospheric magnetic field configuration, the
largest ones from $\delta$-configurations \citep[see review by][and
references therein]{ben_08}. Note that a $\delta$-configuration denotes
umbrae of opposite polarity existing within a single penumbra.
Only rarely are spotless regions reported to be the source of major flares
\citep{ruz_vrs_89,ser_val_93,li_zho_95}.
M- and X-class flares are considered to be "major'' flares
in this context with SXR peak fluxes on the order of $10^{-5}$~W\,m$^{-2}$ and
$10^{-4}$~W\,m$^{-2}$, respectively.
\cite{ruz_vrs_89}, however, mentioned an enhanced occurrence rate of a
specific type of eruption in spotless regions, namely
two-ribbon flares, involving
two progressively spreading bands of chromospheric emission.

As noted by \cite{schr_09}, an eruption might
not be triggered per se due to flux emergence but only if a sufficient amount
of magnetic flux emerged.
The imbalance of the emerging magnetic flux does not seem to be crucial
for an eruption to occur, although it might play a role.
\cite{cho_ven_02} noted that only roughly one third of their analysed
cases showed a flux asymmetry of more than
10\%. A super-AR, however, was found to exhibit a flux imbalance of
$\approx40$\% which, they admitted, could have partly been caused by
instrumental effects. It is worth noting that, such analyses often suffer the
problem of projection effects which is noticeable in the degree
of flux imbalance systematically increasing with distance from the disk
center \citep{gre_dem_03}. \cite{che_wan_12} recently revisited the
importance of flux-imbalance for the eruptive nature of ARs; they only
considered ARs that fulfilled certain selection criteria, including being
observed within $\pm30^\circ$ from the central meridian and being
associated to major flaring activity.
They analysed 14 ARs and found a significant imbalance of magnetic
flux: more than half of the ARs showed a flux imbalance of $\gtrsim20$\%
\citep[see also][]{tia_liu_02,2007A&A...474..633R}.
However, thermal effects on the photospheric line profiles can influence the
determination of magnetic flux. This can lead to a seeming imbalance between
the two polarities of an AR if, \eg, one polarity of the AR is
composed more of sunspots, while the other polarity has more plage.
Consequently, this topic needs further study.

\subsubsection{Magnetic shear and twist}

An eruption may also be triggered if, owing to the motion of its photospheric
footpoints, a coronal loop arcade is sheared too much
\citep{pri_for_02}. It has long been known
that the combination of strong fields and strong shear seem an essential
ingredient to, for example, ribbon flares \citep[][]{1986AdSpR...6....7H}. Strong
shear may develop due to sunspot motions, converging giant convection
cells, as well as due to the emergence, cancellation and submergence of
magnetic flux \citep[see][and references therein;  see also
\citealt{1994SoPh..155..285W}]{1984AdSpR...4...71H}.
Note, however, that it might not always be possible to judge
whether horizontal flows caused the observed shear or if an
arcade already emerged as a sheared configuration
\citep{2005A&A...433..701W}.
\cite{1982SoPh...79...59K} compared the evolution of the transverse
magnetic field component to the motions within a flare-productive AR
observed in white light. They found that over a period of several days the
observed shearing motion resulted in an ever increasing field alignment
with the PIL and was associated to the onset and increase of flaring
activity, and which has been supported by the MHD model results
of \cite{1984SoPh...90..117W}. Force-free modelling results
have initiated the search for a possible critical amount of shear
which, once reached, would inevitably trigger an eruption
\citep{1977ApJ...212..234L,1977ApJ...217..988L}, but for which
observational support seems sparse.

\cite{1984SoPh...91..115H} estimated the critical shear
to lie in the range $80^\circ\lesssim\theta\lesssim85^\circ$.
They defined the shear angle, $\theta$, as the angular difference
between the orientation of the transverse field component and the
direction perpendicular with respect to the PIL.
Consequently, $\theta=0^\circ$ implies
a field with no shear and $\theta=90^\circ$ would account for a transverse
field parallel to the PIL.
\cite{1992SoPh..138..353S} followed a different approach: they used H$\alpha$
observations to determine the orientation of the major axis of filaments
(as an indication for the direction of the PIL)
with respect to an approximation to the orientation of the potential
field azimuth. The latter has been defined by the direction perpendicular to a
straight connection between the two main
spots of the associated bipolar sunspot group. Their study revealed that
flares occurred only when the angle between the principal filament
direction (\ie, the PIL) and the direction of the potential field azimuth
exceeded $\approx$\,85$^\circ$ (using the same
notation for the shear angle as above).

Another source of instability is the amount of twist of a coronal structure,
as induced by systematic motions of the photospheric footpoints
of the field. For instance, an equilibrium coronal loop becomes
instable if its length is stretched beyond a critical value, or if it is
twisted by photospheric motions \citep{pri_78}.
For twisted magnetic field configurations, different critical thresholds
were found within numerical experiments, including uniformly twisted
toroidal or periodic, line-tied cylindrically symmetric configurations
\citep{hoo_pri_81,bat_hey_96} and locally twisted configurations
\citep{bat_hey_96,mik_schn_90,fan_gib_03,toe_kli_03}.
Note, the results of all of these model experiments agree fairly well on the
critical amount of twist: the field lines must perform more than one full
turn ($T>2\pi$) about the center of the flux tube for it to become
kink-unstable, which would be observed as an eruption for under
coronal conditions.
These numerical considerations also revealed that the critical twist,
tends to rise with increasing aspect ratio,
that is the ratio of the loop length to its cross-sectional diameter
\citep{ein_vanh_83,toe_kli_04}. On the other hand it
 is smallest for loops of a strong magnetic field, for given
temperature and density values \citep{pri_78}.

\subsection{Magnetic energy budget}
\label{ss:menergy_ar}

\subsubsection{Energy build-up and storage}

Magnetic flux
continuously emerges from where it is generated through the photospheric
layers and punches into the pre-existing coronal magnetic field.
At the same time, small- and large-scale convective motions in the
photospheric layers, shuffle around bundles of frozen-in magnetic field lines.
The energy
required to perform the work of moving magnetic structures against the
ambient magnetic field contributes to the magnetic energy content of the
coronal field \citep{par_87}. The two major
contributions to the transport of magnetic energy to
the corona are the emergence of material through the photosphere from
below, dragging magnetic field along, and horizontal motions of the
photospheric material that increase the shear of the magnetic field
above \citep[][]{wan_95}.
The footpoint motions of coronal loops result in an energy flux of
$\approx$\,$10^4$~W\,m$^{-2}$ and $\approx$\,$10^6$~W\,m$^{-2}$,
for typical photospheric quiet-Sun and active-region conditions, respectively,
to the corona. This is at least ten times larger than the corresponding
radiative and conductive losses in chromosphere and
corona \citep{bel_06,2012RSPTA.370.3217P}.

The random displacements of the line-tied footpoints of the coronal
field-lines naturally lead to ``braided'' magnetic fields
\citep[][]{2014ApJ...787...87V}.
Such braiding is a measure of how often different field
lines cross each other in space \citep{ber_asg_09} and
is supported by MHD models
\citep[][]{gud_nor_05,rap_val_08,vba_asg_11}.
These braided small-scale field structures efficiently
heat the coronal environment by the dissipation of electric currents,
where oppositely directed magnetic fields are found in
close vicinity to each other
\citep[][see also \citealt{cir_gol_13,2014ApJ...780..102T}]{par_83}.
The sub-surface dynamics controlling the temporal evolution of the
coronal magnetic field, also launch waves propagating upwards.
Currently, from all of the excited types of waves, only
Alfv{\'e}n waves may be capable of actually reaching coronal
heights, whereas other types of waves are efficiently damped already
in the low atmosphere and thus do not contribute to the heating of
the coronal plasma to the observed temperatures
\cite[see the recent review by][]{2012RSPTA.370.3217P}.
A discussion about
the corresponding coronal heating processes is outside the scope of our
review and interested readers might consult also the in-depth discussion
by \cite{kli_06} and dedicated chapters in \cite{cra_09} and \cite{rea_10}.

The time scale on which magnetic energy is transported through the
photosphere is much larger than that of the transport through the corona
itself. For typical coronal conditions, the travel time over a characteristic
coronal length scale is on the order of minutes.
For typical photospheric values, on the other hand, one finds characteristic
travel times of hours. This implies that the time span during which substantial
amounts of magnetic energy are stored in the corona can be long, because
sufficiently intense current distributions need to develop.

\begin{figure}
   \centering
   \includegraphics[width=\textwidth]{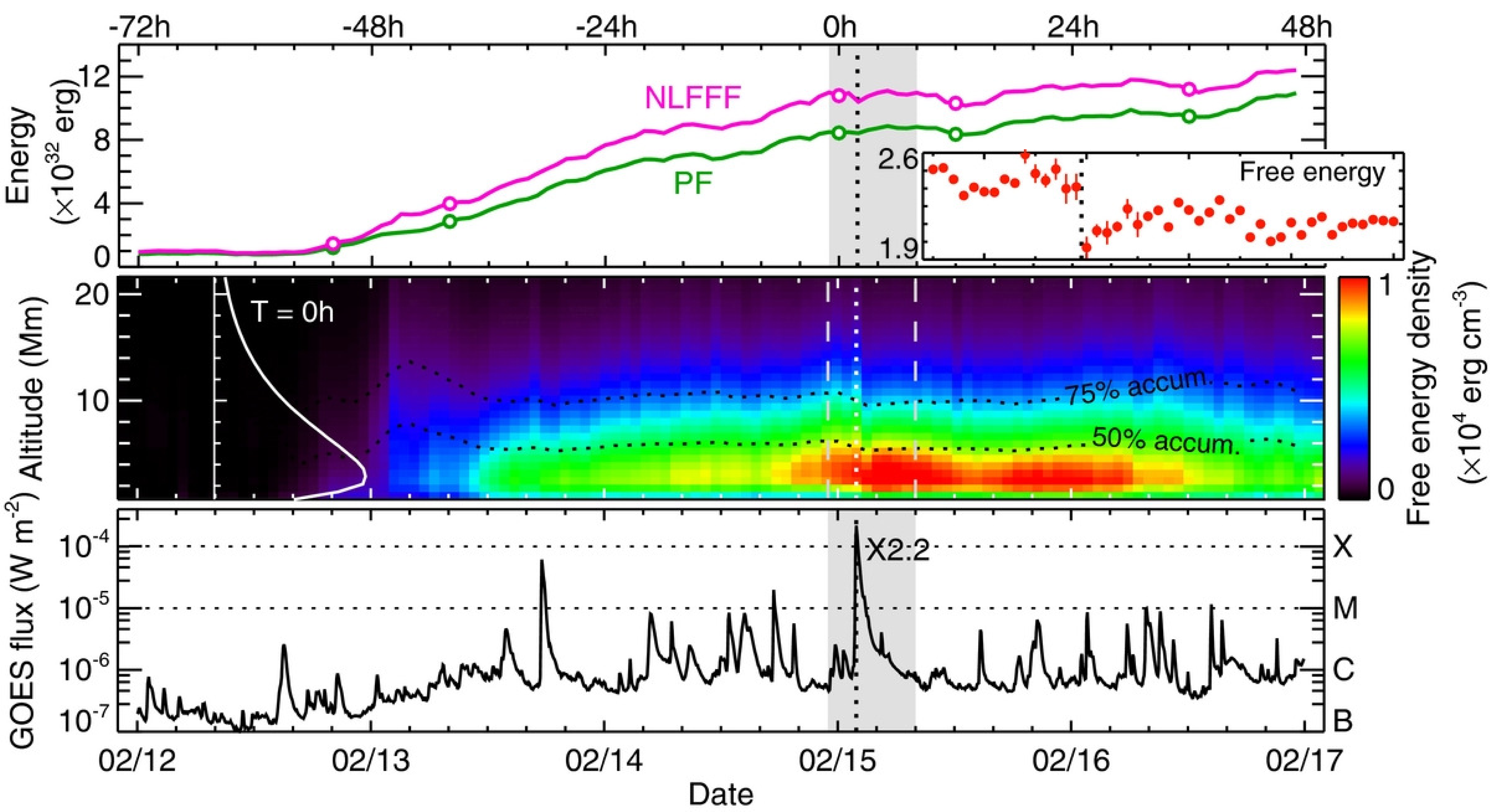}
   \put(-305,165){\bf\sf(a)}
   \put(-305,112){\bf\sf\white(b)}
   \put(-305,60){\bf\sf(c)}
   \caption{
   (a) Temporal evolution of the total (pink solid line; estimated from a NLFF
   model) and potential (green solid line) magnetic energy stored
   in the coronal volume above AR~11158 during five days. The inset shows
   the evolution of the free magnetic energy (that is the excess energy over
   a current-free field) around the time of an X-class flare (indicated by the
   vertical dotted line).
   One can clearly see the more or less continuous increase in magnetic
   energy, predominately caused by an emerging flux rope
   ($\approx10^{26}$~J during $\approx$\,48~hours).
   (b) Height profile of the associated average free magnetic energy.
   The white curve shows a sample
   altitude profile of the average free energy on February 15 at 00:00~UT.
   The course of each profile over time and over height is visualized
   by the color code shown. Black, more or less horizontally running, dotted
   lines indicate the iso-contours of 50\% and 75\% of the total free energy.
   (c) Measured full-Sun \goes\ 5-min SXR flux in the 1~\AA\ to 8~\AA\ channel for
   the same time range.
   The vertical, white-dashed lines in (b) as well as the gray-shaded
   area in (a) and (c) mark the duration of an X2.2 flare. The peak time of
   the flare, on February 15 at 01:56~UT, is indicated by a vertical dotted
   lines.
   (Adapted from Figures 4 of \citealt{2012ApJ...748...77S}. \copyright~AAS. Reproduced with permission.)
   }
   \label{fig:sun_hoe_apj_12_fig4_part}
\end{figure}

\subsubsection{Relation to flare productivity}

Depending on the magnetic field topology and its
temporal evolution, very different amounts of magnetic energy are
stored in time intervals of various lengths. Force-free modelling
suggests energy storage rates of a
few $10^{23}$~J\,h$^{-1}$ to a few $10^{25}$~J\,h$^{-1}$
\citep[][and see also Figure~\ref{fig:sun_hoe_apj_12_fig4_part}a]
{tha_wie_08a, 2012ApJ...748...77S,2014ApJ...783..102M,
2014JGRA..119.3286H}.
Besides, such modelling suggests that magnetic energy
is stored mainly in the low solar atmosphere, possibly only a few Mm
above photospheric levels \citep[][also cf.
Figure~\ref{fig:sun_hoe_apj_12_fig4_part}b]{tha_wie_08a,2012ApJ...748...77S}.

The size and frequency of eruptive events appears to be related to the
portion of the previously stored magnetic energy that is actually
available for release at a given time
\citep{cho_che_00,jin_tan_10,tzi_geo_12}.
This so-called ``free magnetic energy'' is the energy difference
between a NLFF and a current-free field.
One generally finds a higher free energy content prior to larger eruptive
events and a small amount of free energy ($\lesssim10^{23}$~J)
prior to the weakest flaring activity \citep{gil_whe_12},
where ``weakest'' activity refers to smaller than C-class events.
A free energy of $\approx$\,$10^{24}$~J to $\approx$\,$10^{25}$~J has been reported
prior to weak flaring activity \citep{ble_ama_02,reg_pri_07b,
tha_wie_08b,2012ApJ...757..149S}, where ``weak'' means C-class flares.
Moderate to strong (M- to X-class) events
tend to occur only when free energies of some $10^{25}$~J to some $10^{26}$~J are
present \citep{reg_fle_05,met_lek_05,tha_wie_08a,jin_che_09,
2012ApJ...757..149S,fen_wie_13}. Note, however, that the build-up of a
sufficient amount of energy on its own does not guarantee
the occurrence of a subsequent eruption and
the release of part of this energy \citep{gil_whe_12}.

\subsubsection{Energy release}
\label{ss:erelease}

The release of magnetic energy is believed to take place in current
sheets, very limited regions where magnetic field and plasma are locally
decoupled.
This allows the magnetic field to diffuse and change its topology by means of
magnetic reconnection (see section~\ref{sss:cs_mr}).
It has been contended by \cite{moo_lar_95} and \cite{lin_li_09}
that the width of these regions is $\approx$\,0.1~Mm to 1~Mm.
Narrower current sheets, as proposed by \cite{woo_neu_05}, would
result in even smaller diffusion time scales. Clearly,
unlike the global magnetic diffusion time-scale of several months, the diffusion
time scale in currents sheets is very small ($\tau_d\approx$\,1~s to 10~min).
In other words, the presence of current sheets in the corona, permits the
efficient release of magnetic energy on active-region scales, in contrast to
the non-efficient magnetic diffusion on global scales.
Due to the rarity of vector magnetic field measurements in the corona,
direct observational detections of current sheets are restricted basically
to the upper chromosphere. From observations, only an upper limit of
roughly 1~Mm can be given on the width of current sheets
\citep{2003Natur.425..692S}.

Force-free models of ARs indicate that the height
where the gross part of the magnetic energy is released is located
$\lesssim20$~Mm above the photospheric level
\citep{tha_wie_08a,2012ApJ...748...77S}.
Only a fraction of the released free energy goes into the acceleration of
particles. Heated and/or accelerated, particles propagate from the
reconnection site, near the apex of the coronal loop, downwards along
the legs of the loop to the chromosphere, where they loose their energy
due to collisions with the denser chromospheric material. That, in turn, is
heated and convected upwards and fills the newly reconnected field lines
with chromospheric and transition region plasma heated to coronal temperatures,
emitting in SXRs \citep{neup_68}.

Just as the free energy content prior to observed eruption
is related to the flare class, one would expect
that the actually released amount of free energy relates to the flare
class too. Recent studies, however, do not show a clear tendency, \ie,
no clear trend of more energetic events consuming more
of the previously stored energy. So far,
the estimated upper limits for the decrease of free magnetic
energy range from 5\% to 50\% \citep[][]{tha_wie_08a,tha_wie_08b,jin_che_09,
2012ApJ...748...77S, fen_wie_13}.
Cases were also reported without a clear decrease
\citep{2012ApJ...757..149S} or even an increasing free
energy \citep[][and references therein]{met_lek_05}.
However, there have been doubts about the reliability of the latter
owing to very uncertain estimates.
Often also a decrease of the free energy occurs already before the
time of the peak SXR flux of a flare, and
in other cases this has been seen even before
the flare onset \citep{jin_che_09,2014JGRA..119.3286H}.

Recently, \cite{fen_wie_13} investigated the partition of magnetic energy during an
X-class and CME event.
The radiative energy release estimated from the recorded flare emission
was found to be at least one order of magnitude lower than the upper limit of
available free magnetic energy; this also held for the kinetic
and potential energy of
the associated CME and confirmed the general consensus that by far not all
of the free energy is released even during the most energetic events, but that a
considerable portion remains available to power successive eruptions
(see also Figure~\ref{fig:tzi_geo_13_fig11}).
As discussed for the helicity budget on a global scale (see
section~\ref{ss:hbudget_global}),  a non-potential magnetic field is
never found to relax to a corresponding current-free configuration,
in the course of an eruption.
Despite expected in terms of helicity conservation, a non-potential coronal
field may never entirely relax to a corresponding constant-alpha (preserved
helicity) state due to the line-tied nature of the coronal structures
\citep{ant_dev_99b}. Besides, newly emerging
flux also permanently disturbs the relaxation process.

\subsubsection{Coronal implosion and photospheric response}

\begin{figure}
   \centering
   \includegraphics[width=\textwidth]{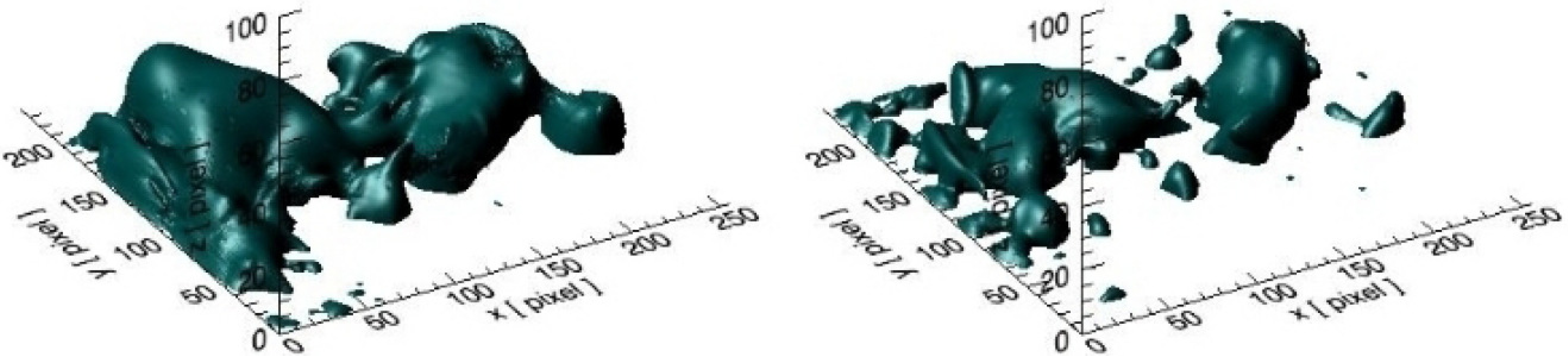}
   \put(-340,80){\bf\sf(a)}
   \put(-170,80){\bf\sf(b)}
   \caption{NLFF magnetic field model of the active-region field above NOAA
   AR 10540 (a) before and (b) after a large M-class flare. Iso-surfaces of a
   absolute magnetic field strength of 200~G are shown. The iso-surfaces give
   an impression of the magnetic pressure, sustaining different parts of the AR.
   It is evident that various areas within the AR react with different sensitivity to
   the flare, \ie, show differently strong signatures of implosion.
   (Adapted from Figure 3.7 of \cite{tha_10}. Reproduced with permission.)}
   \label{fig:tha_10_fig37}
\end{figure}

The release of magnetic energy naturally leads to a deviation from the previous
balance of the magnetic forces in an AR. The magnetic pressure is directed
towards weaker magnetic fields, \ie, generally pointing upwards towards
higher altitudes and competes with the magnetic tension that tries to reduce
the curvature of the magnetic field lines.
The rapid release of energy during an eruption may cause a reduction of
magnetic pressure, accompanied by a deflation of the magnetic field,
termed ``coronal implosion''
\citep[][and see Figure~\ref{fig:tha_10_fig37}]{hud_00}.
Model calculations by \cite{jan_low_07}
confirmed that if the post-eruption thermal pressure cannot compensate for
the eruption-related magnetic pressure reduction, a magnetic system will
implode.

\begin{figure}
   \centering
   \includegraphics[width=\textwidth]{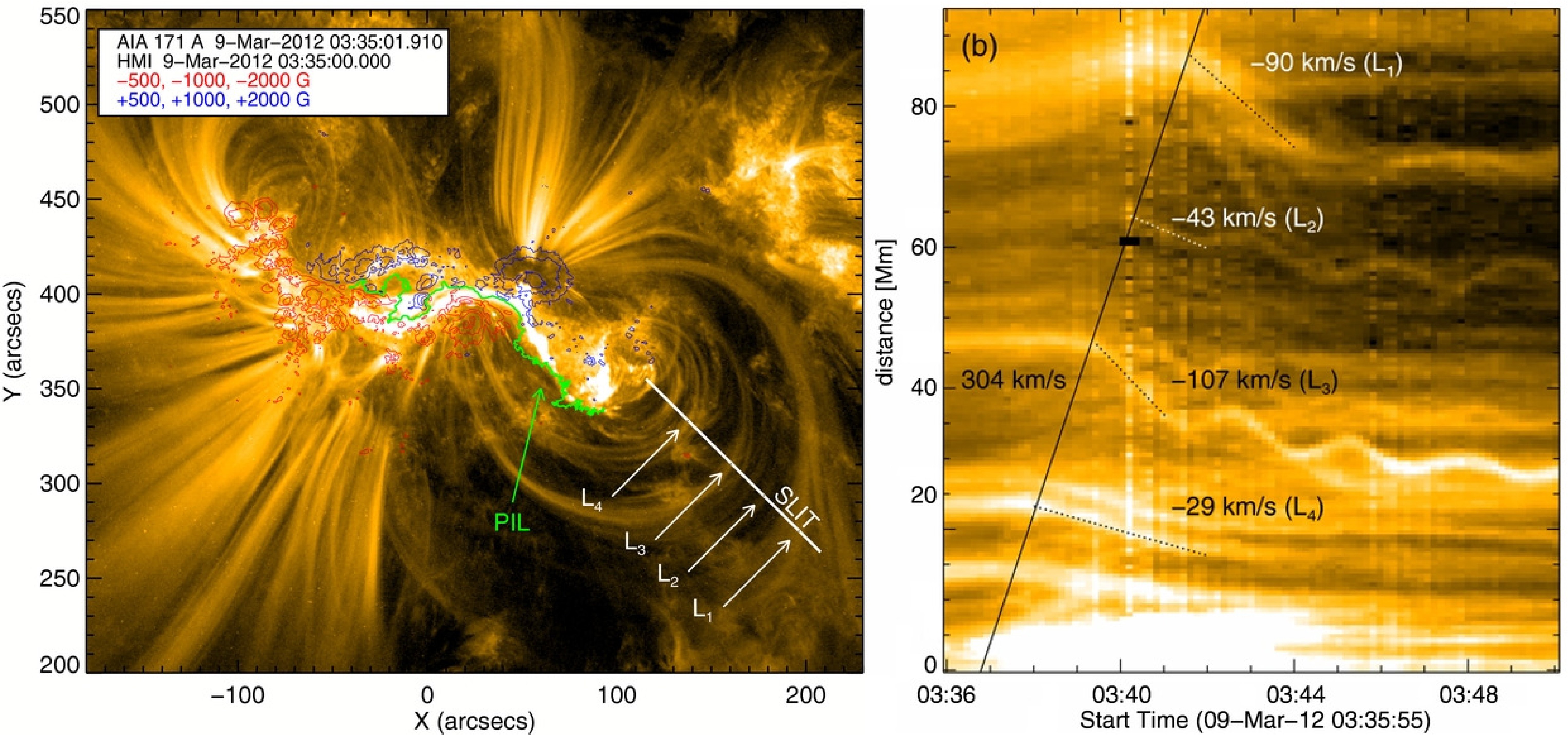}
   \put(-170,150){\bf\white\sf(a)}
   \caption{(a) \sdo/AIA image of the 171~\AA\ emission of NOAA AR~11429
   on 2012 March 9. The blue/red contours resemble the co-temporal
   negative/positive LOS component of the photospheric magnetic field
   measured with \sdo/HMI. The green line marks the PIL. The white line
   indicates the slit along which the motion of four well-defined bundles of
   coronal loops (L$_1$ to L$_4$) is investigated. (b) Time-position diagram
   during the impulsive phase of an M6.4 flare. Shown is the projected speed
   of the coronal loop bundles which contract with different apparent speeds.
   A delay of the implosion onset with height is clearly visible and indicated by
   the inclined black straight line.
   (Adapted from Figures 1 and 2 of \cite{sim_fle_13}. \copyright~AAS. Reproduced with permission.)
   }
   \label{fig:sim_fle_13_fig1_fig2b}
\end{figure}

Frequently, observational evidence for such implosions is found in the form of
the contraction of observed coronal loops
\citep{liu_wan_09a,liu_wan_09b,2012ApJ...748...77S,gos_12,sim_fle_13}.
\cite{liu_liu_12} claimed that the apparent loop contraction is not just a
projection effect.
This conclusion, they argued, was possible due to their observation of the
contracted coronal loops to reside at lower heights than those of the pre-flare
system \citep[see Figure~\ref{fig:sim_fle_13_fig1_fig2b} and also][]{liu_wan_10}.
Importantly, the magnetic field lines after an eruption never regained their
pre-flare heights. Almost all of the studied eruptions showed both
contracting overlying large-scale loops that are observable in
cooler EUV channels,
and expanding components underlying, newly-reconnected field lines,
preferentially visible in warm/hot EUV channels.
However, even before the conventional large-scale expanding motions
could be seen
 for particular flares, contracting motions during the impulsive phase
of flares were observed in the form of descending motions of loop-top sources
\citep{kru_hur_03,jos_ver_09} and EUV flaring loops
\citep{li_gan_06,liu_wan_09a}, along with converging H$\alpha$ ribbons
or HXR footpoints. \cite{ji_hua_07} explained the observational aspects
related to the initial contraction in terms of the relaxation of a modelled
highly-sheared core field. They formed new field lines, still highly sheared
and containing much of the non-potential energy.
The dissipation of free
energy causes the sheared field to relax and contract; their effect is much
stronger than the expansion effect by reconnecting arcade field lines
during the initial phase.

The deflation of coronal loops, caused by the release of magnetic energy,
is often also accompanied by changes of the shear of the horizontal
component of the photospheric magnetic field. The sign of this change
remains inconclusive, however, since both, an increase of the shear
\citep{wan_92,che_wan_94,wan_ewe_94,liu_den_05,pet_12,
wan_liu_12,wan_liu_12a},
as well as a decrease has been found during flares \citep{wan_06}.
To explain such contradictory results, \cite{dun_kur_07} proposed that the
measures for the non-potentiality of a magnetic field may take on different
values from one portion of an AR to another.
Findings may then depend on the specific choice of the analysed
photospheric area. Moreover, projection effects due to the location of the
investigated area on the solar disk might also result in changes of the
observed longitudinal field, similar to that expected for flare-related
reconfigurations \citep{1989SoPh..119...77W,1989SoPh..122..215V,
spi_yur_02}.
We also point out the need to check carefully to what extent the
determined magnetic field configuration is affected by the thermal and
velocity structure of the atmosphere \citep[see review by][]{sol_93}
by weak blends in the employed spectral lines, or by instrumental effects.
Finally, as recently speculated, the magnetic field may behave differently at
different altitudes with varying domains of increasing and decreasing shear
\citep{jin_wie_08,2012ApJ...748...77S}. No unique mechanism
accounting for all of these observed aspects of magnetic shear has so
far been identified.

\subsection{Magnetic helicity budget}
\label{ss:helicity_ar}

The storage of magnetic energy in the coronal volume above ARs
is accompanied by an accumulation of
magnetic helicity
(see section~\ref{ss:hbudget_global} for the Sun's global helicity budget
and section~\ref{QS:helicity} for a corresponding discussion on quiet-Sun
fields).
Often, also the current helicity is investigated, this quantifies how
much the fields are locally twisted \citep{dem_07}.
In contrast to magnetic helicity, the current helicity
is not a conserved quantity; the general relationship
between the two is not known \citep{vdg_dem_03}. Within this work, we
restrict ourselves to the discussion of magnetic helicity.

\subsubsection{Helicity build-up and storage}
\label{ss:hstorage_ar}

Any magnetic helicity gained
by a dynamo action somewhere within
the convection zone during
the generation of magnetic field is transported into the solar atmosphere by the
emergence of helical magnetic flux tubes in the photosphere
\citep{see_90}.
The flux of magnetic helicity at photospheric levels is the
result of the combined effect of the motions of magnetic structures due to
convective plasma flows and the subsequent advection of these
helicity-carrying structures. For a review on observations
and modelling of the helicity at photospheric levels see
\cite{dem_par_09}.
The photospheric motions that twist and/or shear the magnetic
flux tubes result in the coronal magnetic field to gain even more helicity.

Investigating the helicity injection rate in ARs by shearing
motions only requires the knowledge of the normal component of the
magnetic field and the horizontal velocity of the magnetic elements within
an AR \citep{ber_84,dem_ber_03, cha_moo_04}. The resulting
estimate is thus a mixture of the enhancement of the helicity content by shearing
motions and the portion of helicity emerging from below the photosphere.
\cite{2003ApJ...586..630M} proposed that the helicity
contribution from the vertical flows might only dominate early phases
of flux emergence, followed by a dominant contribution from the shearing
flows in the later stages of an evolving AR. This is in
the opposite of what has
been presented recently by \cite{2014ApJ...785...13L} who analysed
the ARs during their emergence phases and found that
shearing motions contribute the most to the ARs' helicity content.

And also other contributions to the helicity budget within an AR were
detected, including the rotational
motion of sunspots around their own axis,
which results in enhanced
magnetic twist. The rotation of the center of one polarity around
the center of the reversed polarity center
also adds to the writhe of a magnetic flux tube's axis
\citep[][]{tia_ale_06,2008ApJ...684..747T,kum_par_13}.
\cite{liu_zha_06} estimated the accumulation
of helicity associated to the rotation of a pair of bipolar sunspots (during
about 31~hours) to be $\approx2\times10^{42}$~Mx$^2$. They noted another
increase of $\approx3\times10^{43}$~Mx$^2$ during a following period of
$\approx96$~hours, while strong shearing motions were observed.
\cite{zha_liu_08} estimated the helicity injected
by the rotational motion of sunspots around their centers within an
AR to be $\approx$\,$10^{43}$~Mx$^2$, which was on the same order as
the combined contribution by shearing motions and flux emergence.

The above discussion underlines that further detailed research on
evolving ARs is needed in order to disentangle the contribution of
emerging, shearing and rotational motions and to
assess their importance for the total helicity content of ARs.
Importantly, \cite{vem_amb_12b} showed that such estimates depend on the
time interval between successive magnetic vector maps and the
window size within which the horizontal footpoint velocities are tracked.
Correspondingly, they guessed the general uncertainty of calculated
helicity injection rates and total helicity accumulations to be
$\approx$\,10\%.
\cite{vem_amb_12a} found evidence that individual portions of an AR
contributed differently to the total helicity content. While in one part
an accumulation of negative helicity was recorded,  an
injection of positive helicity was found in another part.
\cite{vem_amb_12b} recognized that
major flaring activity only occurred at times when helicity was injected
in the AR with a sign opposite to the dominant sign of the AR
\citep[see also][]{2002ApJ...577..501K}.
Similar findings have  been presented for CME-
\citep[][]{2004ApJ...615.1021W} and MC-productive \citep[][]{cha_par_10}
ARs. For these too, the sign of the helicity
inherent to the newly emerging magnetic flux systems may well be
opposite to that of the dominant helicity in the preexisting active-region
magnetic field.

\subsubsection{Relation to flare productivity}

\begin{figure}
   \centering
   \includegraphics[width=\textwidth]{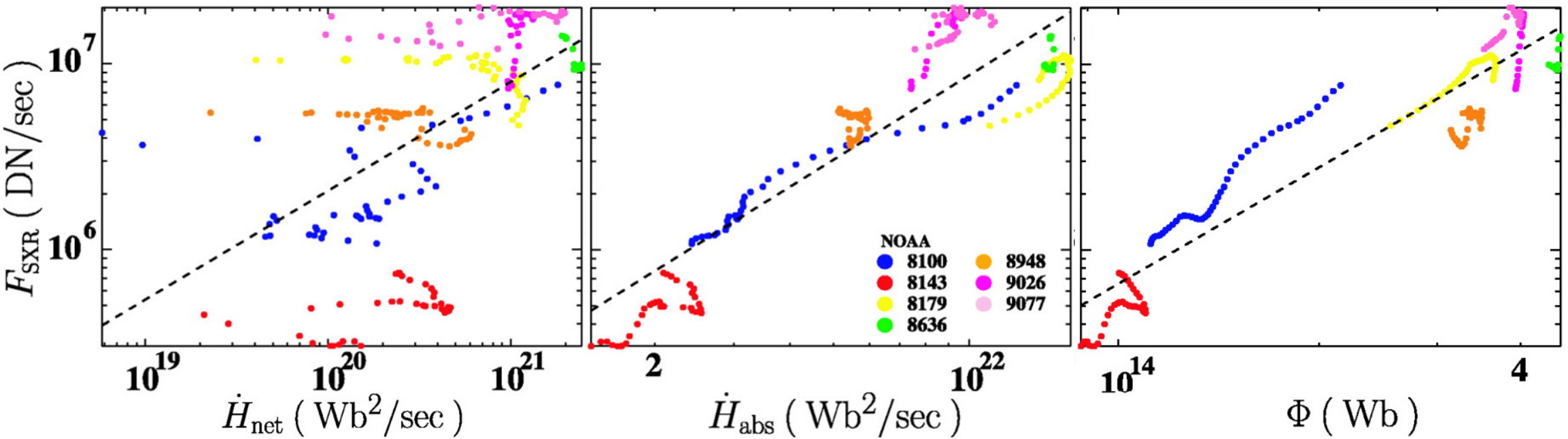}
   \caption{
   Time-averaged (a) net helicity flux, $\dot{H}_{\rm net}$, (b)
   absolute magnetic helicity flux, $\dot{H}_{\rm abs}$, and (c)
   unsigned magnetic flux, ${\rm \Phi}$,
   as a function of the SXR flux, $F_{\rm SXR}$, of seven ARs.
   The SXR flux is estimated from the measured
   intensity \yohkoh/SFD images by integration over the area used for
   helicity calculations and corresponds to the \yohkoh\ data number (DN)
   per second, where 1~DN corresponds to 100 electrons in the CCD
   detector. Time averages are taken over 24~h. Values
   corresponding to NOAA ARs 8100, 8143, 8179, 8636, 8948, 9026, and
   9077 are represented by blue, red, yellow, green, orange, magenta, and
   pink dots, respectively. The dashed line represents a regression line,
   obtained from a least-squares method. One can see that the absolute
   helicity flux correlates better with the SXR flux than does the net helicity
   flux for the individual ARs. Note that for all ARs except 8143 and 8636
   repeated flaring activity (C-, M- or X-class) has been reported.
   (Adapted from Figures 3 and 8 of \cite{2005ApJ...620.1069M}. \copyright~AAS. Reproduced with permission.)
   }
   \label{fig:mae_kus_05_fig3_fig8_part}
\end{figure}

The rates of injected magnetic flux and helicity are necessarily related
to each other \citep[][]{lab_geo_07,2008ApJ...673..532T,par_lee_08,
2009SoPh..258...53C,tzi_geo_12}.
The flare productivity, however, may be more closely related to magnetic
helicity injection than to the amount of
injected magnetic flux (or stored magnetic energy).
More precisely, if positive and negative helicity is injected at
the same time at approximately the same rate in flare-productive ARs,
the absolute helicity flux (transported into the atmosphere by emerging motions)
correlates well with the SXR flux \citep[][and see
Figure~\ref{fig:mae_kus_05_fig3_fig8_part}]{2005ApJ...620.1069M}.
\cite{par_cha_10a} based a similar conclusion on the observation that for
flaring ARs with a large unsigned
magnetic flux, the average helicity injection rate was twice that of non-flaring
regions. In fact, a systematic study by \cite{lab_geo_07} on the helicity
injection in ARs showed that all
X-class flares were associated to peak helicity fluxes greater than
$6\times10^{36}$~Mx$^2$\,s$^{-1}$ and consequently proposed a causal
link of the peak helicity flux to the ability of an AR to produce an X-class
flare.

This proposal raises the question whether or not there is a ``critical'' amount
of helicity which, once reached, favours the onset of an eruptive process.
\cite{par_lee_08} found a nearly constantly increasing coronal helicity,
preceding all of the considered 11 X-class flares. Importantly, they suspected an
individual critical limit of accumulated helicity for each of the eruptive ARs,
since they could not identify a
common threshold for all  events.
\cite{par_cha_10a} compared the magnetic properties and flaring activity for
a sample of 378 ARs. They note that prior to all flares with a
peak SXR flux of more than $5\times10^{-5}$~W\,m$^{-2}$ a significant
amount of helicity ($10^{42}$~Mx$^2$ to $10^{43}$~Mx$^2$) was monotonically
accumulated during 12~h to 48~h.
Analysing more than 150 ARs, \cite{tzi_geo_13} found that a relative
helicity of $\approx10^{42}$~Mx$^2$ and $\approx10^{43}$~Mx$^2$ is
accumulated prior to M- and X-class flaring, respectively.
Note that whenever we speak of ``relative'' helicity, the helicity
of a non-potential field with respect to that of the corresponding potential field
is meant.
Earlier already
\cite{tzi_geo_12} had estimated the instantaneous magnetic energy and helicity
budgets of ARs and found that those with a free energy of
$\lesssim4\times10^{24}$~J and a relative helicity of
$\lesssim2\times10^{42}$~Mx$^2$ were not flare-productive.
Also, the flare class of the observed events was successively smaller (X-
to C-class) for ARs with successively less free energy and relative helicity
budgets. Early estimates of the helicity budget
of flare and CME-productive ARs, based on force-free field models,
range from $\approx10^{34}$~Mx$^2$
to $\approx10^{43}$~Mx$^2$, with the general tendency that ARs with a
higher helicity content are more flare productive
\citep[][]{reg_ama_02,reg_can_06,reg_pri_07b}.

In recent years, considerable effort has been invested into developing
methods that infer the coronal helicity budget based on the reconstructed
NLFF coronal magnetic field. This is far from  trivial, since it often
involves the computation of the magnetic vector potential that has to
satisfy specific gauge properties
\citep[][]{rud_mys_11,tha_inh_11,val_dem_12,yan_bue_13}.
So far, these models have not been applied  to real solar cases,
but have been tested by use of semi-analytic force-free magnetic field
solutions.
Only \cite{val_gre_12} investigated the helicity content of an AR during
flux emergence and found a relative helicity of $\approx10^{42}$~Mx$^2$
prior to background, \ie, B-level flaring activity.
The hesitation to apply
the developed methods certainly arises from the strong influence
the choice of a particular gauge as well as its numerical implementation
may have on the model outcome, especially when applied to real data.
Only when the reliability of such methods has been assessed, will it become
clear, whether summing over the helicity
injected at a photospheric level, \ie, estimated from routine direct photospheric
magnetic field measurements, represents the safest estimate of the coronal
helicity content \citep[][]{dem_par_09}.

\subsubsection{Helicity dissipation and transport}

\begin{figure}
   \centering
   \includegraphics[width=\textwidth]{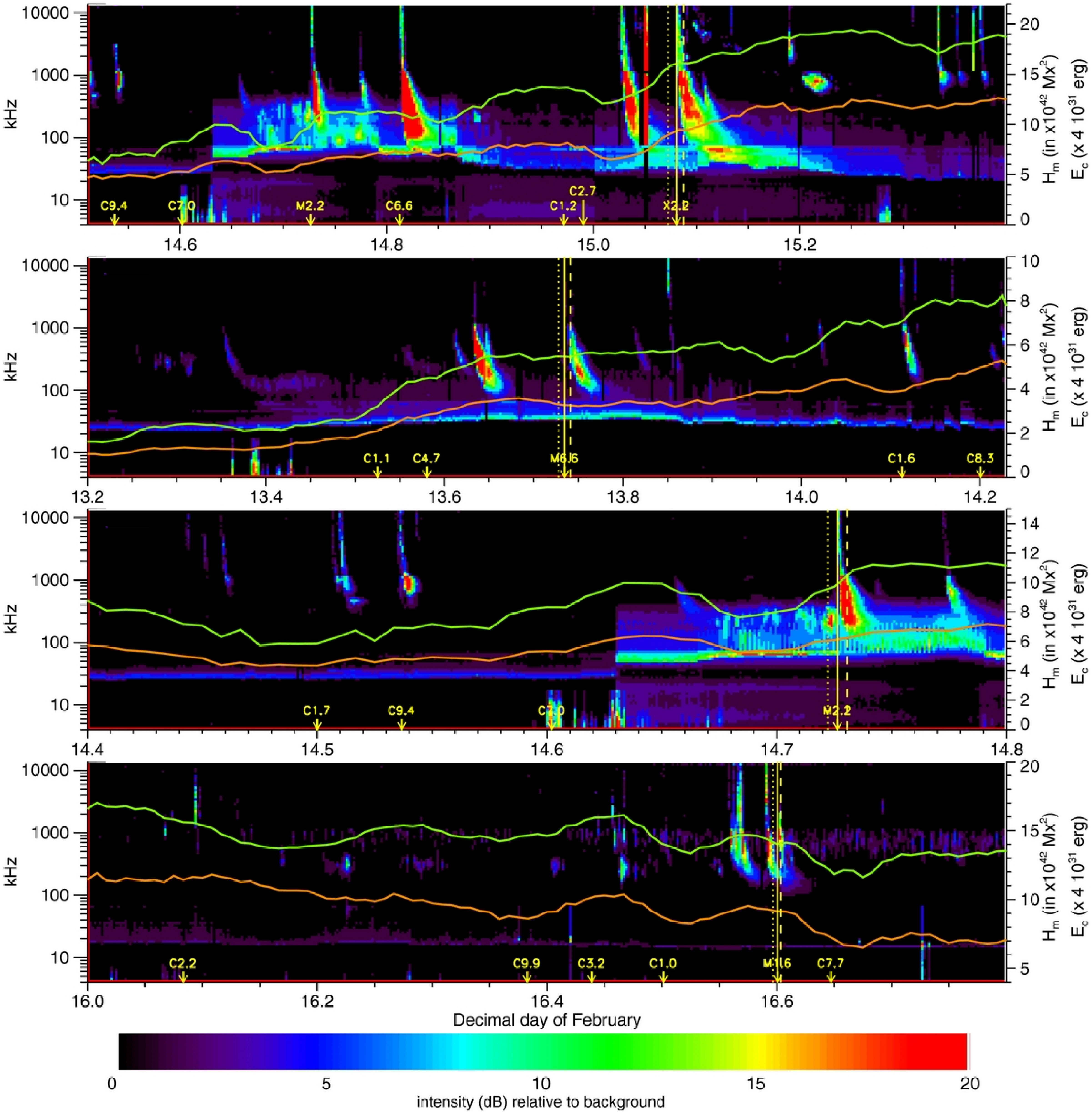}
   \put(-305,335){\bf\white\sf(a)}
   \put(-305,255){\bf\white\sf(b)}
   \put(-305,176){\bf\white\sf(c)}
   \put(-305,97){\bf\white\sf(d)}
   \caption{Relative helicity (green curves) and free magnetic energy (orange
   curves) of AR 11158 in February 2011, around the times of four large
   eruptive events (with values are averaged over 72 minutes): (a) an X2.2
   flare of February 15, (b) an M6.6 flare on February 13, (c) an M2.2 flare on
   February 14, and (d) an M1.6 flare of February 16. The flare onset, peak and
   end times (defined by the \goes\ SXR flux) are indicated by dotted, solid and
   dashed vertical lines, respectively. The color-coded background resembles
   the co-temporal \wind/WAVES frequency-time radio spectra. It can well be
   seen that a decrease of the magnetic energy and relative helicity magnitude
   starts before the onset of the flares and displays a gradual, rather than
   instantaneous character.
   (Figure 11 of \cite{tzi_geo_13}. \copyright~AAS. Reproduced with permission.)
   }
   \label{fig:tzi_geo_13_fig11}
\end{figure}
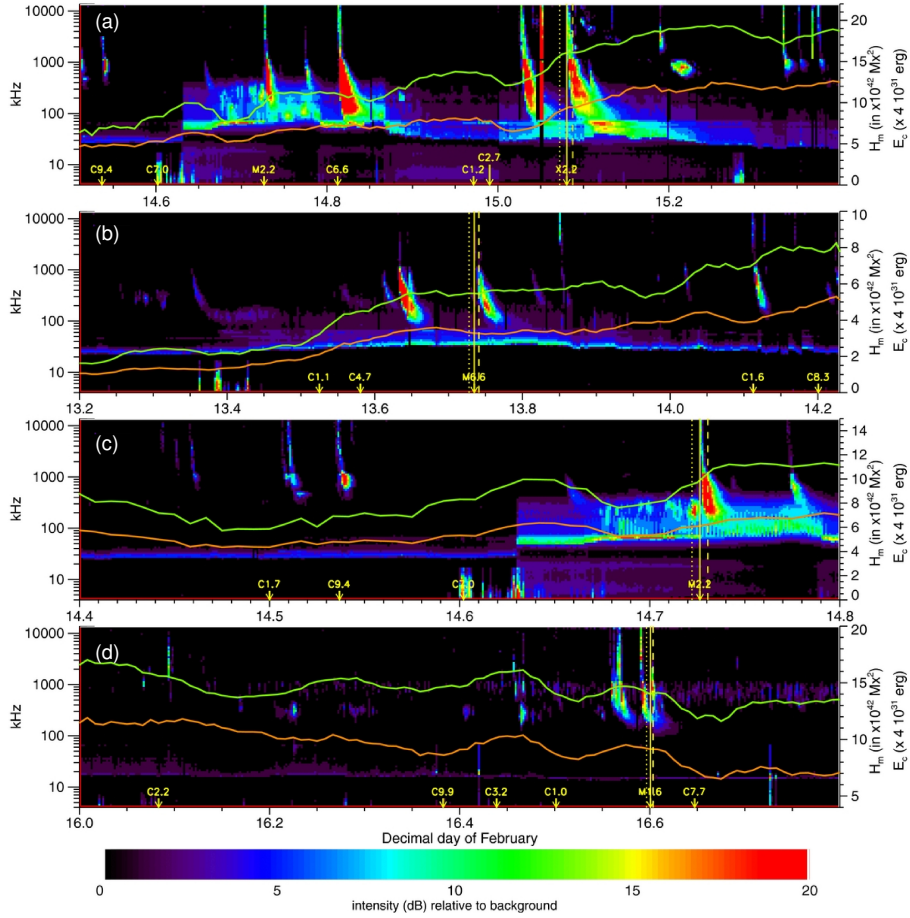

In contrast to dissipating  magnetic energy, magnetic reconnection
is a very inefficient process for removing magnetic helicity. This was
shown by \cite{ber_84}, who had tested the magnetic helicity decay in a
coronal loop, by assuming classical anomalous conductivity in a circular loop
with a length of 100~Mm, a radius of 10~Mm, and an uniform temperature
of 1~MK.
He found a diffusion time scale of $\approx10^{9}$~s to $10^{13}$~s, which is
far too long to be relevant for time scales of impulsive (eruptive) processes
such as flares, where typical time scales are $\approx10^3$~s.
Instead, during magnetic reconnection,
and on time scales relevant for it, the magnetic helicity is approximately
conserved. Only its source, the mutual linking of different flux systems or
the internal twist of a magnetic structure, is transformed
\citep[][]{hor_ras_97}. This provides a constraint on the relaxation of a
current-carrying magnetic-field configuration by energy dissipation: it can
only relax to another lower-energy, still current-carrying state but not to a
current-free one \citep[][]{wol_58,tay_74,bel_99,bel_06}.
Importantly, the decrease of both magnetic helicity and of
magnetic energy due
to an eruption is often temporary and sets in before the onset of the
eruptive event, as defined by the impulsive rise of SXR
emission \citep{jin_che_09,tzi_geo_13}. Both are often found to return to, or even
exceed the pre-eruptive level within hours, thus allowing for subsequent
flaring activity (see Figure~\ref{fig:tzi_geo_13_fig11}).

As a consequence of the conservation properties of magnetic helicity, a
change of the coronal magnetic helicity content can only be caused by
an expulsion of a magnetic structure and its inherent helicity, such as
a CME to interplanetary space. A typical CME contains a magnetic
helicity on the order of $10^{42}$~Mx$^2$, as this is also a the typical
magnetic helicity content of a MC \citep[][and see also
section~\ref{ss:hbudget_global}]{dev_00, geo_rus_09}.
These estimates agree very well with the helicity budget of ARs prior to
major eruptions \citep[$\gtrsim2\times10^{42}$~Mx$^2$; see][and see
section~\ref{ss:hstorage_ar}]{tzi_geo_12}. MCs  often show
the same chirality as their source region, as inferred from
the twist of the flux tubes they originate from \citep{rus_94, man_dem_04}.
But events have also been reported where this has not been the case. For
instance, \cite{cha_par_10}
investigated a case in which observed features (sunspot whorls and
flare ribbons) as well as a corresponding LFF magnetic field suggested
a predominantly negative helicity within an AR. In contrast, the associated
MC was attested a positive helicity, which contradicted helicity
conservation. Only a very accurate analysis of the helicity injection revealed
that a strong, local injection of positive helicity served as the source for the
observed positive helicity of the MC.

Note that already \cite{2004ApJ...610..537K} outlined the importance
of neighbouring flux systems of helicity with opposite signs within a single AR for
its flare productivity.
And \cite{lea_can_04} attested only models invoking magnetic reconnection
between the small-scale active-region magnetic fields and their overlying
large-scale envelope field, where the latter ejected an MC,
the ability to relate their helicity aspects correctly.

\section{Quiet-Sun magnetic fields}
\label{s:quiet}

Traditionally the ``quiet Sun'' regions got their name because
they were assumed to lack activity. However, while large-scale
eruptions are mainly associated with ARs (see section~\ref{s:active}),
the QS is not quiet either. Dynamic processes that
occur on smaller scales had simply not been resolved by early missions
and instruments.
Already \cite{1979SoPh...61..283L} pointed out that the total magnetic
flux of the QS exceeds the flux contained in ARs
by far and that it drives activity, such as spicules. These were first observed
in H$\alpha$ and described as ``small spices of chromospheric material''
\citep{1945ApJ...101..136R}. Ephemeral ARs (see section~\ref{sss:emergence})
were also found to be centers of activity \cite{har_mar_73}.
As in their active-region counterparts, the
photospheric magnetic field in the ephemeral regions drives activity,
such as X-ray bright points,
\citep{1974ApJ...189L..93G,1977SoPh...53..111G,1994GMS....84....1P}
in the upper solar atmosphere.

\subsection{3D magnetic field structure in the quiet Sun}

Because of the often mixed magnetic polarities on
smaller scales in the QS, we mainly find magnetic loops which are
very short and hardly reach into higher layers of the solar atmosphere.
With modern high-resolution measurements, however, we can use
magnetic field modelling techniques to model the 3D structure in the QS,
similar to those described in section
\ref{model:extrapol} and which are routinely applied to the active Sun
(see section \ref{s:active}). Nevertheless, some care has to be taken,
as is discussed in section \ref{qs_forcefree}.

% \subsubsection{QS potential field models}
Because of their easier computational load and because
only photospheric LOS measurement are required as boundary
condition, the topology of the quiet Sun's outer atmosphere
and its temporal evolution is usually investigated by use of
potential field models.
However, one has to keep in mind that any static-model approach
provides only a snapshot of the, in reality dynamic, magnetic carpet.
\cite{schr_tit_03} modelled the fraction of magnetic flux that connects
from network elements into the upper solar atmosphere and found it to
scale with the flux density of the underlying internetwork field.
For relatively strong internetwork flux densities, on the order of
$5\times10^{-3}$~Wb\,m$^{-2}$, only about 30\% to 40\% of the
flux connects to the corona while the greater part connects back down
in the form of closed loops.

\subsubsection{Photospheric quiet-Sun loops}

The structure and dynamics of the quiet-Sun magnetic field can be
investigated with spectro-polarimetric measurements.
\cite{mar_col_07}, using data from VTT, found low-lying photospheric
magnetic loops connecting about 10\% to 20\% of the visible magnetic
flux of opposite polarity. Note that emerging small-scale loops
are observed in the form of newly detected horizontal fields well before
vertical flux is detected \citep[see][for a corresponding study
based on \hinode/SOT-SP data]{cen_soc_07}. In subsequent studies,
\cite{mar_bel_09} and \cite{2010ApJ...714L..94M}
presented observations of emerging ${\rm\Omega}$-loops  by
combining magnetic and imaging data from \hinode/SOT and \dotscp.
With an even higher resolution, based on the balloon-borne
\sunrise/IMaX instrument, \cite{dan_bee_10} investigated
a large number of transverse photospheric features.
The high-resolution IMaX data
revealed that 97\% of these features are smaller than the mean size
of such structures ($\approx$\,1$\farcs$0 squared) found in earlier studies.

Already \cite{2008ApJ...679L..57I} had detected emerging
$\rm\Omega$-shaped loops. They showed that such structures,
after emergence, reach the chromosphere
and reconnect with the mainly vertical fields there. Thus they contribute
to the heating of the chromospheric plasma
there and to generating coronal MHD waves. These high-frequency waves may
then contribute to coronal heating and the acceleration of the solar wind particles.
For a description of the kinetic processes involved in coronal heating,
we refer to specialized reviews such as that by \cite{mar_06}.

\subsection{Associated photospheric fields and the response on the upper atmosphere}

QS structures are quite dynamic, as
new magnetic flux permanently emerges, usually in bipolar form.
The corresponding
loops connecting the two polarities emerge into the solar atmosphere,
while the loop footpoints move apart
due to the shape of the rising loop. The footpoints also undergo
displacement due to horizontal photospheric motions.
The opposite polarity footpoints keep separating until
eventually one or both cancel with older photospheric magnetic flux
\citep[][]{martin:etal85}. As a consequence, the magnetic flux in the
quiet-Sun photosphere is replaced every $\approx$\,14~hours
\citep[][]{2001ApJ...555..448H}. An important question is, how
the corona responds to
these changes in the photospheric field and how we can get insight
into the quiet-Sun field in the upper solar atmosphere.
This was addressed
by \cite{2004ApJ...612L..81C,2005SoPh..231...45C} using
potential field modelling based on photospheric \soho/MDI LOS magnetograms.
They found that the coronal magnetic field becomes redistributed
even faster, with a typical time of 1.4~hours.

\subsubsection{Doubts about the concept of a magnetic canopy}
\label{ss:canopy_qs}

The concept of a magnetic canopy \citep[][and see
section~\ref{intro_canopy}]{gab_76}, in the sense of a horizontal magnetic
field that fills the upper chromosphere and overlies a most field-free
atmosphere, is still a subject of debate.
One problem is that there are multiple definitions
of a canopy. One is that it is nearly horizontal field overlying
largely field-free regions. Another is that it is a horizontal
boundary between the $\beta>1$ and  $\beta <1$ field.
Instead of vertically more or less well divided
domains of high or low $\beta$, such domains may well be mixed up,
even ``islands'' may exist \citep[][]{ros_bog_02}.

A number of comparative studies of magnetic field structures and/or
bright emission from photospheric levels, and corresponding signatures
from higher atmospheric layers, revealed no obvious expansion of the
magnetic network with height (see section~\ref{ss:coupling}).
\citeauthor{2000SoPh..196..269Z} (\citeyear{2000SoPh..196..269Z},
\citeyear{2000SoPh..194...19Z}), for instance, compared
quiet-Sun magnetic signatures inferred from Fe\,{\sc{i}} 5324~\AA\
(photosphere) and H$\beta$ 4861~\AA\ (chromosphere) observations
and found no significant change of the magnetic configuration over a
height of $\approx$\,1.5~Mm.
They emphasized, however, that the H$\beta$ line exhibits
high noise levels so that the expansion of the magnetic features might simply
be hidden by the noise. \cite{pet_02}, among others,
presented \soho/SUMER C\,{\sc i} continuum (chromosphere)
and O\,{\sc vi} 1037~\AA\ (transition region) brightness observations
which revealed network elements of nearly the same size.
This observation can be construed to contradict
the concept of a nearly horizontal canopy at a height of $\approx$\,1~Mm
above the photosphere, or alternatively, it implies that the brightness
structures may expand less rapidly with height than the magnetic field.
Using \trace\ 171~\AA\ observations, already \cite{zha_zha_99} had found that
the width of the emission associated to fibrils was nearly the same at
different heights: at their
photospheric root as well as $\approx$\,30~Mm higher up in the atmosphere.
It was speculated as a possible explanation that, either the observed
emission might not be sensitive to the detection of the desired canopy
structure -- as recently supported by \cite{pet_bin_12}, see
section~\ref{sss:morphology_corona} for details -- or
that network-fields in the QS could, by some mechanism, be prevented
from fanning out \citep{zha_05}.
These claims would imply that the need for an extra force, at least as
strong as the horizontal pressure gradients, that prevents the magnetic
field from expanding with height. At least in photospheric layers, magnetic
structures and the bright points associated with them are seen to expand
with height (\cite{1995A&A...299..596B} and B{\"u}hler \etal, 2014, \aap,
submitted).

\cite{schr_vba_05} modelled the quiet-Sun network magnetic field at
heights of $\lesssim$\,20~Mm above chromospheric levels. They found
an average plasma-$\beta$ close to unity
in the entire simulation domain, indicating
that the quiet-Sun corona is not be force-free, \ie, that the magnetic
field there is not unaffected by the plasma.
A magnetic canopy also suffers at the hands of atmospheric flows.
Thus, \cite{pie_cam_11} demonstrated on the basis of 3D radiation
MHD models that in particular downflows pull U-loops out of a
magnetic canopy into the lower lying atmosphere, giving rise to
some of the horizontal fields seen in the quiet photosphere.
With time the canopy loses its clear identity due to this process.

The canopy separates the fields originating from the network
and supposedly opening into the atmosphere
from closed fields, originating and closing
in the weak internetwork. In a numerical experiment, \cite{schr_tit_03}
showed that strong internetwork fields allow flux densities capable of
considerably modifying the dynamic geometry of the magnetic coupling
between the photosphere and outer layers. They argue that bunches of
field lines likely connect from strong internetwork regions well up to the
corona while field lines from the network close back onto internetwork
fields. They estimated that the magnetic flux which
originates from internetwork fields and connect to the corona
to be $\approx50$\% of the total
flux of the quiet-Sun coronal magnetic fields.
It may well be that in the mixed-polarity QS, canopies play a
less important role than in the more unipolar environment of CHs
(see Figure~\ref{fig:Ito10_fig11} in section~\ref{PCH:QS}),
which is where they are generally placed in theoretical studies.

\subsubsection{Network and internetwork magnetic loops}

\begin{figure}
      \centering
      \includegraphics[width=\textwidth]{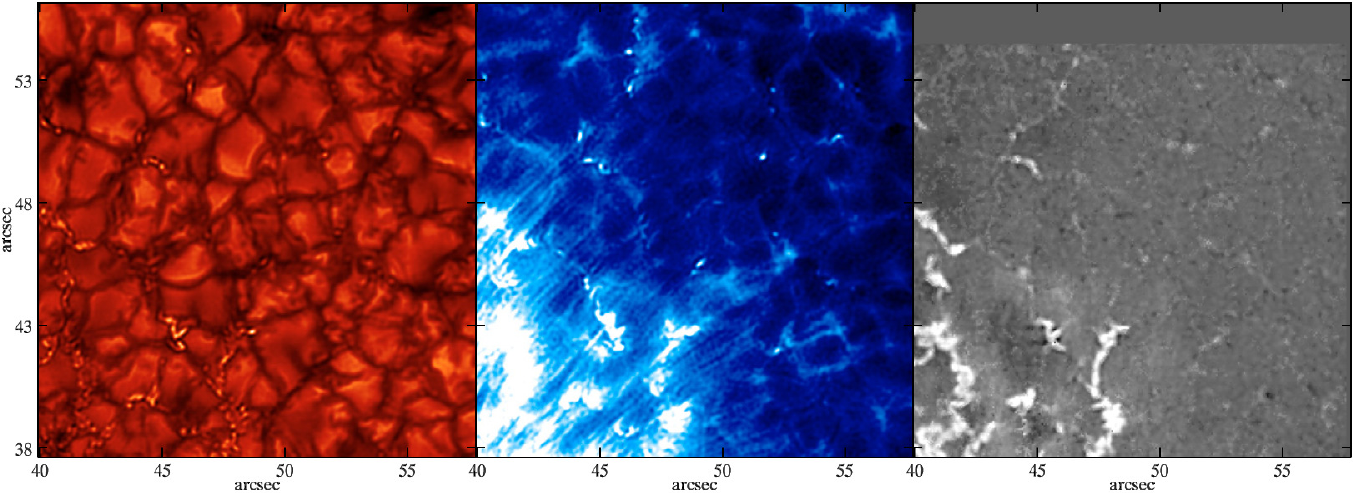}
      \put(-325,112){\bf\sf\white(a)}
      \put(-215,112){\bf\sf\white(b)}
      \put(-105,112){\bf\sf\white(c)}
      \caption{(a) Continuum image, (b) Ca\,{\sc ii}~K intensity and (c)
      longitudinal magnetic flux (ranging from -800~G to 1~kG) on
      August 9, 2007, covering a plage region and the QS seen as bright and
      dark areas in (b), respectively. The continuum
      image displays a clear pattern of dark intergranular lanes, often
      exhibiting emission from more or less tightly packed bright photospheric
      filigree. Comparison with the Ca\,{\sc ii}~K image shows that many of the
      observed thin chromospheric fibrils originate from the locations
      populated with filigree. Longer fibrils, originating near the boundary
      between the bright plage and surrounding (darker) QS areas indicate
      a canopy-like magnetic connection between areas of stronger, \ie, plage,
      and weak, \ie, QS, magnetic fluxes.
      (Figure~8 from \citealt{pie_hir_09}. Reproduced with permission, \copyright~ESO.)}
      \label{fig:canopy_plage}
\end{figure}

\cite{wie_sol_10} used a potential and LFF magnetic field model
and found that the majority of the model field lines connect
strong internetwork features below equipartition
field strength with network elements.
They were also able to show that for most of the field lines
reaching chromospheric and coronal heights, one of the footpoints
is located in the strong internetwork fields.
In other words, while the idea of a canopy, formed by horizontal fields above
activity centers, seems well established it is more controversial for the
atmosphere above quiet-Sun regions.
More recently, however, observational evidence was again found
by \cite{pie_hir_09} that chromospheric fibrils
formed a canopy over parts of a plage region and the quiet-Sun network
(see Figure~\ref{fig:canopy_plage}).
They point out that this might not even
contradict the finding of \cite{schr_tit_03} -- namely,
that internetwork fields disable
the formation of a large-scale canopy -- but instead, that the apex of
small-scale loops of the internetwork may not always
reach a height in the
atmosphere where a canopy might exist. And indeed, potential and
force-free magnetic field extrapolations carried out by \cite{wie_sol_10}
showed that the number of short field lines (with apex heights of
$\lesssim$\,500~km) outnumber the field lines reaching the
corona (with apex heights $\gtrsim$\,2.5~Mm). It has to be
noted, however, that a rather small portion of the solar atmosphere was
investigated in this study (based on the limited \sunrise/IMaX FOV),
so that a possible bias cannot be ruled out.

\subsubsection{Magnetic carpet}

\begin{figure}
      \centering
      \includegraphics[width=0.9\textwidth]{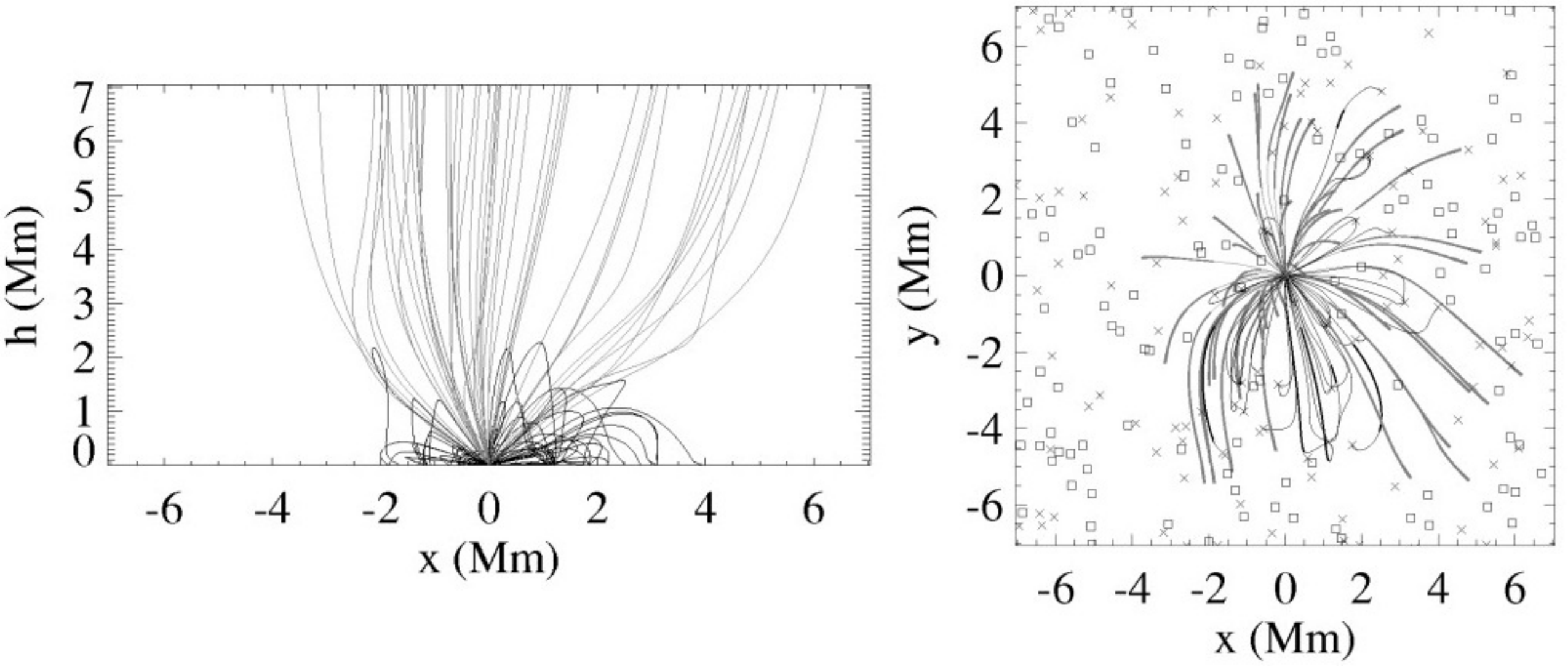}
      \put(-310,120){\bf\sf(a)}
      \put(-130,120){\bf\sf(b)}
      \caption{Illustration of the magnetic carpet, computed from a
      potential field model, when viewed (a) from the side and (b) from
      top. The plot depicts a photospheric
      magnetic flux concentration with a flux of $3\times10^{8}$~Wb,
      surrounded by a small-scale mixed-polarity field (marked by
      squares and crosses for opposite polarities).
      The FOV reaches halfway to the neighbouring network sources.
      Field lines with one footpoint in the central flux concentration
      (closed field lines shown as black and open field lines shown as
      gray curves) suggest that a collar of field lines that emanates from the
      network and closes back onto the internetwork field within several Mm,
      implying that most of the coronal field is likely anchored in the
      internetwork field, rather than in the network.
      (Adapted from Figure~1 of \cite{schr_tit_03}.
      \copyright~AAS. Reproduced with permission.)}
      \label{fig:schr_tit_fig1_2003}
\end{figure}

The dynamics of the QS is quite complex since the atmosphere above the
magnetic network is permanently and everywhere filled with small-scale magnetic flux
in the form of magnetic loops emerging from below.
This small-scale magnetic mesh is usually referred to
as ``magnetic carpet''
\citep[][and see Figure~\ref{fig:schr_tit_fig1_2003}]{tit_schr_98}.
The structure of this mixed-polarity network is driven by processes
such as magnetic flux emergence, flux cancellation as well the
coalescence and fragmentation
of magnetic elements \citep[][]{2001SoPh..200...23P}.
Widely used for the statistical analysis of the properties of magnetic features
associated with the magnetic carpet
are potential and LFF field models (see section~\ref{pfss}).
Both methods are based on the observed
LOS photospheric magnetic field and deliver similar results concerning
the length of field lines, the height of their apex in the atmosphere and the
approximate field strength \citep{wie_sol_10}.

Therefore, although the evidence for a common canopy-like field (nearly
horizontal field overlying a nearly field-free region) is relatively strong, the
criticisms of the canopy as a boundary between open and closed field lines
as well as between high and low-beta plasma are better founded. Therefore,
such views of the canopy do need revising.

\subsubsection{Role of magnetic fields in coronal and chromospheric heating}

The fact that the plasma temperature in the solar corona is a
factor of $\approx 10^3$ higher than in the photosphere motivates the search for
plausible mechanisms yielding the needed amount of energy involved.
Throughout the last decade, a number of authors, including
\cite{wal_ire_03,asch_05,mci_dep_11,wed_scu_12,win_wal_13},
demonstrated that one of several mechanisms, involving small-scale
X-ray jets, bright points, micro- and nanoflares as well as Alfv{\' e}n and MHD
turbulence and waves, are able to provide
partial explanations for the extreme heating of coronal plasma.
Mass and energy-transport between the lower atmosphere (photosphere and
chromosphere) and the outer atmosphere (corona) is, in general, a challenging concern.
Not only because of the complicated and highly dynamic structure of these layers, but
also because of the highly non-linear physical processes involved. Nevertheless,
some progress was made already during the \skylab-era
\citep[see][for a review]{1977ARA&A..15..363W}. The discussion of coronal
and chromospheric heating is well outside the scope of the present paper,
however and we therefore refer the
interested reader to the specialized review by \cite{kli_06}.

\begin{figure}
      \centering
      \includegraphics[width=0.6\textwidth]{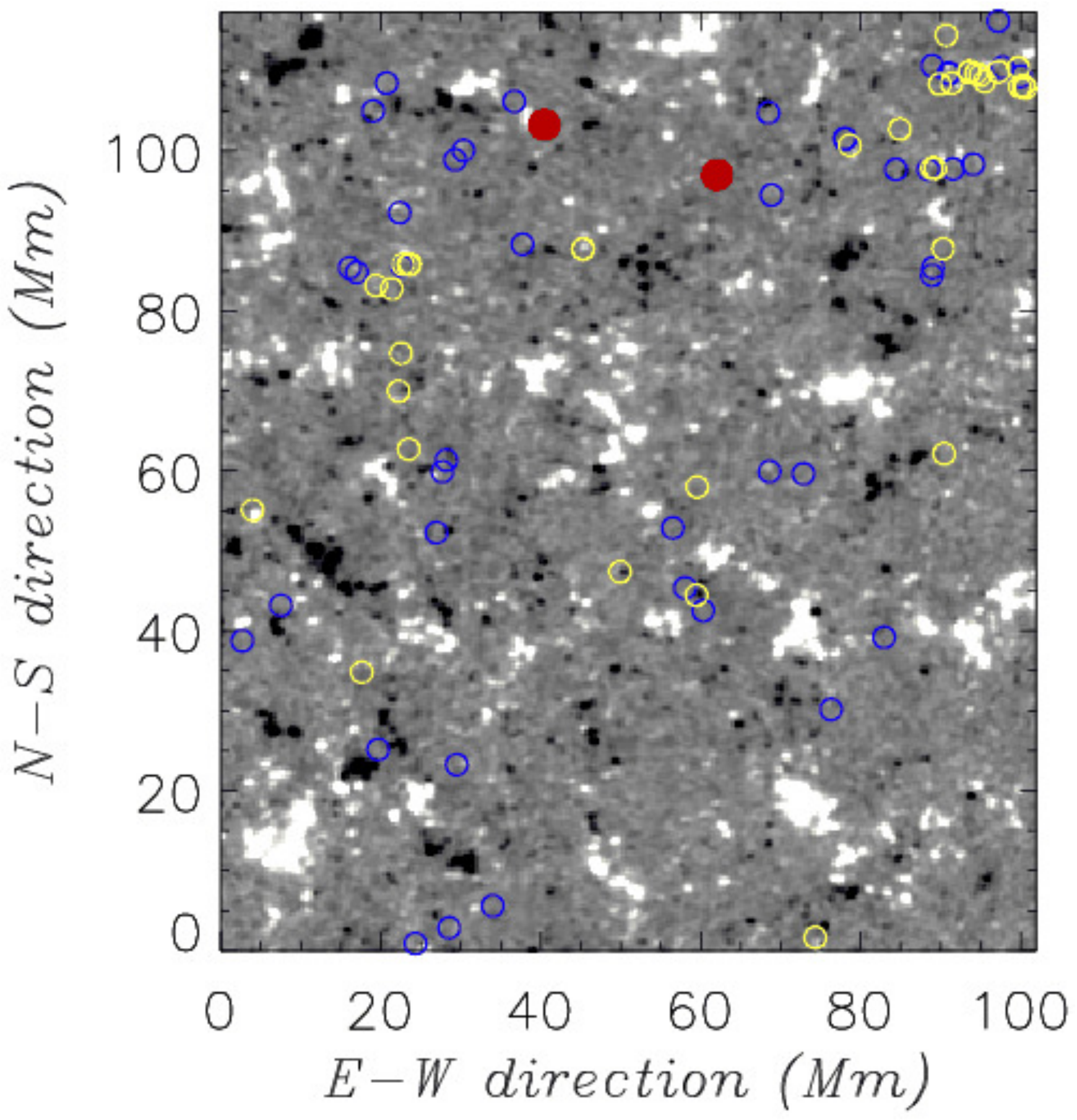}
      \caption{
      Spacial distribution of the projected location of magnetic null points, computed from a
      potential field model. Blue, yellow and red circles mark the locations of photospheric,
      chromospheric, and coronal nulls, respectively. The quantity of coronal nulls is
      outnumbered by that of the photospheric and chromospheric nulls.
      Photospheric magnetic field in a quiet-Sun area observed with \hinode/SOT
      on 24 June 2007 at 22:09~UT. The gray-scale background depicts the LOS magnetic field
      component, in the range $\pm$\,50~G. Black/white represents negative/positive polarity.
      (Figure~1 of \cite{2008A&A...484L..47R}. Reproduced with permission, \copyright~ESO.)
      }
      \label{fig:regnier_2008_fig2}
\end{figure}

Numerical experiments, carried out by \cite{2002SoPh..207..223S} showed that
low-lying chromospheric fields are well outlined by the magnetic connection
between magnetic elements observable, \eg, in the form of H$\alpha$ fibrils.
Although a complex magnetic field structure was found, which contained multiple
null points representing a potential source for reconnection,
they were found to reside in the low atmosphere
only and thus were not likely to significantly contribute to coronal heating.
\cite{2008A&A...484L..47R} used a potential magnetic
field model based on \hinode/SOT-SP magnetograms to investigate the relation
of coronal and photospheric null points and confirmed a shortage at coronal heights.
In particular, they found that 98\% of the null points present in the
model where located at photospheric or chromospheric heights, and
only 2\% were found to be situated at coronal heights
(see Figure~\ref{fig:regnier_2008_fig2}). Thus, they
confirmed earlier findings based on numerical experiments which stated that
magnetic reconnection at coronal null points is unlikely to be the source
for coronal heating.

\cite{2013SoPh..283..253W} investigated a 22-minute time series
of potential field equilibria extrapolated from \sunrise/IMaX data and
found that the magnetic connectivity in the upper
solar atmosphere changes rapidly, with a recycling time of $\approx$\,3~min --
a short time compared to $\approx$\,14~min in the photosphere. An estimation of the
upper limit for the free magnetic energy, which might be converted
to heat by magnetic reconnection was still somewhat too small
to be the dominant source for chromospheric and coronal heating
in the QS.
The same conclusion was reached by \cite{2014ApJ...793..112C},
based on a magneto-resistive computation starting from series of
\sdo/HMI and from \sunrise/IMAX magnetograms.

\subsubsection{Force-freeness of quiet-Sun magnetic fields}
\label{qs_forcefree}

Going beyond the potential field approach and
applying force-free extrapolation methods (see section~\ref{model:extrapol}),
to picture the magnetic field in the outer solar
atmosphere above quiet-Sun regions, is tempting.
This is, however, challenging, since force-free models are based on
the magnetic field measurements at low atmospheric heights, \ie, routinely at
photospheric, or occasionally at low chromospheric heights.
Additionally, in the QS one has to deal with a poor signal-to-noise ratio of
the transverse magnetic field measurements.
Nevertheless, \cite{2009RAA.....9..933Z} aimed to compare the measured
horizontal field of \hinode/SOT-SP vector magnetograms with the horizontal field component
of an associated potential field model. They confirmed that the quiet Sun's
magnetic field is non-potential, that it carries significant electric currents, and that it is
sheared, on average by about $40^\circ$ with respect to a potential field.
A different approach is to employ the magneto-resistive method of van Ballegooijen,
which returns force-free fields without the need of making use of the measurements
of horizontal photospheric magnetic field components. This approach has been taken
by \cite{2014ApJ...793..112C}.

The difficulty in modelling QS-related coronal magnetic fields using different NLFF
methods has recently been demonstrated by \cite{2011SoPh..270...89L}: the
results gained from the different modelling techniques deviated considerably.
The problems may run deeper, since even the assumption that the
magnetic field in the QS is force-free is questionable.
Following, \cite{schr_vba_05} it is unlikely
that the quiet Sun's coronal magnetic field is force-free or even
quasi-steady. One of the necessary conditions a force-free environment
has to meet is that, at a given height,
the surface integral $\int_S B_{{\rm vertical}}^2 - B_{{\rm horizontal}}^2 dS$
vanishes \citep[see][for details]{aly_89}. This condition cannot
be fulfilled if, on average, the (unsigned) horizontal field is
much larger than the vertical one.

This means that a predominantly
horizontal field cannot exist at the base of a force-free
environment, and if the horizontal flux is on average about a factor of five higher
\citep[as claimed by][]{2008ApJ...672.1237L}
the quiet-Sun photosphere is even farther away from
a force-free state than photospheric active-region fields are.
It should be mentioned though, that \citet{2010A&A...513A...1D}
found that a ratio closer to 1 is also compatible with observations.
Obviously, Aly's condition is not
fulfilled in the internetwork photosphere where the horizontal field
is on average much stronger than the vertical field, which has
important implications for the reliability of force-free magnetic field
reconstructions in the QS. The fact that these fields are definitely
far from being force-free in the photosphere and that most of the
weak (nearly) horizontal fields are the tops of very short and
low-lying loops \citep{mar_col_07,cen_soc_07,mar_bel_09,dan_bee_10}
which don't reach up to layers where the field is closer to being
force-free, the impact of the plasma on the field cannot be neglected.

\subsubsection{Magnetic energy and helicity budget}
\label{QS:helicity}

\begin{figure}
      \centering
      \includegraphics[width=0.9\textwidth]{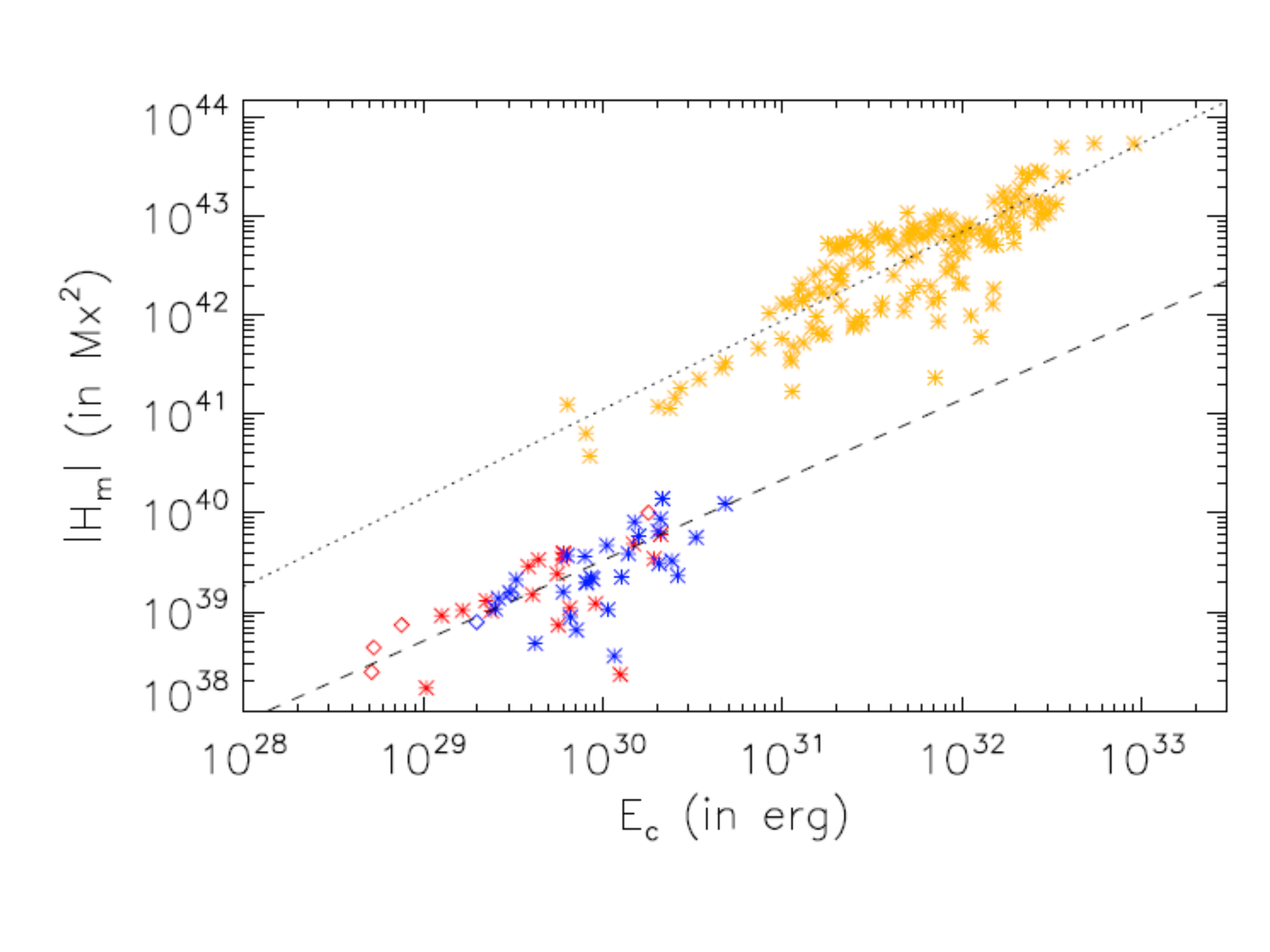}
      \caption{Free magnetic energy ($E_c$; horizontal axis) in relation to
      the magnitude of the relative magnetic helicity ($H_m$; y-axis).
      The yellow stars in the upper right part of the graph correspond
      to the values calculated for a number of ARs by \citep[][]{tzi_geo_12}.
      Blue/red plot symbols correspond to a positive/negative total
      helicity budget of solar quiet regions, respectively.
      Asterisks and diamonds mark regions
      centered within $\pm$\,30$^\circ$ heliographic latitude and higher latitudes
      than that, respectively. The dashed and dotted lines represent least-squares
      fits to the values calculated for the quiet-Sun and active-region areas,
      respectively.
      (Figure~6 of \cite{2014A&A...564A..86T}. Reproduced with permission \copyright~ESO.)
      }
      \label{fig:tzi_6}
\end{figure}

Despite the difficulties and limitations of applying force-free models
to quiet-Sun areas (see section~\ref{qs_forcefree}), NLFF models have
recently been applied to investigate the magnetic energy and helicity budget
in the QS. \cite{2014A&A...564A..86T} performed a statistical study
with a large number of quiet-Sun vector magnetograms from \hinode/SOT-SP.
They found that the QS has no dominant sense of helicity
(in contrast to ARs) and that both the
free magnetic energy and the relative magnetic helicity are correlated
with the area occupied with network elements. Consequently, the free
magnetic energy and the relative helicity are correlated as well
(see blue/red stars and diamonds in Figure~\ref{fig:tzi_6}, respectively).
In an earlier work, \cite{tzi_geo_12} investigated these quantities for
ARs (shown as yellow stars in Figure~\ref{fig:tzi_6}, for comparison).
Both energy and helicity are much lower in the QS,
compared to ARs, which is naturally caused by
the lower magnetic flux they contain. The authors also noted that
the investigated quiet-Sun areas are closer to a potential field than
the ARs, which is another reason for a lower free energy in the QS.
See also sections \ref{ss:hbudget_global} and \ref{ss:helicity_ar}
for magnetic helicity investigations on global scales and in ARs,
respectively.

\subsection{Small-scale dynamics}

The observations of the coronal magnetic field show numerous dynamic phenomena
on small as well as larger (active-region) scales.
In particular, we now know that the term ``quiet Sun'' is not really
adequate to describe the coronal field even outside of ARs, since numerous
dynamic phenomena on small scales are known to occur there
\citep[][]{1983SoPh...89..287S,par_88,1992PASJ...44L.173S}.

\cite{2013ApJ...778L..17C} found the
first evidence for sigmoidal (\ie, S-shaped) loops in the QS using \sdo/AIA
data. These were interpreted as a clear indicator for non-potential magnetic fields.
So far S-shaped loops had been observed to connect within ARs or
to connect ARs on either side of the solar equator (\ie, in the form of TELs) only.
S-shaped loops in the QS indicate
that similar dynamic phenomena as in ARs might
occur here, but on smaller scales. In fact, events similar to small versions of
flares and CMEs are found in the QS \citep{2009A&A...495..319I,2010A&A...517L...7I},
which mostly attract less attention
than their impressive, and sometimes globally disturbing, active-region counterparts.
In fact, outside of ARs the Sun displays a broad range of smaller-scale, transient
phenomena which involve only a fraction of the energy that is involved in the
dynamics associated to the predominant large-scale activity sites.

\subsubsection{Jets}
\label{qs:jets}

Already in the 80s of the last century, \cite{1983ApJ...272..329B} reported
observations of high-energy supersonic jets in quiet-Sun areas. They found
that the associated shock waves reached to heights of about 4~Mm to 16~Mm.
The authors applied a cloud model for the solar wind and assumed that
the entire kinetic energy of the wind is provided at the Suns surface.
This involves the presence of emerging flux and
a field strength of 15~G and
a flux emergence of $1.3\times10^{15}$~Mx is required to power the
corresponding high speed jets.
In their opinion, magnetic forces are the most
likely cause for the jets and a correlation to active areas in the QS was found.
Moreover, the influence of the individual jets was discernible as fine structures in
the solar wind flow at a distance of one~AU.
Later, X-ray jets were observed by \cite{1992PASJ...44L.173S,1994ApJ...431L..51S},
although not necessarily in quiet-Sun regions.

\begin{figure}
      \centering
      \includegraphics[width=0.6\textwidth]{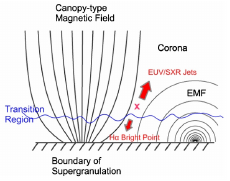}
      \caption{Schematic view on the reconnection occurring in the solar corona upon
      emerging magnetic flux from the center of a supergranule below with a
      pre-existing magnetic field. Near the reconnection X-point, the temperature is
      enhanced and outflows are observed presumably as
      EUV/X-ray jets in the corona and H$\alpha$/Ca surges in the chromosphere.
      (Adapted from Figure~10 of \cite{jia_fan_12}. \copyright~AAS. Reproduced with permission.)}
      \label{fig:jia_fan_2012_fig10b}
\end{figure}

\cite{1995Natur.375...42Y} carried out 2D resistive MHD simulations
including a simplified convection zone, photosphere, chromosphere and corona.
They found that magnetic reconnection plays an important role for
activity in the upper solar atmosphere and can be a common cause
of both hot X-ray jets, as well as cooler loop brightening
that become visible in H$\alpha$. In the former case plasma is heated up to
temperatures between about 4~MK and 10~MK and is accelerated to the
Alfv{\'e}n speed of about 100~km\,s$^{-1}$.
In the latter case, cool ($\approx$\,$10^4$~K) chromospheric material
is accelerated almost without plasma heating.
The authors concluded that the combined
energy of cool-loop brightening and hot-jet acceleration might be
significantly larger than previously thought. Also the energy in
loop brightenings might have been underestimated from previous observations.

The role of magnetic reconnection
for heating and dynamics of the solar corona was revisited
by \cite{2007Sci...318.1591S}, this time observationally
with Ca\,{\sc ii}~H data from \hinode/SOT. So called ``anemone jets'' were found
both in the corona and in the chromosphere.
The authors interpreted the frequently observed
small chromospheric anemone jets as a manifestation of ubiquitous
magnetic reconnection. The coexistence of cooler chromospheric material,
that becomes visible in H$\alpha$ and the hot jets expected from
numerical experiments by
\cite{1995Natur.375...42Y} was observationally confirmed. The combined
energy of these events was more than two orders of magnitude too small to
explain the heating the active corona. Nonetheless, the authors pointed out that the
jets observed by \hinode/SOT are only the largest events and a great number of
smaller jets might have gone undetected, but may also contribute to the heating.

Recently \cite{jia_fan_12} presented 2.5D MHD simulations of the
processes involved in canopy-type magnetic configurations triggered by newly
emerging magnetic flux into a pre-existing field configuration at the boundary
of a supergranule. They found hot ($\approx$\,$10^6$~K) and cold ($\approx$\,$10^4$~K)
jets originating from coronal microflares and associated them to what is usually
observed in coronal and chromospheric images as EUV or HXR jets and
H$\alpha$/Ca surges, respectively (Figure~\ref{fig:jia_fan_2012_fig10b}).
That magnetic reconnection between newly emerging magnetic flux and the
ambient field plays a major role for jets has been confirmed in numerical
experiments recently by \cite{2013ApJ...771...20M}.
For a recent review on small-scale, jet-like events at chromospheric levels, we
refer to \cite{tsi_tzi_12}, as well as to section \ref{CH:jets}
for polar jets.

\section{Coronal holes}
\label{s:chs}

\begin{figure}
      \centering
      \includegraphics[width=0.9\textwidth]{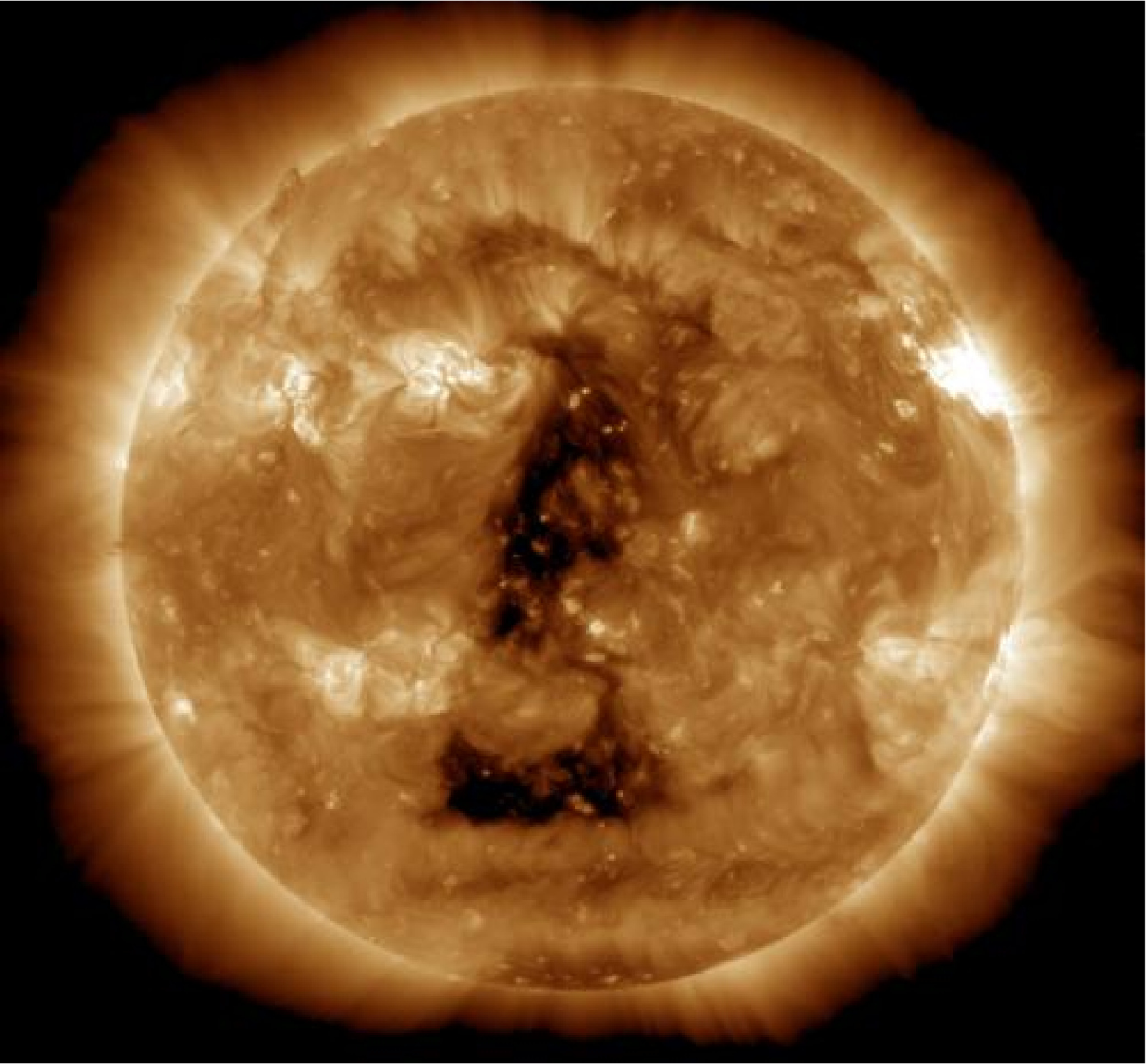}
      \caption{A large coronal hole observed with \sdo/AIA at 193~\AA\ on 13 March 2012.
      It can easily be distinguished from its quiet-Sun surrounding as the region of lowest EUV emission,
      while ARs exhibit strongest emission. Source:
      {\scriptsize{http://phys.org/news/2012-03-huge-coronal-hole-solar.html}}
      }
      \label{1-hugecoronalh}
\end{figure}

CHs are regions of strongly reduced emissivity
at coronal temperatures and, consequently seen as
dark features in coronal images (which is why they are called ``holes'';
see Figure \ref{1-hugecoronalh}). The magnetic field of CHs is mainly open,
\ie~organized in magnetic funnels having their footpoints in the network.
Along open fields, charged particles can escape the outer solar
atmosphere, leaving behind plasma of strongly reduced
density and slightly reduce temperature, if compared to the quiet Sun's corona.
The main motivation for the research of CHs since the 1970s was to understand
the role of CHs in terms of the mass and energy flow between the
Sun's surface and the heliosphere. For a recent overview we refer to
\cite{cra_09}. One distinguishes between polar CHs (PCHs) and equatorial
CHs \citep[ECHs; see][]{1999JGR...104.9753D}.

For techniques used to analyse observations of CHs,
it makes a difference whether the
holes are observed on the disk or above the solar limb. Discussing
the corresponding challenges and issues, however, lies outside the scope of this work
and we refer to the review by \cite{cra_09}. Here, we focus on the magnetic field
structure in CHs. As other magnetic structures on the Sun, CHs and the corresponding
solar wind streams change over the solar cycle
\citep{1979SSRv...23..139H}. For changes of the magnetic structure of CHs
in the course of the solar cycle see section~\ref{ss:global_cyclic}.

\subsection{Properties of coronal holes}

\subsubsection{Properties of photospheric fields associated with coronal holes}

Almost continuous observations of CHs and measurements
of the associated photospheric magnetic fields are available since the
\skylab\ era in the 1970s. To our knowledge, a first review on
observations of CHs and the
underlying solar magnetic field was presented by \cite{1979SSRv...23..139H}.
Somewhat larger CH-averaged values, ranging from 0.8~G to 17~G, with an
average value of nearly 8~G, were found by \cite{wiegelmann:etal04}.
Conclusions on the magnetic
structure of CHs were at that time mainly based on the photospheric
LOS magnetic field observations (\eg, from \kpno). Unfortunately, at that time,
the magnetic field data were subject to major uncertainties owing to systematic
errors (a factor of $\gtrsim2$).
Nevertheless, it already became clear that the photospheric field in which
CHs are rooted is dominated by fields of a single polarity, \ie, either positive
or negative.
In ECHs, for instance, the main polarity of the holes dominated
in the range of 58\% to 95\% and, on average, 77\%\,$\pm$\,14\%
\citep[][]{wiegelmann:etal04}. Equatorial quiet Sun regions where
approximately flux-balanced
in the range of 2\% to 29\% and, on average, 9\%\,$\pm$\,9\%.
The magnetic field strength of these photospheric fields,
averaged over the whole CH, was found to lie in the range 0.5~G to 7~G
\citep[][and references therein]{1979SSRv...23..139H}.
It is noteworthy that there is some scatter in the values found, depending on
the observations used for analysis.

As in ARs and in the QS, reliable magnetic field measurements
are mainly available at photospheric levels.
Because of the poor signal-to-noise ratio in weak-field regions, it
is difficult to measure the weak transverse field with high accuracy.
As a consequence, observational studies and coronal magnetic field models
are often based on LOS magnetic field measurements.
An additional problem arises for the observation of CHs in polar regions,
where foreshortening effects cannot be neglected (since
the spatial resolution in the direction
to the limb decreases) and where limb darkening can also become an
issue.

\subsubsection{Modelling of coronal-hole magnetic fields}

Similar to the QS and global magnetic fields, potential field
models based on photospheric field measurements are used
as a first approximation of the magnetic field in CHs.
Note, however, that also when applied to model the
magnetic field in coronal-hole areas potential field models have
their limitations (see section~\ref{qs_forcefree} for details).
The magnetic field structure of CHs is often also studied with the help of
PFSS models (see sections~\ref{pfss} and \ref{global_top}).
In their output, coronal-hole areas are easily identified by their
open magnetic fields (\ie, field lines connecting
the photosphere with the source surface). A powerful tool for
CH-investigations is the combination of flux-transport models in
the photosphere and coronal field modelling with PFSS
(see section ~\ref{model:mhd:flux} for a description of the
method and section \ref{model_ech} for an application to ECHs).

\cite{2009SSRv..144..383W} pointed out that coronal-hole areas
can be reasonably well reproduced over the entire solar cycle when assuming
the source surface of a PFSS model to be located at a constant height of
2.5 solar radii (\ie, $\approx$\,$1.74\times10^{9}$~m) above the photosphere.
The main contribution to the open flux in PFSS models arises from the
dipole and quadrupole moment of the Sun's global magnetic field --
the latter, especially around solar maximum.
Higher order multipoles decrease too rapidly with height in order to contribute
significantly at the source surface.
Note that taking the quadrupolar field into account does not substantially
change the resulting rotational properties of the coronal-hole associated magnetic
field: the photospheric field (represented by the high-order multipole
moments) rotates differently while the coronal field
(characterized by the low-order multipoles) rotates almost rigidly.
% This indicates that in open flux regions, such as CHs, the lower-order multipoles dominate.

\subsection{Polar coronal holes}
\label{pchs}

During solar activity minimum, large CHs occur in the polar
regions as a consequence of the dominating dipolar
moment of the Sun.
One has to keep in mind, however, that all measurements concerning
near-polar areas, are carried out from the ecliptic, \ie,
from a highly inclined viewpoint with respect to the
target region. With the launch of the \so\ mission, scheduled
for 2017, this drawback will finally be overcome.

\begin{figure}
      \centering
      \includegraphics[width=0.9\textwidth]{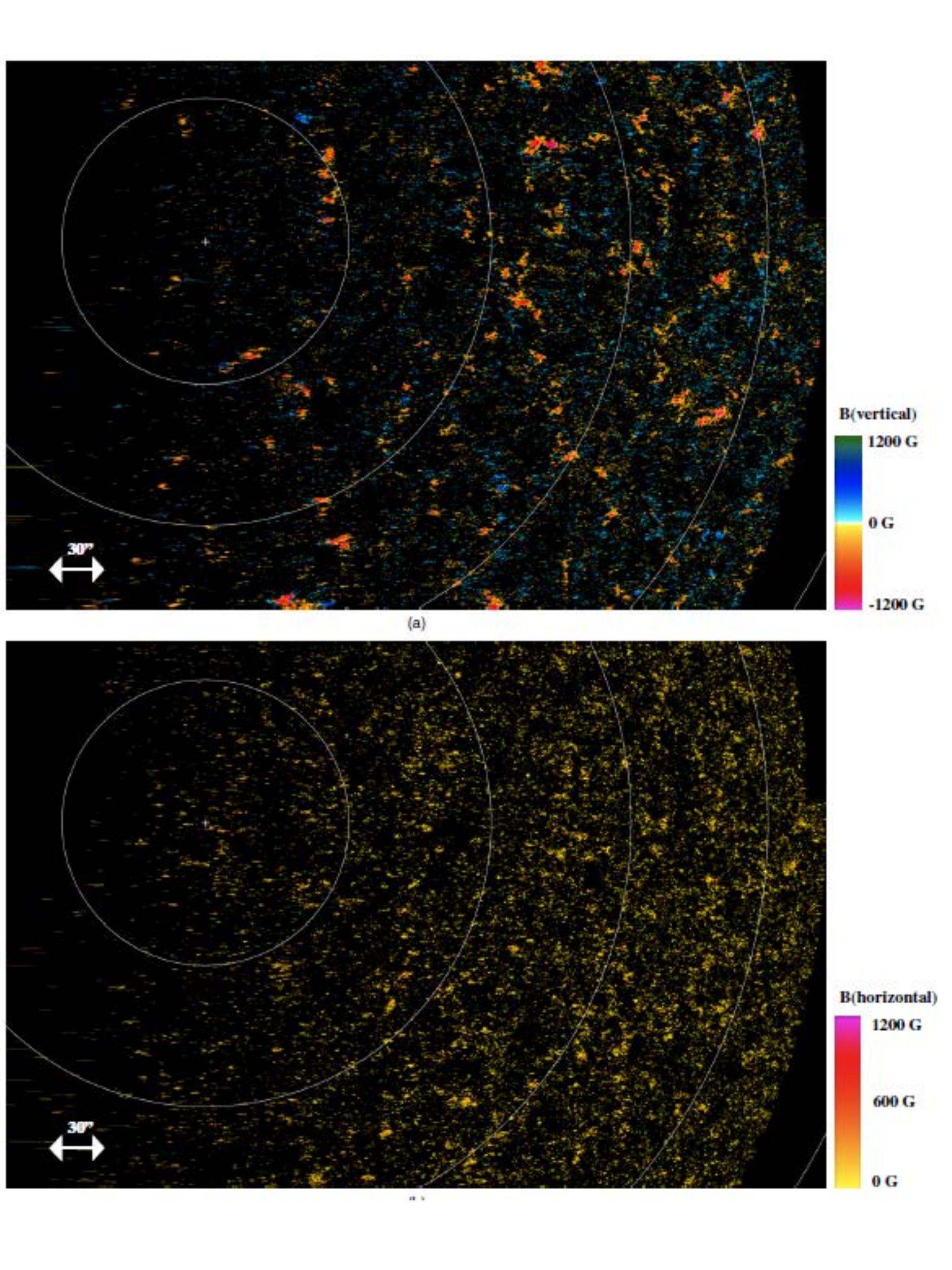}
      \put(-25,375){\bf\sf(a)}
      \put(-25,190){\bf\sf(b)}
      \caption{Maps of the signed strength of the (a) vertical and (b) horizontal
      components of the magnetic field vector on 25 September 2007, converted to a view
      from above the north pole of the Sun. East/west is to the left/right.
      The pixel size is $0\farcs16$. Black areas mark locations where
      the magnetic field strength has not been obtained, because the
      associated polarization signal did not exceed a given threshold
      above the noise level. The color-code in (a) represents the signed
      strength of the vertical magnetic field, with red/blue representing
      negative/positive polarity. Many of the horizontal field concentrations
      in (b) are co-located with the vertical field patches in (a).
      (Figure~4 from \cite{2010ApJ...719..131I}.
      \copyright~AAS. Reproduced with permission.)}
      \label{fig:Ito10_fig4}
\end{figure}

\subsubsection{Magnetic flux in polar coronal holes}

\hinode/SOT-SP provided the capability of studying
the full photospheric magnetic-field vector around the Sun's south polar
region \citep[][]{2008ApJ...688.1374T}.
Unipolar, vertical flux tubes with kG fields,
scattered all over the PCH (ranging from 70$^\circ$ to 90$^\circ$ solar
latitude) and ubiquitous horizontal fields were found.
The strong (kG) vertical spots
are unipolar, while the average field strength
of the entire FOV is only about 10~G, depending on the filling factor.
In total, \cite{2008ApJ...688.1374T} found the horizontal magnetic flux
in the polar region, around solar minimum,
to be almost twice the vertical flux, namely
$4.0\times10^{21}$~Mx and $2.2\times10^{21}$~Mx, respectively.

\cite{2011ApJ...732....4J} revisited \hinode/SOT-SP vector magnetic field
measurements in a PCH and compared it to measurements of quiet-Sun
areas close to the solar limb and the disk center, as well as with low-latitude
CHs. They also uncovered an imbalance of vertical and horizontal field in
the polar region, given an average
vertical and horizontal flux density of $\approx$\,100~G and
$\approx$\,1~kG, respectively.
They estimated that such patches of strong vertical field
occupy $\approx$\,7\% of the CH area
investigated in their study. The authors also revealed a significant
amount of magnetic flux with a polarity opposite to the dominant polarity.
That implied that only about one third of the magnetic flux in the analysed
polar region could be assigned to be actually open magnetic flux.
They also found that the ratio of dominant to minor flux for low-latitude and
PCHs is similar, but that the total vertical magnetic field strength in polar
holes is about 60\% higher.

In general, we expect that as an increasingly larger fraction of the magnetic
flux within CHs is detected, the percentage of open flux relative to the total
flux in a CH will continue to decrease. However, we also expect that much
of the hidden flux is weak and ordered on very small spatial scales.
Consequently, much of this flux is expected to be restricted to very low-lying
loops that do not reach into the corona, as found by
\cite{wie_sol_10} in the quiet Sun.

\subsubsection{Comparison to the quiet-Sun magnetic field}
\label{PCH:QS}
\begin{figure}
      \centering
      \includegraphics[width=0.95\textwidth]{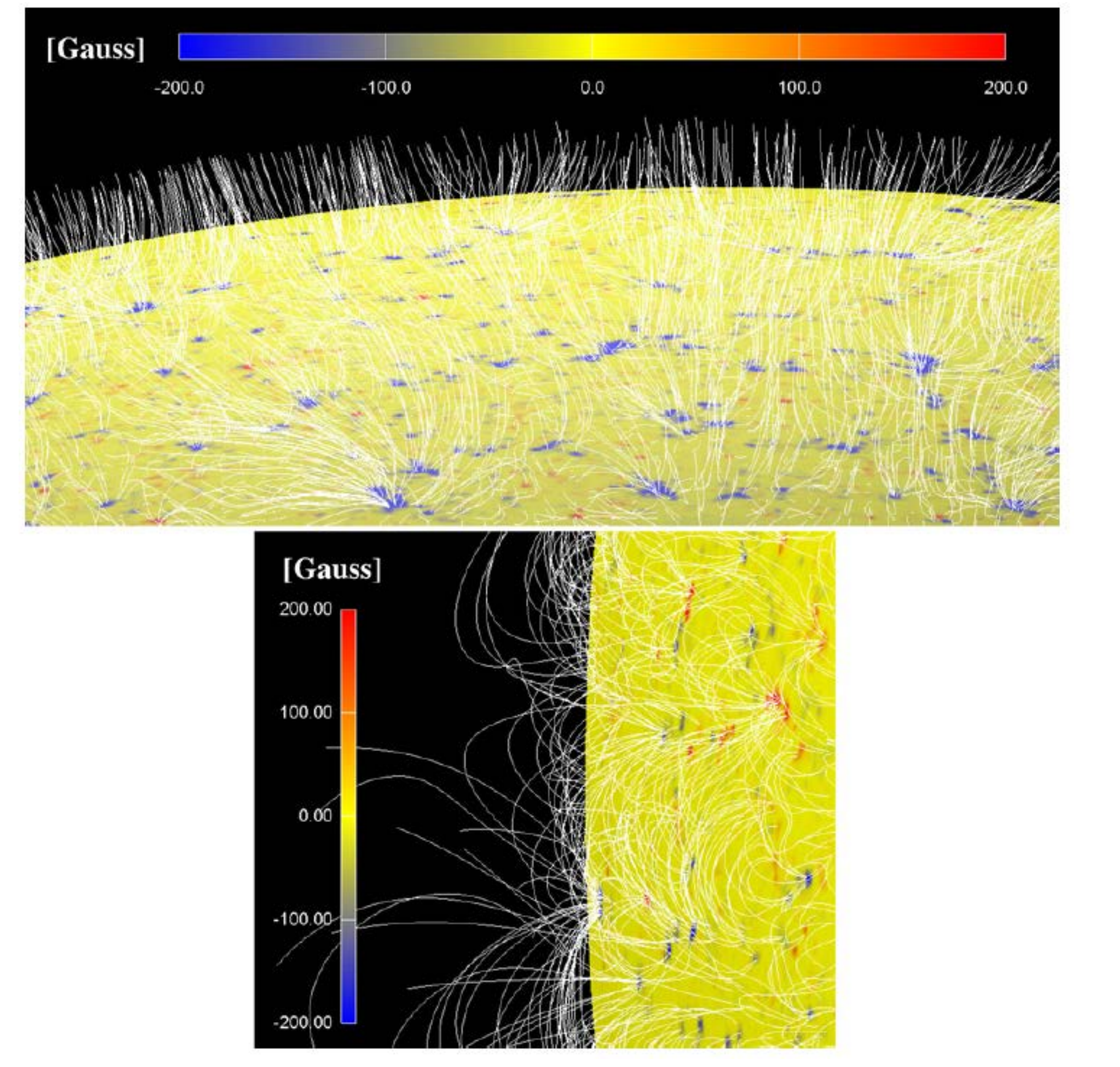}
      \caption{Selected field lines, calculated from a PFSS model,
      outlining the coronal magnetic field structure associated to a CH (a)
      around the north pole of the Sun and (b) in a quiet-Sun area near the
      East limb. It can be seen that the majority of field
      lines that originate in kG-features fan out just above the photosphere
      (in a canopy-like fashion) and that they are open. The color-coded surface
      indicates the vertical magnetic field, where blue/red represents
      negative/positive polarity. It is evident that the polar region is dominated
      by a specific (negative, in this case) polarity, while the QS represents
      an environment of mixed polarity.
      (Figure~11 from \cite{2010ApJ...719..131I}.
      \copyright~AAS. Reproduced with permission.)}
      \label{fig:Ito10_fig11}
\end{figure}

Using \hinode/SOT measurements, \cite{2010ApJ...719..131I} compared a polar
CH region with a quiet-Sun region near the East limb at an epoch close to solar activity
minimum (Figure~\ref{fig:Ito10_fig4}). The total magnetic
flux and area covered by kG-fields was found to be larger in the north polar region
than in the QS near the East limb. Also, while the vertical magnetic field was
found to be nearly balanced in the considered quiet-Sun, near-limb area,
one polarity clearly dominated the polar regions. In particular,
around 90\% of the features with a magnetic field magnitude of more than
500~G were of the same polarity.
A PFSS model revealed mainly open magnetic fields in the
north-polar CH and predominantly closed structures in a quiet-Sun region
near the East limb (see (a) and (b) in Figure~\ref{fig:Ito10_fig11}, respectively).
Because of one dominating polarity in the hole, the opposite flux cancels out
by low-lying loops and above a certain height all field lines are open.
Note that this is a general property of CHs which has also been
found for ECHs (see section~\ref{ECH:QS}).
The authors also found that the horizontal fields are similar
in the QS and at the poles and interpreted this as evidence for a
local dynamo. A global dynamo would create different fields in polar and
equatorial regions, they argued; see also section \ref{qs:dynamo}.

\subsubsection{Polar plumes}

Polar plumes are visible in the form of thin streamers above the solar limb
in the polar regions.
They have been studied intensively, \eg, by a coordinated observing campaign
\citep[\soho\ and several of other space-borne and ground-based instruments;
see][]{1997SoPh..175..393D}. A comprehensive review was given by
\cite{2011A&ARv..19...35W}, including discussion of the 3D structure of polar
plumes, details of their generation and of their interaction with the solar
wind. Here, we concentrate on the magnetic field structure of polar plumes.

Supported
by potential field models, polar plumes are associated to open fields
on large scales \citep[][]{1982SoPh...75..145S}.
While the source region of these plumes is usually not
visible in coronal EUV images, stereoscopic reconstructions by
\cite{2009ApJ...700..292F}
showed that they connected to photospheric magnetic patches.
\cite{1995ApJ...452..457W} and \cite{1998ApJ...501L.145W}
proposed that plumes are produced by the reconnection of
emerging mixed polarity fields with previously existing unipolar fields.
\cite{2008ApJ...682L.137R} reported that
most of the studied coronal jets in polar regions are generally
followed by plumes. The authors pointed out the common feature shared by
plumes and jets: a mixed-polarity field area where their footpoints are located.
In their picture, flux
emergence and subsequent opening of previously closed loops by magnetic
reconnection
provide the energy for jet acceleration and eventually produce the open
(plume) field geometry. For $70\%$ of the observed polar jets, a plume was found to form
within minutes to hours after the appearance of the jet. Such an association
is plausible, given that the formation mechanism of jets is also thought to be
the reconnection of emerging flux with previously present unipolar field
\citep[\eg,][]{1995Natur.375...42Y,1996ApJ...464.1016C}.

\subsubsection{Polar Jets}
\label{CH:jets}

Jets are ubiquitous dynamical features which occur in CHs, ARs and the QS.
(For a discussion of jets occurring in the QS, see section~\ref{qs:jets}.)
For example, \cite{2007Sci...318.1580C} detected about 10 fast
jets per hour in a PCH. Detailed analysis revealed that at least
a sub-set of jets is characterized by two distinct velocities:
one close to the sound speed
($\approx$\,200~km\,s$^{-1}$) and another one on the order of the Alfv\'{e}n
speed ($\approx$\,1000~km\,s$^{-1}$).
This might be explained with the following scenario.
Initially, the outflow triggered by reconnection has a velocity close to the
Alfv\'{e}n speed. Due to the transit of a reconnected magnetic field line
to a relaxed configuration, Alfv\'{e}n waves become excited,
which might also contribute to the acceleration of the fast solar wind.
The reconnection process also has the effect of heating the coronal
plasma by converting magnetic to thermal energy. As a consequence, the
plasma expands, which also leads to a plasma outflow. Its
speed, however, is significantly lower (sound speed) than the outflow
component caused by magnetic reconnection.

Using spectroscopic data from \soho/SUMER and \hinode/EIS,
\cite{2009A&A...502..345K} investigated the formation of
jets in the transition-region and coronal environment of CHs.
They deduced that the open magnetic fields associated with the jets
were rooted in kG vertical fields at photospheric
levels. They also found that both explosive phenomena and cool
up-flows were caused by magnetic reconnection with low-lying
loops in the transition region, confirming and extending earlier results.
\cite{2011ApJ...732L...7Y} investigated the boundaries of the equatorial
extension of a PCH
using co-observations of the photospheric magnetic field from \sdo/HMI and
the coronal plasma EUV emission with \sdo/AIA. A number of jets were
recognized in the EUV images and interpreted as signatures of magnetic
reconnection. The latter was supported by co-temporal emergence
and cancellation of magnetic flux seen in the photospheric magnetic field data.
The coexistence of open and closed fields at CH boundaries
naturally result in energy deposition by multiple reconnection events,
those thought to produce jets.

\subsubsection{Contribution to the solar wind}

\cite{2011ApJ...736..130T} observed high-speed outflows of
$\approx120$~km\,s$^{-1}$ in plume-like structures in polar
and ECHs, as well as quiet-Sun regions
in \sdo/AIA images.
Comparison with a potential coronal magnetic field model based on
\sdo/HMI measurements led the authors to conclude that plume-related
jets do not necessarily contribute to the solar wind.
This confirmed what was found earlier by \cite{2009ApJ...700..292F}
using \stereo/SECCHI data, namely that the contribution of polar plumes
to the fast solar wind is insignificant. Note, however, that studies of outflows
in polar plumes and inter-plume-regions are contradictory. For instance,
\cite{2003ApJ...589..623G} claimed that about half of the fast solar wind at
a distance of 1.1~solar radii originates from plumes. A corresponding
discussion is beyond the scope of this work and we refer to section~4
of \cite{2011A&ARv..19...35W} for a detailed review of the relevant literature.

\subsection{Equatorial coronal holes}

During cycle phases of higher solar activity, CHs are present at all
latitudes. ECHs are smaller and persist for a
shorter time than PCHs. Nevertheless they persist over several solar
rotations. Many low-latitude holes are located close to the edges of
ARs. That helps to understand the changes of CHs in the course
of the solar cycle: because ARs tend to emerge closer and closer to
the solar equator, CHs exhibit a similar trend.
In some cases ARs appear even within
these CHs \citep[see section 3 in][for details]{cra_09}.
As a consequence of the emerging AR, the CH decreases in size.
%\cite{1998JGR...103.6585L}, suggested that CMEs
%play a role in opening up closed magnetic fields and thus the formation of
%CHs.

\subsubsection{Formation and evolution}
\label{model_ech}

In contrast to the differential character
of the solar rotation in the sub-surface layers of the Sun
ECHs at coronal
heights rotate more rigidly \citep{1975SoPh...42..135T}, but still
with a differential character \citep{1995SoPh..160....1I}.
\cite{1974Natur.250..717G} suggested ``magnetic merging'',
or more precisely magnetic reconnection,
of oppositely directed magnetic arches could possibly form CHs
and also explain their more rigid rotation.

\begin{figure}
      \centering
      \includegraphics[width=\textwidth]{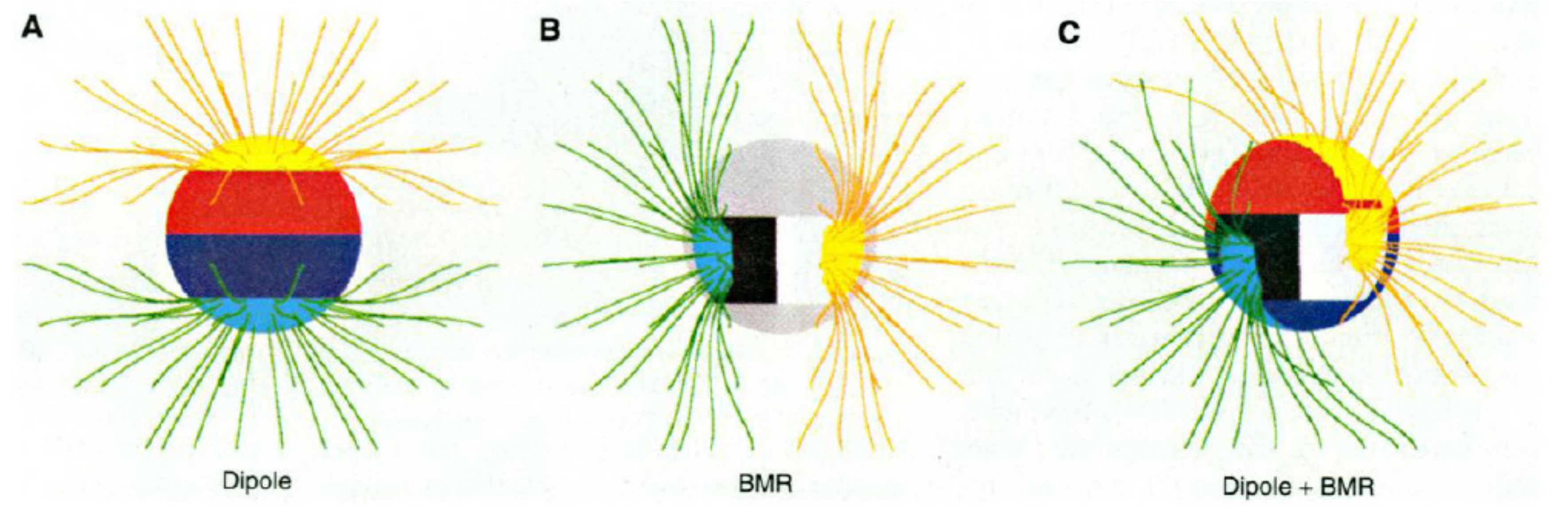}
      \caption{Examples for idealized open magnetic field regions:
      Panel A shows an axisymmetric dipole, panel B a bipolar magnetic
      region (BMR) and Panel C a superposition of dipole and BMR.
      Orange and green field lines originate in positive and negative
      flux regions, respectively. The color coding on the solar surface
      correspond to different regions: white and black to
      positive and negative polarities in ARs, yellow and light blue
      to positive and negative polarities in CHs,  red and dark blue
      to positive and negative background fields.
(Adapted from Figure~4 of \citep{1996Sci...271..464W}.
Reprinted with permission from AAAS.)}
      \label{wang_1996_fig4}
\end{figure}

Associated numerical experiments have been carried out by
\cite{1996Sci...271..464W} and as a possible origin
of ECHs, the interaction of bipolar ARs has been identified,
which is able to explain the imbalance of magnetic flux that is
characteristic for CHs.
Modelling a bipolar magnetic region superposed on an axisymmetric
dipole global field,  \cite{1996Sci...271..464W}
found that both, the local photospheric field
as well as the overlying global coronal field, determine the location of open
field regions. The rotation of the CH, however, is then controlled by
the bipolar magnetic region with which it co-rotates
(see Figure~\ref{wang_1996_fig4}). During this process
the coronal magnetic field has to change its topology by
magnetic reconnection from closed to open, and vice versa when
entering and leaving the CH.
By definition, magnetic reconnection is prohibited in potential field models,
as is the breaking down of the frozen-in condition of the
high-conductivity coronal plasma.

Therefore, physically more advanced approaches, able to
investigate the nature of the
magnetic reconnection involved in the development of CHs, were carried
out by \cite{2005ApJ...625..463L}. They found that while the magnetic
structure of a CH might be relatively stable, magnetic flux may permanently
move in or out
of the region covered by the CH, causing reconnection to
happen. The transitions from closed to open magnetic fields at the boundaries
of CHs were suspected to be the origin of the slow
(with velocities of $\lesssim450$~km\,s$^{-1}$) solar wind
\citep[][]{1996Sci...271..464W}.
% A similar scenario as described above was supported in a recent review by
% \cite{2009SSRv..144..383W} who pointed out that the open flux
% originates from ARs but becomes redistributed to eventually form PCHs.
Interchange reconnection between open and closed fields at the
boundary of CHs are thought to be responsible for, or at least related to
coronal jets and polar plumes.

Recently, \cite{2013ApJ...775..100P} used synoptic magnetograms
from GONG to approximate the corona with a potential-field model
and compared it to synoptic EUV maps from \stereo. They investigated
the relationship between decaying ARs and CHs at low latitudes.
Newly emerging ARs generally result in significant changes of the
global coronal magnetic-field structure and cause the re-shaping of
the streamer belt. Decaying ARs which were found
to evolve steadily and gradually, however, would not yield a considerable
change of the global field, even though they left behind a considerable flux
imbalance. The authors argued that some of the imbalanced flux,
nevertheless needs to (re)connect elsewhere and thus to form the
streamer belt or the open fields.

\subsubsection{Comparison with quiet-Sun magnetic fields}
\label{ECH:QS}

\begin{figure}
      \centering
      \includegraphics[width=\textwidth]{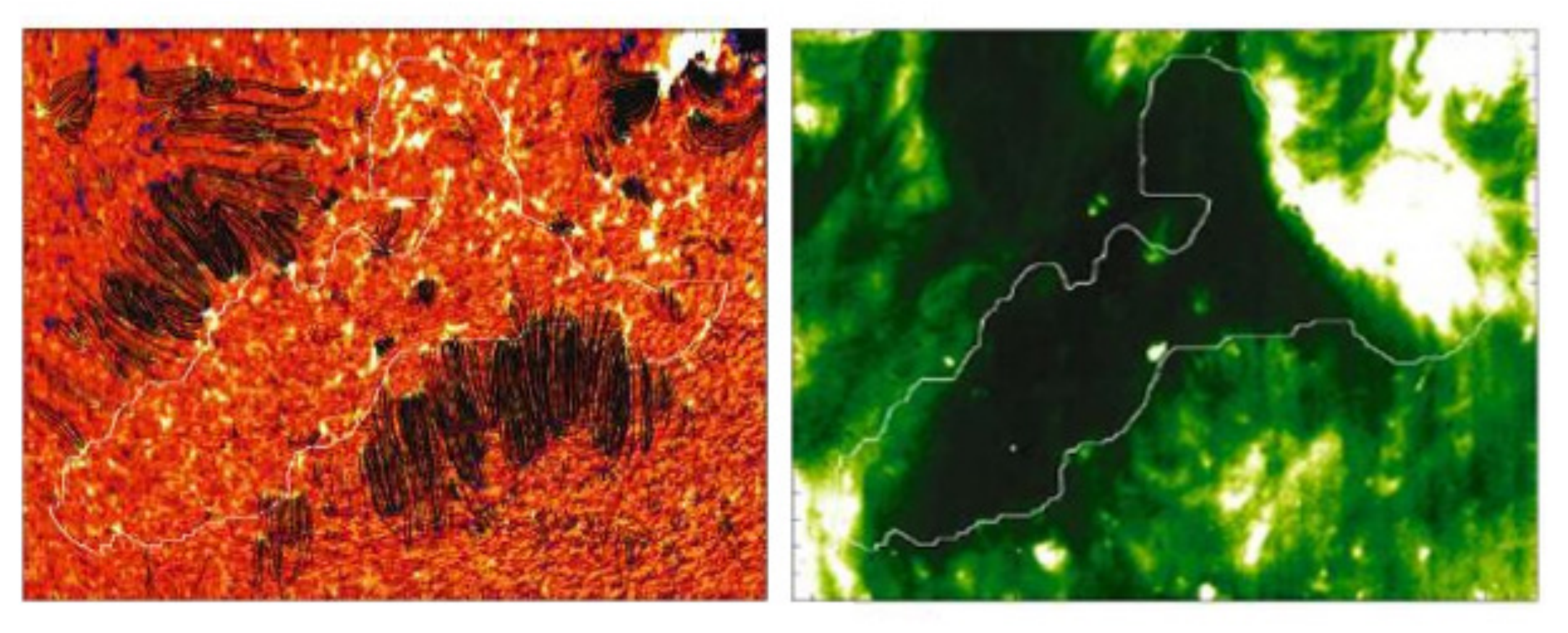}
      \put(-330,120){\bf\sf\white(a)}
      \put(-160,120){\bf\sf\white(b)}
      \caption{(a) \soho/MDI magnetogram and overlaid closed magnetic field lines
      (black) obtained from a potential field model. Only closed
      field lines with a field strength of $\ge$\,20~G are shown. One can see that
      outside of the CH boundary, the number of of closed loops is significantly
      higher and that these loops are also longer. (b) Co-spatial
      EUV image from \soho/EIT at 195~\AA. The boundary of an ECH is outlined
      by a white contour in both images. The CH area is seen as a region of reduced
      emissivity.
      (Adapted from Figure~3 of \citep{wiegelmann:etal04}. With kind permission from Springer Science and Business Media.)}
      \label{fig:twsks04_fig3}
\end{figure}

The local, small-scale magnetic field structure of ECHs in the upper solar atmosphere
has been investigated by \cite{wiegelmann:etal04} under the presumption that
electromagnetic radiation from coronal plasma mainly originates
from closed magnetic loops, while open field
lines remain almost invisible owing to their strongly reduced plasma density.
A statistical study using a potential field model based on \soho/MDI magnetograms
revealed that only small closed magnetic loops exist within the CH.
The apex of these loops
does not reach coronal heights. Above a certain height, all field lines
were found to be open since the minority flux has been cancelled out.
In contrast, quiet-Sun areas are more or less flux-balanced and thus contain
field lines of various apex heights, including numerous field lines
reaching up into the corona (see Figure~\ref{fig:twsks04_fig3}).
For corresponding investigations
of PCHs see section \ref{PCH:QS} and Figure \ref{fig:Ito10_fig11}.
\cite{2011ApJ...726...49Y} compared the magnetic field vector field in two
ECHs and the QS using \hinode/SOT-SP
data. They found that horizontal magnetic fields, inclination angels as well
as the current density and current helicity are stronger in CHs
than in the QS. The authors also concluded that the magnetic field in both
QS and CH is non-potential (see also section \ref{qs_forcefree}).

\subsubsection{Contribution to the solar wind}

\begin{figure}
      \centering
      \includegraphics[width=0.9\textwidth]{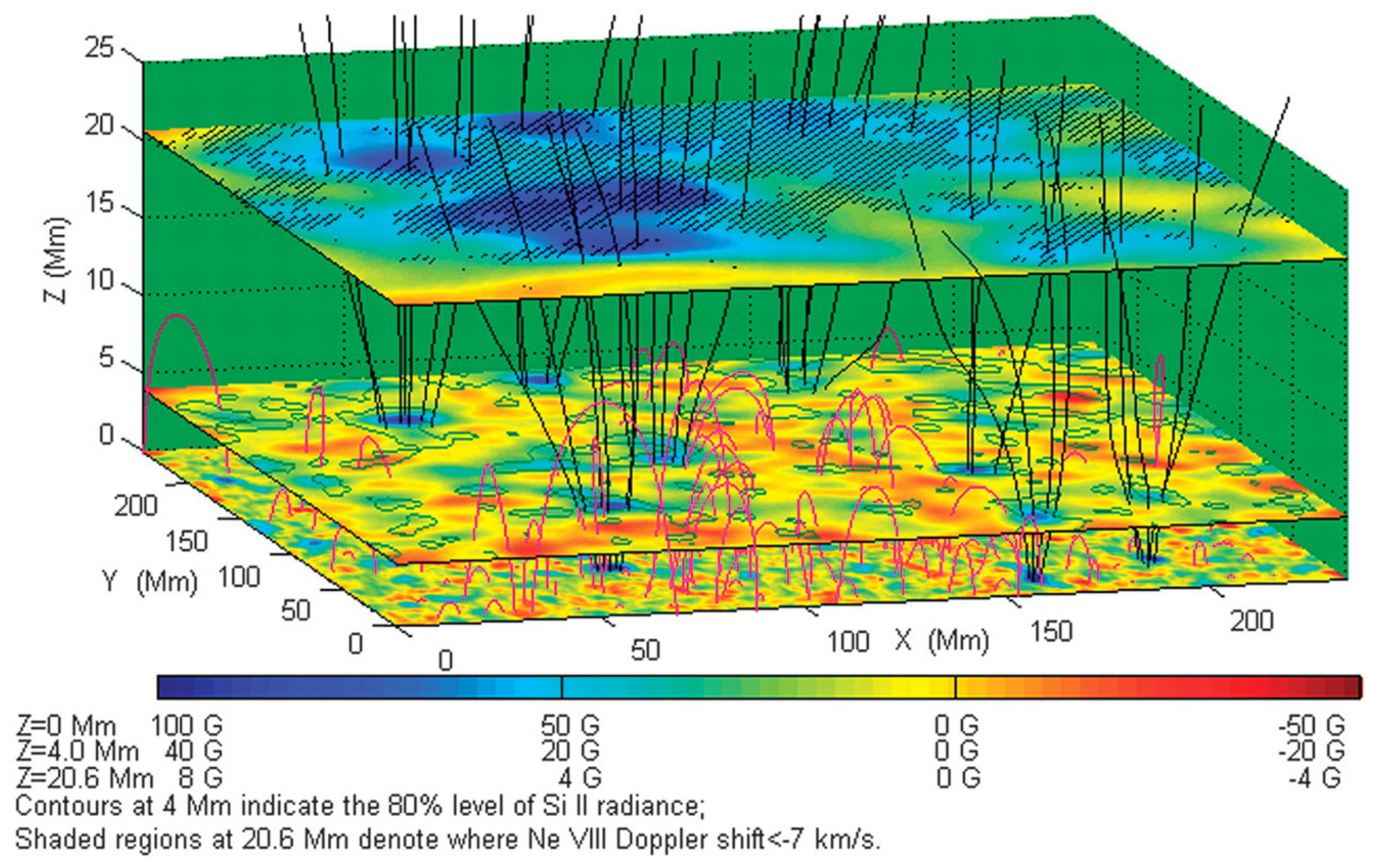}
      \caption{The linear force-free coronal magnetic field model
      shows open (black) and closed (red) field lines. Because of the
      decreasing magnetic field strength, the inserted magnetic maps
      at 0~Mm, 4~Mm and 20.6~Mm correspond to a different color bar.
      At the 4~Mm and 20.6~Mm level the authors made a correlation
      analysis of the extrapolated magnetic field Bz and Doppler
      maps from \soho/SUMER. The shaded area at 20.6~Mm corresponds
      to a detected outflow faster than 7~km\,s$^{-1}$.
      (Figure~5 of \citealt{tu:etal05}. Reprinted with permission from AAAS.)}
      \label{tu2005_fig4}
\end{figure}

\cite{1999Sci...283..810H}, using Doppler maps from \soho/SUMER,
investigated the relationship between the chromospheric magnetic network
and plasma outflows. They showed that the solar wind is rooted at the
boundaries of the cells of the magnetic network.
\cite{tu:etal05} went a step further and
combined data from the same instrument with
a potential coronal magnetic field model based on \soho/MDI data.
A correlation analysis of the modelled field structure, of observed
Doppler-velocity and radiance maps revealed an acceleration of the
solar wind at heights between five and 20~Mm above the Sun's surface,
and originating from coronal funnels (Figure~\ref{tu2005_fig4}).
\cite{2011ApJ...732....4J}, using \hinode/SOT-SP data were able to support
the scenario described by \cite{tu:etal05}, namely that the solar wind streams along
magnetic funnels and the magnetic reconnection of open and closed fields
might provide energy for the acceleration of the particles.

\section{Conclusion and Outlook}
\label{s:summary}

Within this paper, we aimed to review our current understanding of the role
of magnetic fields for the physics of the solar atmosphere,
in particular the corona. In the past years, ground-based and space-borne
instruments have delivered data of unprecedented spatial and temporal resolution.
Together with the ever increasing sophistication of numerical techniques,
this led to new insights into the nature and the dynamic
evolution of the coronal magnetic field. A broad range of spatial scales
is involved in coronal processes: from very small scales,
at which magnetic reconnection may locally reconfigure the magnetic
field, to very large scales, at which the long-term recycling of the
global magnetic field takes place.

There are a number of techniques available for the measurement of the coronal
magnetic field, but none has found widespread and regular use.
Observations at radio wavelengths have provided by far the most
direct measurements of coronal
magnetic fields, augmented by polarization measurements in coronal lines
in the infrared for a relatively small number of cases.
More frequent are observations using magnetically sensitive
chromospheric lines in the infrared, which can sample layers close to the coronal base.
Current limitations on the spatial and temporal
resolution at radio frequencies are expected to be partly lifted
once FASR starts operating. More regular observations of the
coronal magnetic field vector are also envisaged with
the NSO/DKIST telescope. A more indirect approach is taken by coronal
seismology, which permits the magnetic field strength
to be derived from the analysis of oscillating loops.

Nevertheless, by far the
most accurate routine measurements are of the magnetic field in
the photosphere, ranging from global scales with SDO to high resolution
measurements with \sunrise.

The instabilities that might trigger magnetic reconnection are still only
accessible through 3D numerical MHD models. This is also because even the
present imaging and
spectropolarimetric instrumentation with the highest spatial resolution (\eg,
\sunrise\ and Hi-C), might still be far from resolving the relevant spatial
scales.
Besides, high spatial resolution often comes with a
restricted FOV.
Such restrictions are less severe for the analysis of processes
observed on larger active-region scales. Coronal imaging data of
sufficiently high spatial resolution and temporal cadence, in the form of
\sdo/AIA images, have allowed detailed analysis of the coronal dynamics in
recent years. The interpretation of these observations has been increasingly
aided by static force-free magnetic field models which, at the same time,
were validated by the match with structures seen in the same coronal images.
The force-free modelling approach
is well justified in the corona above an active region owing to the low
plasma $\beta$ there, but less suitable to model
the quiet Sun, where $\beta$ is not low and non-magnetic forces
have to be taken into account, e.g., with magneto-static and MHD-models.

For studying the evolution of coronal magnetic fields a strong tool
is the combination of flux-transport models and extrapolation techniques.
Complementary stereoscopic and
tomographic methods significantly improved the ability to reconstruct
the 3D structure of coronal loops by using simultaneously
observed 2D images
from a number of vantage points, such as \soho/EIT, the \stereo\
spacecrafts, \sdo/AIA, and in the future also \so.

A promising approach is to combine extrapolations from photospheric fields
and stereoscopy within one model. Such attempts
are still in their infancy though and
currently non-linear force-free extrapolations of
the photospheric field vector into the corona are the state-of-the-art.
The maturing of such modelling tools for the interpretation of coronal dynamics
has mainly become possible due to the development and operation
of advanced instruments, such as \solis/VSM, \hinode/SOT-SP
and \sdo/HMI that deliver vector magnetic
field maps regularly for the entire solar disk and active-region scans
with both high time cadence and high spatial resolution.
In addition, forward MHS and MHD
modelling techniques proved their strength in the ability to compute
synthetic spectra characteristic of ARs which quantitatively reproduced
spectra recorded with, \eg, \soho/SUMER.
Recent milestones using one or the other tools individually or combination with
each other, were discussed throughout this review. Besides having shed light on
some of the hidden dynamics of the magnetic field,
indirectly accessible to us only by the analysis
of coronal images, the scientific outcome in recent years naturally also
changed the course of the field and uncovered and/or strengthened the
importance of answering still open questions.

On small granulation scales, the permanent restructuring of line-tied,
braided and/or twisted active-region magnetic fields
has been a strong contender for providing a
contribution to the heating of the coronal plasma to the observed
temperatures. Its contribution to the release of parts of the vast amount of
energy stored in the coronal active-region magnetic fields, however, is still
to be determined in detail.
On active-region scales, dynamic phenomena such as flares and
CMEs received great attention and have revealed some of their secrets. We are
increasingly able to relate observed flare-associated energetics to specific
magnetic field structures, both in space and in time. The ability to predict
eruptive phenomena, however, is still in its infancy. Detailed analysis of
the eruption-associated magnetic field topology and its evolution is
highlighting
the complex interplay between photospheric driver and coronal response,
underlining that the task of attempting any forecasting is a challenging one.
Similarly demanding is the prediction of the long-term behaviour of activity and
even more so since the importance of local eruptive processes for the cyclic global
restructuring remains unclear.

Finally, the findings of recent
years showed us that the magnetic structure of the solar corona, even during
times of low solar activity when the global magnetic field is thought to be
best represented by a ``simple'' dipole field, is rather complex. Consequently, much
work is still needed to unravel the nature of the Sun's magnetic
field, on local as well as on global scales, the latter being important in
order to understand its contribution to the solar wind and its effects on our
interplanetary environment.

\begin{acknowledgements}
We thank all colleagues who have provided us with permission to reprint
their figures and have helped with discussions and comments.
This work was supported by the grants
Deutsches Zentrum f\"ur Luft- und Raumfahrt (DLR): 50 OC 0904,
Deutsche Forschungsgemeinschaft (DFG): WI 3211/2-1,
Fonds zur F\"orderung der wissenschaftlichen Forschung (FWF): P25383-N27,
and by the BK21 plus program
through the National Research Foundation (NRF) funded
by the Ministry of Education of Korea.
\end{acknowledgements}
\appendix
\section{Abbreviations}\label{app:abbreviations}

\begin{tabbing}
      \hspace{3.5cm} \= \hspace{10cm} \kill
      AR(s)                   \> Active Region(s) \\
      CH(s)                   \> Coronal Hole(s) \\
      CME(s)                  \> Coronal Mass Ejection(s) \\
      ECH(s)                  \> Equatorial Coronal Hole(s) \\
      EUV                            \> Extreme Ultraviolet) \\
       FOV                     \> Field-Of-View \\
      FASR                          \> Frequency Agile Solar Radiotelescope \citep{Gary:1338733}\\
      \goes                   \> Geostationary Operational Environmental Satellite \\
                                    \> ({\scriptsize\url{http://www.goes.noaa.gov/}})\\
      Hi-C                    \> High-resolution Coronal Imager \citep{gol_cir_06,cir_gol_13} \\
      \hinode                 \> \solb\ \citep{2005JKAS...38..307I} \\
      \hinode/EIS       \> \hinode\ EUV Imaging Spectrometer \citep{cul_har_07}\\
      \hinode/SOT \> \hinode\ Solar Optical Telescope \citep{2008SoPh..249..197S} \\
      \hinode/SOT-SP \> \hinode\ SOT Spectro-Polarimeter \citep{tsu_08,lit_aki_13} \\
       KPNO                   \> Kitt Peak National Observatory (\scriptsize\url{http://www.noao.edu/kpno/}\\
      LFF                           \>  Linear Force-Free\\
      LOS                     \> Line-Of-Sight \\
      MC(s)                   \> Magnetic Cloud(s) \\
      MHD                     \> Magneto-Hydro-Dynamic \\
      M(H)S                     \> Magneto-(Hydro-)Static \\
      \nso                    \> National Solar Observatory (\scriptsize\url{http://www.nso.edu/})\\
      \nso/DIKST              \> \nso\ Daniel K.\ Inouye Solar Telescope (\scriptsize\url{http://atst.nso.edu/})\\
      NLFF                          \>  Nonlinear Force-Free\\
      PCH(s)                  \> Polar Coronal Hole(s) \\
      PIL(s)                     \> Polarity Inversion Line(s) \\
      PFSS                    \> Potential Field Source Surface \\
      QSL(s)                  \> Quasi-Separatrix Layer(s) \\
      QS                      \> Quiet Sun \\
      \sdo                    \> Solar Dynamics Observatory \citep{pes_tho_12} \\
      \sdo/AIA                \> \sdo\ Atmospheric Imaging Assembly \citep{lem_tit_12} \\
      \sdo/HMI                \> \sdo\ Helioseismic and Magnetic Imager \citep{scho_sche_12} \\
      \soho                   \> Solar and Heliospheric Observatory \citep{sche_95} \\
      \soho/EIT         \> \soho\ Extreme ultraviolet Imaging Telescope (Delaboudini\a'{e}re et al. 1995) \\
      \soho/LASCO \> \soho\ Large Angle and Spectrometric Coronagraph \citep{brue_how_95}\\
      \soho/MDI         \> \soho\ Michelson Doppler Imager \citep{sche_95} \\
      \soho/SUMER \> \soho\ Solar Ultraviolet Measurements of Emitted Radiation \\
                                    \> \citep{wil_cur_95} \\
      \solis                        \>  Solar Optical Long-term Investigations of the Sun \citep{kel_har_03a} \\
      \solis/VSM        \> \solis\ Vector SpectroMagnetograph \citep{kel_har_03b}\\
      \so                           \>  \citep{2013SoPh..285...25M}\\
      \stereo                 \> Solar-TErrestrial RElations Observatory  \citep{kai_kuc_08} \\
      \stereo/SECCHI \> \stereo\ Sun Earth Connection Coronal and Heliospheric Investigation\\
                                          \>    \citep{how_mos_02}\\
      \stereo/SECCHI-COR1(2) \> \stereo/SECCHI\ inner (outer) CORonograph \citep{liu_luh_09}\\
      \stereo/SECCHI-EUVI     \> \stereo/SECCHI\ Extreme UltraViolet Imager \citep{2004SPIE.5171..111W} \\
      \sunrise                \> \solc\ \citep{sol_bar_10,bar_gan_11} \\
       SXR(s)                            \>  Soft X-ray(s)\\
      \sunrise/IMaX     \> \sunrise\ Imaging Magnetograph eXperiment ({Mart{\a'{i}}nez Pillet} et al. 2011)\\
      TEL(s)                     \> Trans-Equatorial Loop(s) \\
      \trace                        \> Transition Region and Coronal Explorer \citep{han_act_99} \\
      VTT                                 \> Vacuum Tower Telescope {\scriptsize\url{http://www.kis.uni-freiburg.de/}}\\
      \wind/WAVES                   \> \citep{1995SSRv...71..231B}\\
      \yohkoh                 \> \sola\ \citep{tsu_91} \\
      \yohkoh/SXT       \> \yohkoh\ Soft X-ray Telescope \citep{tsu_91}
\end{tabbing}

\addcontentsline{toc}{section}{References}
%\bibliographystyle{named}
%\bibliographystyle{abbrvnat}
%\bibliography{AARev_bib}

\end{document}